\documentclass{jaa}
\usepackage{graphicx, url, hyperref, natbib, subcaption}
\usepackage{caption}
\usepackage{booktabs}
\usepackage{xcolor}
\usepackage[figuresright]{rotating}

\begin{document}

\title{Structure and Evolution of Multi-Cluster within Galactic Disc: Gaia DR3 Insights into Eight Open Clusters}

\author{A. Ahmed\textsuperscript{1,*}, W. H. Elsanhoury\textsuperscript{2}, D. C. Çınar\textsuperscript{3}, A. A. Haroon\textsuperscript{4,5}, M. S. Alenazi\textsuperscript{2} and E. A. Elkholy\textsuperscript{2,5}}

\affilOne{\textsuperscript{1}Astronomy, Space Science and Meteorology
Department, Faculty of Science, Cairo University, Giza, 12613, Egypt.\\}
\affilTwo{\textsuperscript{2}Department of Physics, College of Science, Northern Border University, Arar, Saudi Arabia.\\}
\affilThree{\textsuperscript{3}Institute of Graduate Studies in Science,
Programme of Astronomy and Space Sciences, Istanbul University, Beyazıt, Istanbul 34116, Turkey.\\}
\affilFive{\textsuperscript{4}Astronomy and Space Science Department, Faculty of Science, King Abdulaziz
University, Jeddah, Saudi Arabia.\\}
\affilSix{\textsuperscript{5}Astronomy Department, National Research
Institute of Astronomy and Geophysics
(NRIAG), Helwan 11421, Cairo, Egypt}
\twocolumn[{
\maketitle

\corres{ahamza@sci.cu.edu.eg}

\msinfo{20 Nov 2025}{08 Feb 2026}

\begin{abstract}
In this study, we present a comprehensive analysis of the structural, astrophysical, and dynamical properties of eight open clusters: NGC 559, NGC 1817, NGC 2141, NGC 7245, Ruprecht 15, Ruprecht 137, Ruprecht 142, and Ruprecht 169, utilizing the precise astrometric and photometric data from $Gaia$ Data Release 3. By fitting King's model to the radial density profiles, we determined their structural parameters, including core and cluster limiting radii, which were found to be in the ranges of 3.07 -- 16.21 arcmin and 9.97 -- 25.97 arcmin, respectively. The fundamental astrophysical parameters were derived by fitting {\sc PARSEC} isochrones to the colour-magnitude diagrams. Our analysis indicates that the clusters' logarithmic ages span from 7.95 to 9.34 years, with metallicities ranging between 0.007 -- 0.015  and heliocentric distances from 1640 to 5203 pc. The total stellar masses for the clusters were calculated to be within the range of 257 -- 1916 $M_{\odot}$. The slopes of the mass function for most clusters were found to be consistent with the Salpeter initial mass function. Furthermore, our dynamical analysis suggests that all investigated clusters except Ruprecht 15 show signs of being dynamically relaxed. Notably, the spatial distribution and bimodal structure observed in the radial density profile of NGC 7245 provide strong evidence that it is a binary cluster candidate. Finally, our kinematic analysis and orbit integration show that the clusters possess dynamical properties that are fully consistent with their membership in the Galactic thin disc.
\end{abstract}

}]


\doinum{12.3456/s78910-011-012-3}
\artcitid{\#\#\#\#}
\volnum{000}
\year{0000}
\pgrange{1--}
\setcounter{page}{1}
\lp{1}


\section{Introduction}

\label{introduction}

Open clusters (OCs) are fundamental building blocks of the Galactic disc, serving as reliable tracers of Galactic structure and chemical enrichment. Also, the study of OCs plays a crucial role in constraining and testing the current models for stellar formation and evolution. It also helps to provide answers for the current debates, such as the predicted mass distributions in the clusters and the observed mass segregation in many stellar clusters. 
The availability of high-precision astrometric and photometric data from the $Gaia$ mission \citep{prusti2016gaia} has enabled a renewed exploration of both well-studied and poorly studied OCs, offering a unique opportunity to investigate their properties homogeneously.
Due to the importance of getting a large database of the measured physical parameters of OCs, many catalogues have been published for thousands of OCs in our Milky Way (MW) Galaxy, e.g., \cite{kharchenko2013global,dias2014proper,sampedro2017multimembership}. However, a detailed investigation of individual clusters provides more careful treatment of these clusters and more precise estimates for their parameters.

A crucial step in any study of OCs is the identification of the true members of the clusters. The measured physical parameters of the clusters can be significantly affected by the exclusion of true members or by the addition of field stars. Consequently, a reliable statistical approach using the observed stellar proper motions and spatial coordinates was a necessity, which was started with the pioneer work of \citet{Vas58}, and was improved in the subsequent works of \cite{San71, Zha90, Bal98}.

This study presents a detailed photometric and dynamical study of eight open star clusters, NGC 559, NGC 1817, NGC 2141, NGC 7245, Ruprecht 137, Ruprecht 142, Ruprecht 15, and Ruprecht 169, using the most recent data of astrometric and photometric data of the $Gaia$ DR3 database \footnote{\url{https:://www.cosmos.esa.int/gaia}} \citep{vallenari2023gaia}. 

\begin{figure}
 \centering
 \includegraphics[width=0.99\linewidth]{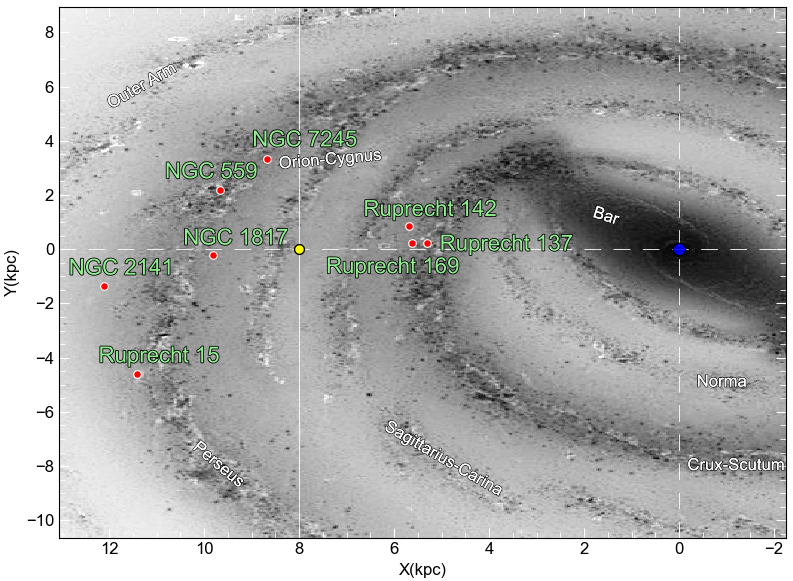}
 \caption{The spatial distribution of the eight target OCs (red dots) projected onto the Galactic plane. The positions of the Sun (yellow dot) and the Galactic Centre (blue dot) are marked for reference against a schematic representation of the MW's spiral arms and bar.}
 \label{fig:galaxy_map}
\end{figure}

By selecting clusters that span a range of ages, metallicities, and distances, our sample is not only diverse but also strategically positioned to trace the structure of the spiral arms of the MW. The distribution of the studied OCs on the Galactic plane is shown in Figure \ref{fig:galaxy_map}. For instance, OCs like NGC 559 and NGC 1817 are located near the Perseus arm, providing insight into recent star formation in this region, while clusters such as Ruprecht 137 and NGC 2141 probe the less studied outer arm environments. This spatial distribution allows us to investigate how cluster properties vary across different Galactic environments, linking local star formation histories to the large-scale spiral structure of the Galaxy. By framing the sample in this way, each cluster serves as a probe of the dynamical and chemical evolution within its respective arm, enabling a more comprehensive understanding of the Galactic structure and evolution.

These clusters have not yet been extensively studied using $Gaia$ astrometry, and a detailed kinematic and structural analysis using $Gaia$ DR3 data remains an important step toward refining their dynamical status and Galactic context.

\section{Targeted OCs and Literature Review}

\subsection{NGC 559}

NGC 559 is an intermediate-age OC located in the constellation Cassiopeia, in the direction of the Perseus arm of the MW at right ascension of $\rm 01^h 29^m 31^s$ and declination $63^{\circ} 18' 14''$, and serves as an important tracer for studying the structure and evolution of the Galactic disc. NGC 559 was the target of many previous photometric and astrometric analyses. \cite{Ann2002} showed that NGC 559 has an age of $\approx$ 400 $\pm$ 100 Myr, and exists at a heliocentric distance of 2.3 kpc with a relatively high line-of-sight reddening of $E(B-V)~\approx~$0.81~ $\pm$ ~0.05 mag and a metallicity of [Fe/H] = $-0.32$ dex.
The most recent study of \cite{Joshi2014} reported a younger age of approximately 224 $\pm$ 25 Myr, with a heliocentric distance of 2.43 $\pm$ 0.23 kpc and a line-of-sight reddening of $E(B-V)~\approx$~0.82 mag, which agrees with the results of \cite{Ann2002}. \cite{Joshi2014} study also indicated a slightly elevated extinction ratio ($R_V~\approx~3.5$) and clear evidence of mass segregation, suggesting that the cluster has reached a state of dynamical relaxation. 
Spectroscopic follow-up by \cite{Carrera2015} provided the first chemical abundance and kinematic measurements, deriving a mean radial velocity of -58.4 $\pm$ 6.8 $\rm km~s^{-1}$ and a metallicity of [Fe/H] $=-0.25~\pm 0.14$ dex, confirming that NGC 559 is mildly metal-poor relative to the Sun. More recently, \cite{Joshi2020} conducted a multi-epoch variability study of the cluster, identifying 70 variable stars, including 30 probable members, encompassing eclipsing binaries, pulsators, and rotational variables. 

\subsection{NGC 1817}

NGC 1817 is a populous, intermediate-age ($\approx$ 0.8–1.2 Gyr) OC located toward the Galactic anti-centre at right ascension of $\rm 05^h 12^m 15^s$ and declination $16^{\circ} 41' 36''$, with a heliocentric distance of approximately 1.5-2.0 kpc, with a reddening of $E(B-V)~\approx$~0.28 and subtle subsolar metallicity ranges between $-0.16$ and $-0.34$ dex \citep{Twarog97,Friel2002,Fri95,Balaguer2004, Salaris2004,Jacobson2009,Donati2014}. 
\cite{Mermilliod2003} confirmed membership for 39 red giants via radial velocity measurements and photometry, revealing a mean cluster velocity of 65.33 $\pm$ 0.09 $\rm km s^{-1}$, a high binary fraction ($\approx$~25 \%), extensive spatial extent (radius$\approx$ 27'), and notable mass segregation among red giant members. \cite{Jacobson2009} performed spectral analysis to get the cluster's metallicity and radial velocity, and their radial velocity measurement matches that of \cite{Mermilliod2003}, 65.2 $\pm$ 1 $\rm km s^{-1}$.
Time-series surveys further demonstrated that NGC 1817 hosts a wealth of pulsating stars—including 19 $\delta$ Scuti variables, several $\gamma$ Dor candidates, and eclipsing binaries, one of which also involves a $\delta$ Scuti pulsator \citep{Arentoft2004,Arentoft2005}. More recently, high-precision observations from the Kepler/K2 mission identified 44 red clump stars (29 with reliable asteroseismic measurements), suppressed dipole modes in some giants, and over 60 main sequence pulsators including $\delta$ Scuti, $\gamma$ Dor, and hybrid stars, highlighting the cluster's value for stellar physics and evolving red clump structure studies \citep{Sandquist2020}. Broadband polarimetric analysis by \cite{Singh2020} measured an average polarization degree $\rm P_{max}\approx~0.93\%$ and presented dust grain size consistent with the general interstellar medium along the sightline ($R_V~\approx$~3.0). Together, these results establish NGC 1817 as a compelling laboratory for exploring intermediate-age stellar evolution, pulsation phenomena, binary demographics, and interstellar medium properties in the Galactic disc.

\subsection{NGC 2141}

NGC 2141 is a rich and relatively old OC located in the anti-centre direction of the Galactic disc at right ascension of $\rm 06^h 02^m 55^s$ and declination $10^{\circ} 26' 48''$, offering a valuable target for studying the chemical and dynamical evolution of intermediate-age stellar populations. Photometric studies, including deep optical and near-infrared observations, have estimated the cluster’s age at approximately 2.5 Gyr, with a heliocentric distance ranges between 3.8 and 4.5 kpc, a reddening of $E(B-V)~\approx$~0.40 mag, subsolar metallicity ranges between $-0.26$ and $-0.43$ dex and clear evidence of mass segregation in its luminosity function, suggesting significant internal dynamical evolution \citep{Rosvick1995,Twarog97,Carraro2001,Salaris2004,Donati2014}, while \cite{Donati2014} estimated a lower age in the range of 1.25 and 1.9 Gyr which depends on the adopted metallicity value. 
On the contrary, high-resolution spectroscopic analysis by \cite{Jacobson2009} revealed that NGC 2141 has a near-solar metallicity ([Fe/H] $=~0.00~\pm ~0.16$ dex), along with light-element abundance patterns (e.g., enhanced Na and Si, depleted O) similar to those observed in other OCs of comparable age. 
More recently, a time-series photometric survey by \cite{Luo2015} identified 10 variable stars in the cluster field, including eclipsing binaries (EA, EB, and W UMa types), an RR Lyrae star, and a candidate blue straggler star (BSS). These variables were evaluated for membership based on their positions in the colour–magnitude diagram (CMD) and spatial distribution. 

\begin{table*}[!ht]
\centering
\caption{Comparison of the estimated parameters in the literature for the investigated clusters:($\alpha$, $ \delta$; sexagesimal), radius ($r$; arcmin), age (Myr), distance ($d$; pc), colour excess ($E(B - V)$; mag), trigonometric parallax ($\varpi$; mas), proper motion components ($\mu_\alpha \cos \delta,~\mu_\delta$; mas yr\textsuperscript{-1}), the number of member stars (N), metallicitiy ([Fe/H]; dex) and the references.
\label{tab:literature}} 
\setlength\tabcolsep{3pt}
\scalebox{0.7}{
\begin{tabular}
{cccccccccccc}
\hline
$\alpha $ & $\delta$ & $r$ & $Age$ & $d$ &\it E(B-V) & $ \varpi$ & $ \mu _\alpha \cos \delta$ & $ \mu _\delta$ & $N$ & [Fe/H] & Ref. \\

(hh:mm:ss) & (dd:mm:ss) & (arcmin) & (Myr) & (kpc) & (mag) & (mas) & (mas $\rm yr^{-1}$) &(mas $\rm yr^{-1}$)& (stars) &(dex) &  \\
\hline\hline
\multicolumn{11}{c}{NGC 559}\\
\hline
01 29 33 & +63 18 14 & 35.60 & 348 & 2.884 & --& 0.334 $\pm$ 0.048 & $-$4.276 $\pm$ 0.077 & 0.191 $\pm$ 0.088& 887 & -- &1 \\
01 29 33 & +63 18 04 & 4.86$^*$ & 257 & 2.884 & --& 0.331 $\pm$ 0.045 & $-$4.277 $\pm$ 0.074 & 0.187 $\pm$ 0.083& 522 & -- &2\\
01 29 33 & +63 18 04 & 4.86$^*$ & -- & 2.852 &--& 0.325 $\pm$ 0.060 &$-$4.258 $\pm$ 0.113 & 0.254 $\pm$ 0.130 & 542 & -- &3 \\
01 29 33 & +63 18 04 & 4.86$^*$ & 257 & 2.884 &--& 0.325 $\pm$ 0.060 & $-$4.258 $\pm$ 0.113 & 0.254 $\pm$ 0.130 & 529 & -- &4 \\
01 29 32 & +63 18 11 & 8.67 & 224& 2.430 &0.82 &--& $-$4.950 $\pm$ 4.340 & 1.600 $\pm$ 3.480 & 182 & -- &5 \\
01 29 35 & +63 18 14 & 6.20 & -- & -- & -- &--& $-$4.580 $\pm$ 2.950 & 1.410 $\pm$ 1.460 & 309 &$-$0.03 $\pm$ 0.08 &6 \\
01 29 31 & +63 18 07 & 18.00 & 631 & 2.200 & 0.60 &--& $-$2.950 & $-$1.070 & -- & -- &7 \\
\hline\hline
\multicolumn{11}{c}{NGC 1817}\\
\hline
05 12 34 & +16 41 35 & 43.33 & 1857 & 1.640 & --& 0.573 $\pm$ 0.049 & 0.430 $\pm$ 0.102 & $-$0.932 $\pm$ 0.093 & 718 & -- &1 \\
05 12 33 & +16 41 46 & 11.22$^*$ & 1122 & 1.799 & --& 0.574 $\pm$ 0.036 & 0.425 $\pm$ 0.075 & $-$0.9350 $\pm$ 0.066 & 412 & -- &2\\
05 12 33 & +16 41 46 & 11.22$^*$ & -- & 1.733 &--& 0.551 $\pm$ 0.056& 0.485 $\pm$ 0.118 &$-$0.890 $\pm$ 0.100 & 460 & -- &3 \\
05 12 33 & +16 41 46 & 11.22$^*$ & 1122 & 1.799 & -- &0.551 $\pm$ 0.056& 0.485 $\pm$ 0.118 &$-$0.890 $\pm$ 0.100 & 413 & -- &4 \\
05 12 15 & +16 41 24 & 12.28 & 407 & 1.972 &0.33 &--& $-$1.700 $\pm$ 4.130 &$-$2.470 $\pm$ 3.600 & 407 & -- &5 \\
05 12 15 & +16 41 24 & 9.00 & -- & -- & -- &--& $-$1.790 $\pm$ 2.000 & $-$2.170 $\pm$ 1.730 & 308 & $-$0.10 $\pm$ 0.02 &6 \\
05 12 20 & +16 40 48 & 15.60 & 794 & 1.504 & 0.35 &--& 1.000 & $-$1.600 & -- & -- &7 \\
\hline\hline
\multicolumn{11}{c}{NGC 2141}\\
\hline
06 02 59 & +10 27 22 & 21.23 & 2783 & 4.087 & --& 0.203 $\pm$ 0.080 & $-$0.083 $\pm$ 0.102 & $-$0.7545 $\pm$ 0.086 & 1324 & -- &1 \\
06 02 56 & +10 27 04 & 4.38$^*$ & 1862 & 5.183 & --& 0.204 $\pm$ 0.075 & $-$0.086 $\pm$ 0.095 & $-$0.754 $\pm$ 0.078 & 729 & -- &2\\
06 02 56 & +10 27 04 & 4.38$^*$ & -- & 4.533 &--& 0.196 $\pm$ 0.116 &$-$0.028 $\pm$ 0.167 & $-$0.767 $\pm$ 0.167 & 831 & -- &3 \\
06 02 56 & +10 27 04 & 4.38$^*$ & 1862 & 5.183 &--& 0.196 $\pm$ 0.116 &$-$0.028 $\pm$ 0.167 & $-$0.767 $\pm$ 0.167 & 729 & -- &4 \\
06 02 55 & +10 26 49 & 8.75 & 1698 & 4.033 & 0.25 &--& $-$0.550 $\pm$ 6.420 & $-$5.470 $\pm$ 6.130 & 506 & -- &5 \\
06 02 55 & +10 26 49 & 6.00 & -- & -- & -- &--& 0.030 $\pm$ 1.480 & $-$4.820 $\pm$ 1.650 & 385 & ~0.30 $\pm$ 0.06 &6 \\
06 02 54 & +10 27 36 & 11.40 & 1758 & 4.364 & 0.31 &--& $-$0.850 & $-$0.780 &--& $-$0.18 $\pm$ 0.15 &7 \\
\hline\hline
\multicolumn{11}{c}{NGC 7245}\\
\hline
22 15 14 & +54 20 20 & 9.14 & 623.2 & 3.352 & --& 0.273 $\pm$ 0.027 & $-$3.942 $\pm$ 0.058 & $-$3.281 $\pm$ 0.059& 241 & -- &1 \\
22 15 15 & +54 20 10 & 3.66$^*$ & 603 & 3.210 &--&0.279 $\pm$ 0.026 & $-$3.948 $\pm$ 0.051 & $-$3.290 $\pm$ 0.054 &225 & -- &2 \\
22 15 15 & +54 20 10 & 3.66$^*$ & -- & 3.342 &--& 0.273 $\pm$ 0.052 & $-$3.952 $\pm$ 0.101 & $-$3.249 $\pm$ 0.095 & 267 & -- &3 \\
22 15 15 & +54 20 10 & 3.66$^*$ & 603 & 3.210 & 0.08 &0.273 $\pm$ 0.052 & $-$3.952 $\pm$ 0.101 & $-$3.249 $\pm$ 0.095 & 230 & -- &4 \\
22 15 11 & +54 20 35 & 3.52 & 447 & 3.467 & 0.45 &--& $-$2.990 $\pm$ 6.550 & $-$1.950 $\pm$ 7.350 & 108 & -- &5 \\
22 15 11 & +54 20 35 & 4.50 & -- & -- & -- &--& $-$1.980 $\pm$ 3.580 & $-$1.760 $\pm$ 3.190 & 191 & -0.01 $\pm$ 0.01&6\\
22 15 14 & +54 19 59 & 9.60 & 355 & 4.000 & 0.48 &--& $-$5.620 & $-$3.200 & -- &-- &7 \\
\hline\hline
\multicolumn{11}{c}{Ruprecht 15}\\
\hline
07 19 38 & $-$19 37 34 & 2.20 & 500 & 1.845 & 0.65 &--& $-$0.840 $\pm$ 0.100 &6.700 $\pm$ 0.090 & 265 & -- &8 \\
\hline\hline
\multicolumn{11}{c}{Ruprecht 137}\\
\hline
18 00 13 & $-$25 06 32 & 13.04 & 429 & 2.162 & --& 0.428 $\pm$ 0.011 & 1.286 $\pm$ 0.058 & $-$1.918 $\pm$ 0.074 & 14 & -- &1 \\
18 00 16 & $-$25 13 41 & 4.96 & 794 & 1.450 & 0.67 &--& 0.140 $\pm$ 5.930 &$-$2.030 $\pm$ 6.050 & 306 & --& 5 \\
18 00 16 & $-$25 13 41 & 3.80 & -- & -- & -- &--& $-$0.070 $\pm$ 3.010 & $-$2.390 $\pm$ 2.400 & 262 & -- & 6 \\
17 59 54 & $-$25 12 54 & 9.60 & 624 & 1.405 & 1.12 &--& 0.450 & $-$1.890 & -- &--& 7 \\
18 00 16 & $-$25 13 41 & 2.80 & 800 & 1.450 & 0.67 & -- & -- & -- & -- &--& 9 \\
\hline\hline
\multicolumn{11}{c}{Ruprecht 142}\\
\hline
18 32 11 & $-$12 13 48 & 4.00 & 398 & 1.735 & 0.91 &--& $-$1.720 $\pm$ 8.040 &$-$0.680 $\pm$ 7.220 & 197 & --& 5 \\
18 32 11 & $-$12 13 48 & 4.30 & -- & -- & -- &--& $-$0.810 $\pm$ 3.810 & $-$1.080 $\pm$ 2.340 & 222 & --& 6\\
18 32 13 & $-$12 12 58 & 6.00 & 339 & 1.802 & 1.00 &--& $-$2.880 & $-$5.670 & -- & --& 7 \\
18 32 11 & $-$12 13 48 & 3.30 & 400 & 1.735 & 0.91 & -- & -- & -- & --& -- & 9\\
\hline\hline
\multicolumn{11}{c}{Ruprecht 169}\\
\hline
17 59 22 & $-$24 46 01 & 4.56 & 1000 & 1.390 & 0.66 & --& $-$1.700 $\pm$ 7.080 & $-$3.040 $\pm$ 7.160 & 276 & --& 5 \\
17 59 22 & $-$24 46 01 & 3.60 & -- & -- & -- &--& $-$1.340 $\pm$ 3.070 & $-$2.600 $\pm$ 3.930 & 219 & --& 6 \\
17 59 22 & $-$24 46 01 & 2.60 & 1000 & 1.390 & 0.66 & -- & -- & -- & -- & --& 9 \\
\hline 
\end{tabular}}
\\
\noindent{\footnotesize %
Ref.: (1) \cite{Hunt2024}; (2) \cite{Pog21}; (3) \cite{Cant20a}; (4) \cite{Cant20b}; (5) \cite{sampedro2017multimembership}; (6) \cite{dias2014proper}; (7) \cite{kharchenko2013global}; (8) \cite{Tadross2012}; (9) \cite{Tadross2008}.
\par 
$^*$ These values are $r_{50}$ or the radii containing half the members.
}
\end{table*}

\subsection{NGC 7245}

NGC 7245 is a relatively young OC situated in the outer regions of the Galactic disc at right ascension of $\rm 22^h 15^m 08^s$ and declination $54^{\circ} 20' 36''$, and although it has been the subject of several photometric studies, it remains relatively under-explored in the context of modern astrometric and spectroscopic surveys. Early CCD photometric analyses by \cite{Viskum1997} and \cite{Subramaniam2007} estimated its fundamental parameters, placing the cluster at a heliocentric distance of 2.8-3.8 kpc, with an age of 320-400 Myr and a reddening of $E(B-V)~\approx$~0.40 mag. 
These values were further supported by the work of \cite{Glushkova2010} and \cite{Kharchenko2005}, who refined cluster distances and extinction estimates using deeper photometry and compiled catalogues. \cite{Carrera2015} reported a mean radial velocity of $V_r = -$ 65.3 $\pm$ 3.2 $\rm km s^{-1}$ and a metallicity of [Fe/H] $\approx~-0.15~\pm~0.18$ dex, representing the first spectroscopic confirmation of the cluster’s kinematic and chemical properties. 

\subsection{Ruprecht 15}

Ruprecht 15 is a sparsely studied OC located near the Galactic plane, existing at right ascension of $\rm 07^h 19^m 38^s$ and declination $-19^{\circ} 37' 00''$, with limited available literature. The first detailed astrophysical investigation was conducted by \cite{Tadross2012} using data from the Positions and Proper Motions–Extended–Large (PPMXL) catalogue \citep{roeser2010ppmxl}, which combines astrometry with optical and 2MASS photometry to estimate the cluster’s age, reddening, distance, and structural parameters. \cite{Tadross2012} found that the cluster has a radius of 2.2 arcmin, a distance of 1845 pc, a reddening of 0.65 mag, total mass of 390 $M_\odot$, and an age of $\approx$ 500 Myr.
An earlier broad survey by \cite{Bonatto2010} included Ruprecht 15 among a sample of Ruprecht clusters studied using Two Micron All Sky Survey (2MASS) near-infrared photometry \citep{Skrutskie2006}. They derived preliminary values for distance, 1.5-2.0 kpc, and age of $\approx$ 500 Myr.

Despite these early efforts, no dedicated analysis based on $Gaia$ DR2 or DR3 astrometry \citep{evans2018,Vallenari2023} has yet been performed for Ruprecht 15, leaving its kinematic membership, internal structure, and precise evolutionary status largely unconstrained in the $Gaia$ era.

\subsection{Ruprecht 137, Ruprecht 142 \& Ruprecht 169}

The clusters Ruprecht 137 ($\alpha=\rm 08^h 00^m 17^s$ \& $\delta=-25^{\circ} 14' 00''$), Ruprecht 142 ($\alpha=\rm 19^h 37^m 36^s$ \& $\delta=-15^{\circ} 45' 00''$) \& Ruprecht 169 ($\alpha=\rm 17^h 37^m 24^s$ \& $\delta=-31^{\circ} 56' 36''$) were firstly investigated in the study of \cite{Tadross2008} who derived the basic parameters of the clusters using 2MASS near-infrared photometry. Also, they were included in the catalogues of \cite{sampedro2017multimembership}, \cite{dias2014proper}, and \cite{kharchenko2013global}. Ruprecht 137 has an age range between $\approx$~400-800 Myr, a heliocentric distance 1.4-2.2 kpc, and a mean reddening of $\approx$~0.95 mag. Ruprecht 142 has an age $\approx$~400 Myr, a heliocentric distance 1.7-1.8 kpc, and a reddening range between 0.67-1.12 mag. Ruprecht 169 has an age 1 Gyr, a heliocentric distance 1.39 kpc, and a reddening 0.66 mag. 
Similarly, almost no dedicated analysis based on $Gaia$ DR2 or DR3 astrometry \citep{evans2018,Vallenari2023} has yet been performed for these clusters. For comparison, the clusters' parameters included in published catalogues were retrieved and listed in Table \ref{tab:literature}.

\section{Data}\label{Data_info}

\begin{figure*}
    \centering
  \centering
    \includegraphics[width=0.24\linewidth]{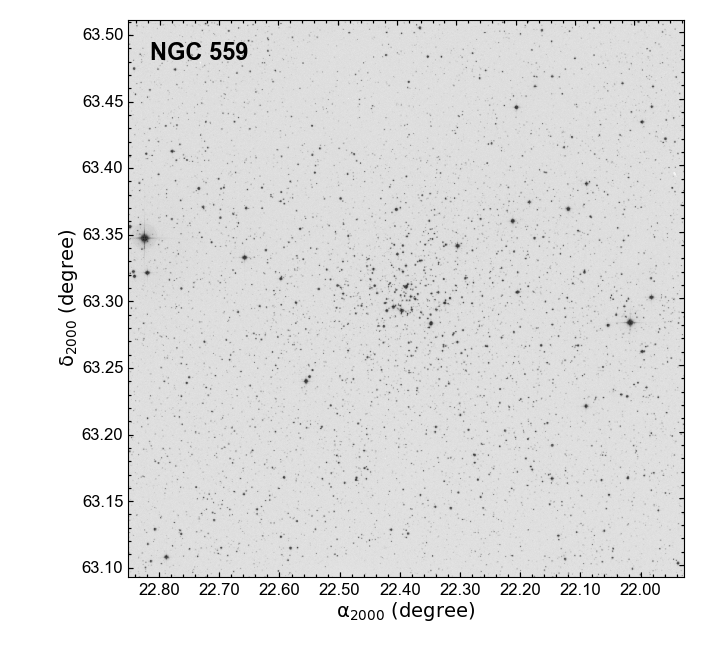}
        \includegraphics[width=0.25\linewidth]{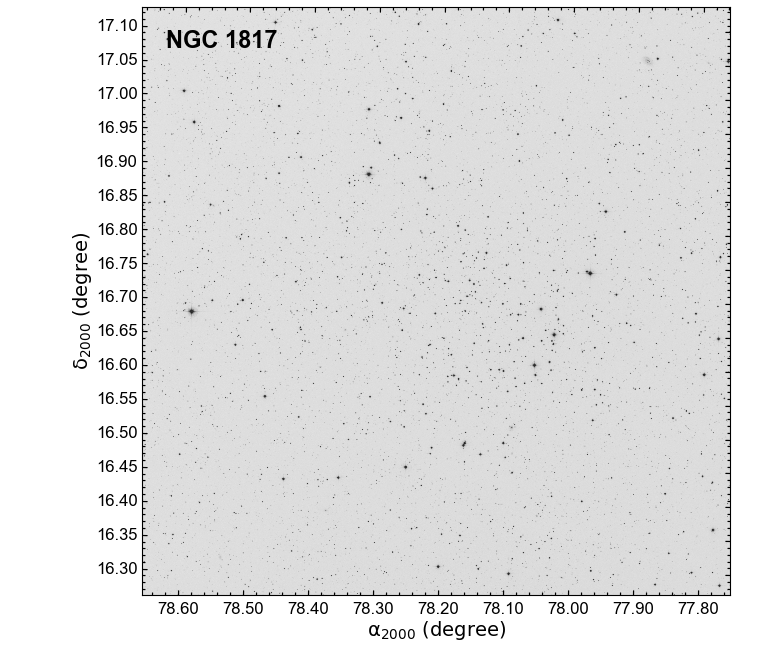}
            \includegraphics[width=0.23\linewidth]{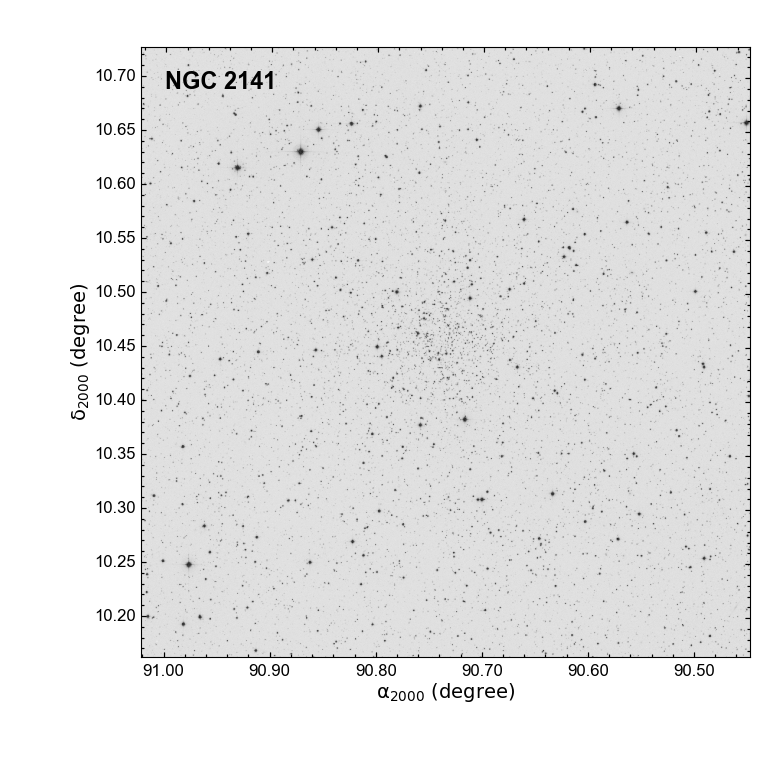}
        \includegraphics[width=0.23\linewidth]{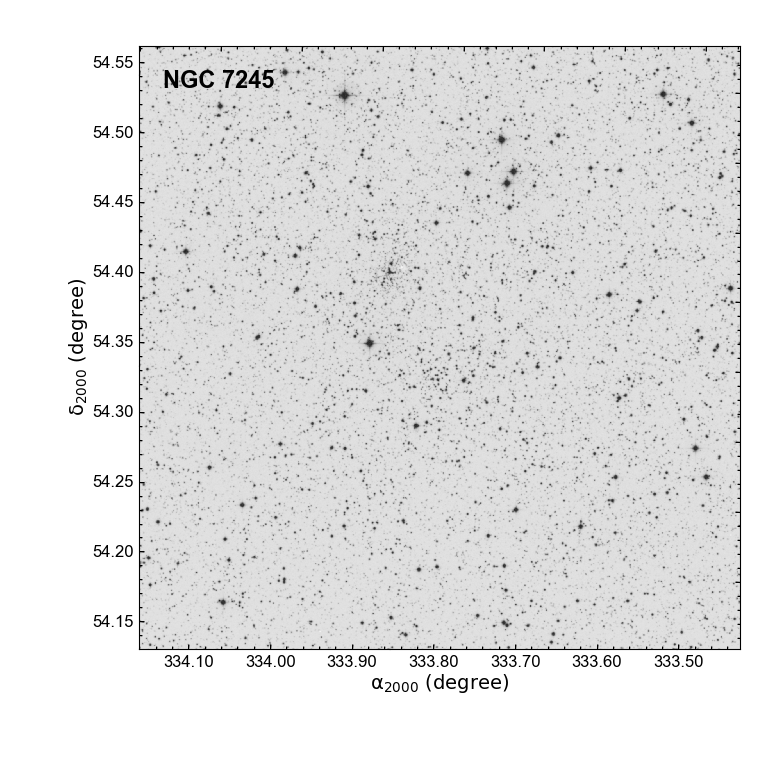}\\
            \includegraphics[width=0.24\linewidth]{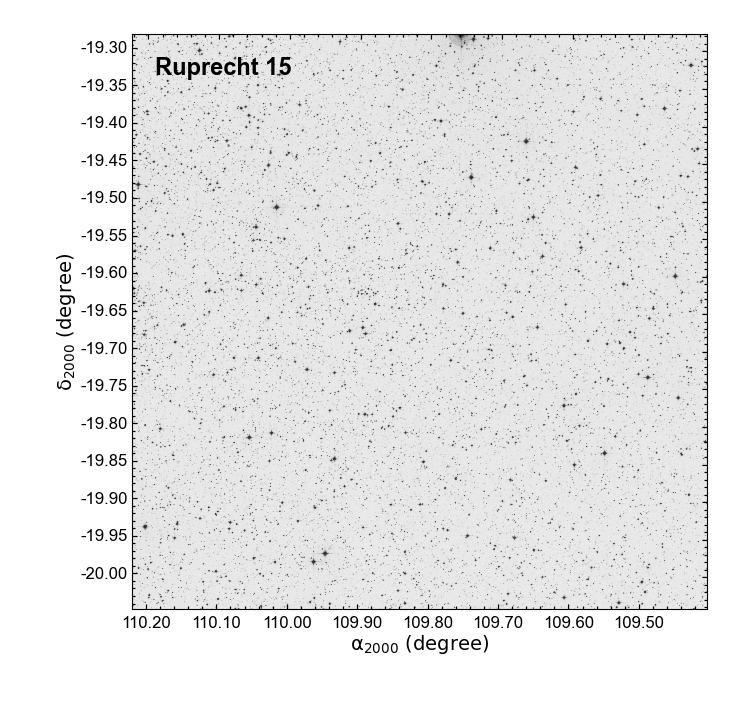}
        \includegraphics[width=0.24\linewidth]{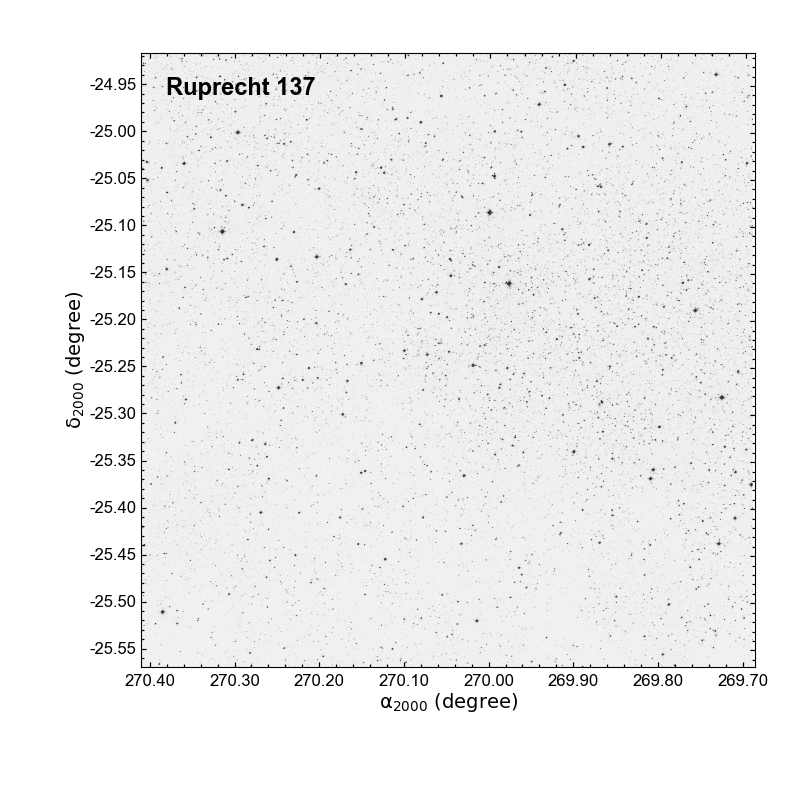}
            \includegraphics[width=0.24\linewidth]{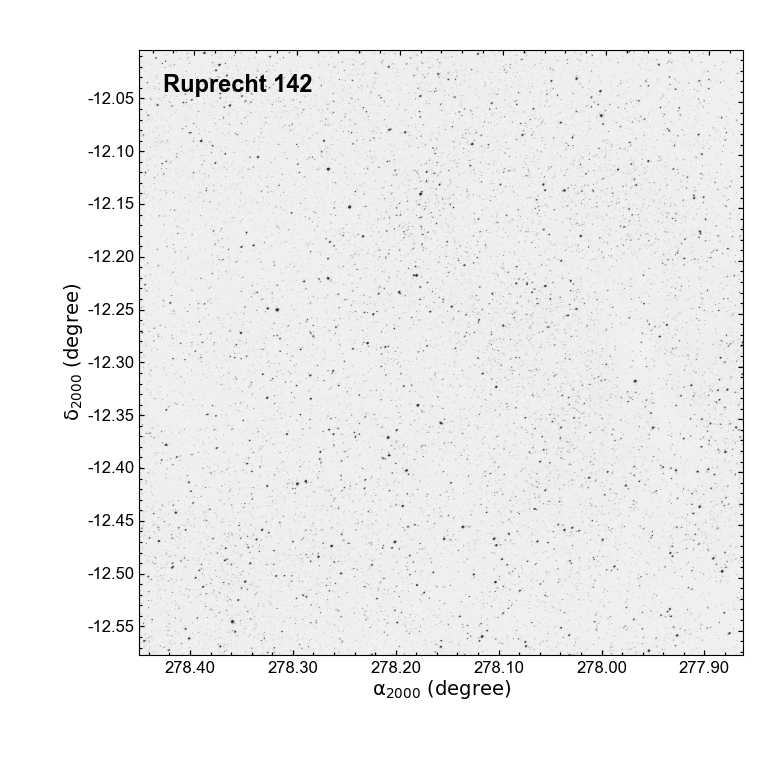}
        \includegraphics[width=0.23\linewidth]{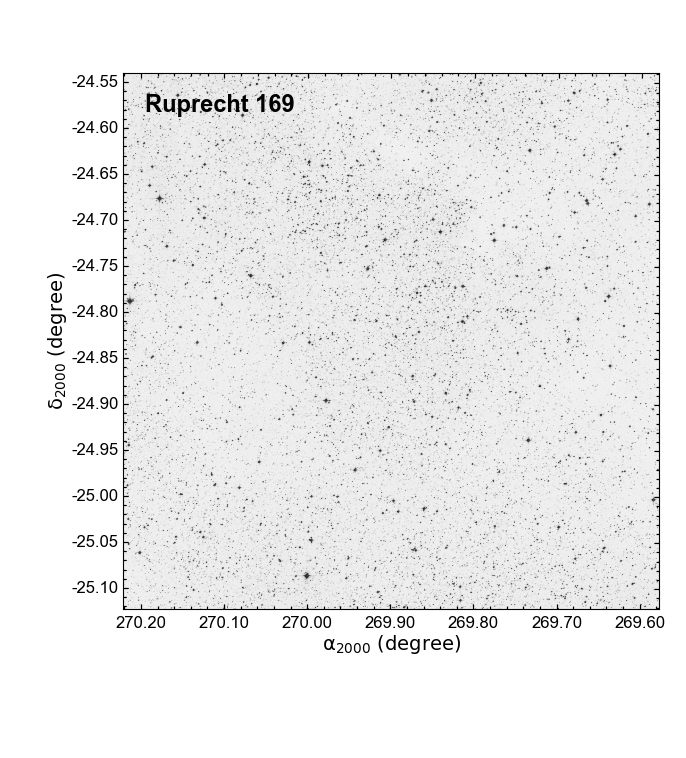}
     \caption{Finding charts of the investigated OCs in the equatorial coordinate system. The coordinate frame is oriented such that north increases upward, while east increases to the left.}
    \label{fig:findercharts}
\end{figure*}

\begin{table*}[!ht]
\centering
\small
\caption{Mean photometric uncertainties ($\sigma_G$ and $\sigma_{G_{\rm BP}-G_{\rm RP}}$) for $Gaia$ DR3 data in the fields of the eight target OCs, binned by $G$ magnitude intervals. $N$ represents the number of stars in each bin.}
\resizebox{\linewidth}{!}{%
\begin{tabular}{c|ccc|ccc|ccc|ccc}
\toprule
& \multicolumn{3}{c|}{NGC 559} & \multicolumn{3}{c|}{NGC 1817} & \multicolumn{3}{c|}{NGC 2141} & \multicolumn{3}{c}{NGC 7245} \\
\cmidrule(lr){2-4} \cmidrule(lr){5-7} \cmidrule(lr){8-10} \cmidrule(lr){11-13} 
$G$ (mag) & $N$ & $\sigma_{{\rm G}}$ & $\sigma_{{G_{{\rm BP}}-G_{{\rm RP}}}}$ & $N$ & $\sigma_{{\rm G}}$ & $\sigma_{{G_{{\rm BP}}-G_{{\rm RP}}}}$ & $N$ & $\sigma_{{\rm G}}$ & $\sigma_{{G_{{\rm BP}}-G_{{\rm RP}}}}$ & $N$ & $\sigma_{{\rm G}}$ & $\sigma_{{G_{{\rm BP}}-G_{{\rm RP}}}}$ \\ \hline \hline
6--14 & 1032 & 0.0029 & 0.0062 & 1526 & 0.0028 & 0.0053 & 2599 & 0.0029 & 0.0059 & 5937 & 0.0029 & 0.0058 \\ 
14--15 & 1247 & 0.0028 & 0.0055 & 1509 & 0.0029 & 0.0055 & 2902 & 0.0029 & 0.0057 & 7354 & 0.0029 & 0.0059 \\ 
15--16 & 2522 & 0.0028 & 0.0060 & 2761 & 0.0029 & 0.0066 & 5340 & 0.0029 & 0.0070 & 14590 & 0.0029 & 0.0064 \\ 
16--17 & 4987 & 0.0029 & 0.0090 & 4951 & 0.0030 & 0.0105 & 10102 & 0.0030 & 0.0114 & 27446 & 0.0029 & 0.0090 \\ 
17--18 & 8989 & 0.0030 & 0.0169 & 7887 & 0.0031 & 0.0214 & 16821 & 0.0032 & 0.0222 & 47336 & 0.0031 & 0.0157 \\ 
18--19 & 14675 & 0.0033 & 0.0376 & 11372 & 0.0036 & 0.0503 & 23965 & 0.0038 & 0.0500 & 76898 & 0.0034 & 0.0329 \\ 
19--20 & 22541 & 0.0042 & 0.0791 & 14515 & 0.0051 & 0.1043 & 31962 & 0.0055 & 0.1089 & 108243 & 0.0045 & 0.0725 \\ 
20--21 & 32003 & 0.0086 & 0.1815 & 18184 & 0.0120 & 0.2546 & 37687 & 0.0123 & 0.2604 & 134378 & 0.0102 & 0.2013 \\ 
21--23 & 5410 & 0.0236 & 0.4080 & 1206 & 0.0296 & 0.3834 & 1268 & 0.0288 & 0.4218 & 9020 & 0.0255 & 0.4143 \\ 
\hline
Total/Error & 93406 & 0.0064 & 0.1132 & 63912 & 0.0067 & 0.1163 & 132647 & 0.0066 & 0.1175 & 431202 & 0.0062 & 0.0982 \\ \hline \hline
& \multicolumn{3}{c|}{Ruprecht 15} & \multicolumn{3}{c|}{Ruprecht 137} & \multicolumn{3}{c|}{Ruprecht 142} & \multicolumn{3}{c}{Ruprecht 169} \\
\cmidrule(lr){2-4} \cmidrule(lr){5-7} \cmidrule(lr){8-10} \cmidrule(lr){11-13}
$G$ (mag) & $N$ & $\sigma_{{\rm G}}$ & $\sigma_{{G_{{\rm BP}}-G_{{\rm RP}}}}$ & $N$ & $\sigma_{{\rm G}}$ & $\sigma_{{G_{{\rm BP}}-G_{{\rm RP}}}}$ & $N$ & $\sigma_{{\rm G}}$ & $\sigma_{{G_{{\rm BP}}-G_{{\rm RP}}}}$ & $N$ & $\sigma_{{\rm G}}$ & $\sigma_{{G_{{\rm BP}}-G_{{\rm RP}}}}$ \\ \hline \hline
6--14 & 15694 & 0.0029 & 0.0056 & 2444 & 0.0030 & 0.0071 & 2866 & 0.0031 & 0.0073 & 2646 & 0.0030 & 0.0067 \\ 
14--15 & 17454 & 0.0029 & 0.0056 & 3479 & 0.0030 & 0.0071 & 4713 & 0.0031 & 0.0087 & 3683 & 0.0030 & 0.0067 \\ 
15--16 & 33910 & 0.0029 & 0.0064 & 7476 & 0.0031 & 0.0109 & 10679 & 0.0032 & 0.0123 & 7551 & 0.0030 & 0.0095 \\ 
16--17 & 58400 & 0.0029 & 0.0088 & 15407 & 0.0033 & 0.0216 & 23605 & 0.0033 & 0.0207 & 14750 & 0.0032 & 0.0169 \\ 
17--18 & 93804 & 0.0030 & 0.0154 & 30524 & 0.0036 & 0.0451 & 45603 & 0.0036 & 0.0392 & 27185 & 0.0034 & 0.0356 \\ 
18--19 & 147359 & 0.0033 & 0.0336 & 56489 & 0.0043 & 0.0906 & 79176 & 0.0044 & 0.0820 & 46568 & 0.0041 & 0.0785 \\ 
19--20 & 211451 & 0.0042 & 0.0718 & 91787 & 0.0065 & 0.1803 & 131337 & 0.0070 & 0.1710 & 70951 & 0.0063 & 0.1660 \\ 
20--21 & 282970 & 0.0087 & 0.1742 & 115807 & 0.0130 & 0.3210 & 140343 & 0.0138 & 0.3298 & 88801 & 0.0128 & 0.3096 \\ 
21--23 & 51576 & 0.0231 & 0.3898 & 3653 & 0.0286 & 0.4922 & 1194 & 0.0300 & 0.4732 & 3021 & 0.0282 & 0.4923 \\
\hline
Total/Error & 912621 & 0.0062 & 0.1007 & 327066 & 0.0081 & 0.1910 & 439516 & 0.0080 & 0.1781 & 265156 & 0.0077 & 0.1725 \\
\bottomrule
\end{tabular}}
\label{tab:photometric_uncertainies}
\end{table*}

To illustrate the spatial distribution of the studied systems, we generated an identification map showing the locations of the eight open clusters: NGC~559, NGC~1817, NGC~2141, NGC~7245, Ruprecht~15, Ruprecht~137, Ruprecht~142, and Ruprecht~169. The star charts, obtained using the \texttt{POSS2UKSTU\_RED} filter\footnote{\url{https://archive.stsci.edu/cgi-bin/dss\_form}}, clearly reveal each cluster as a distinct stellar overdensity within its surrounding field. The final identification charts are presented in Figure~\ref{fig:findercharts}. The astrometric and photometric data of the eight OCs were extracted from $Gaia$ DR3 database \citep{vallenari2023gaia}. 

\begin{figure*}[ht]
    \centering
   \includegraphics[width=0.8\linewidth]{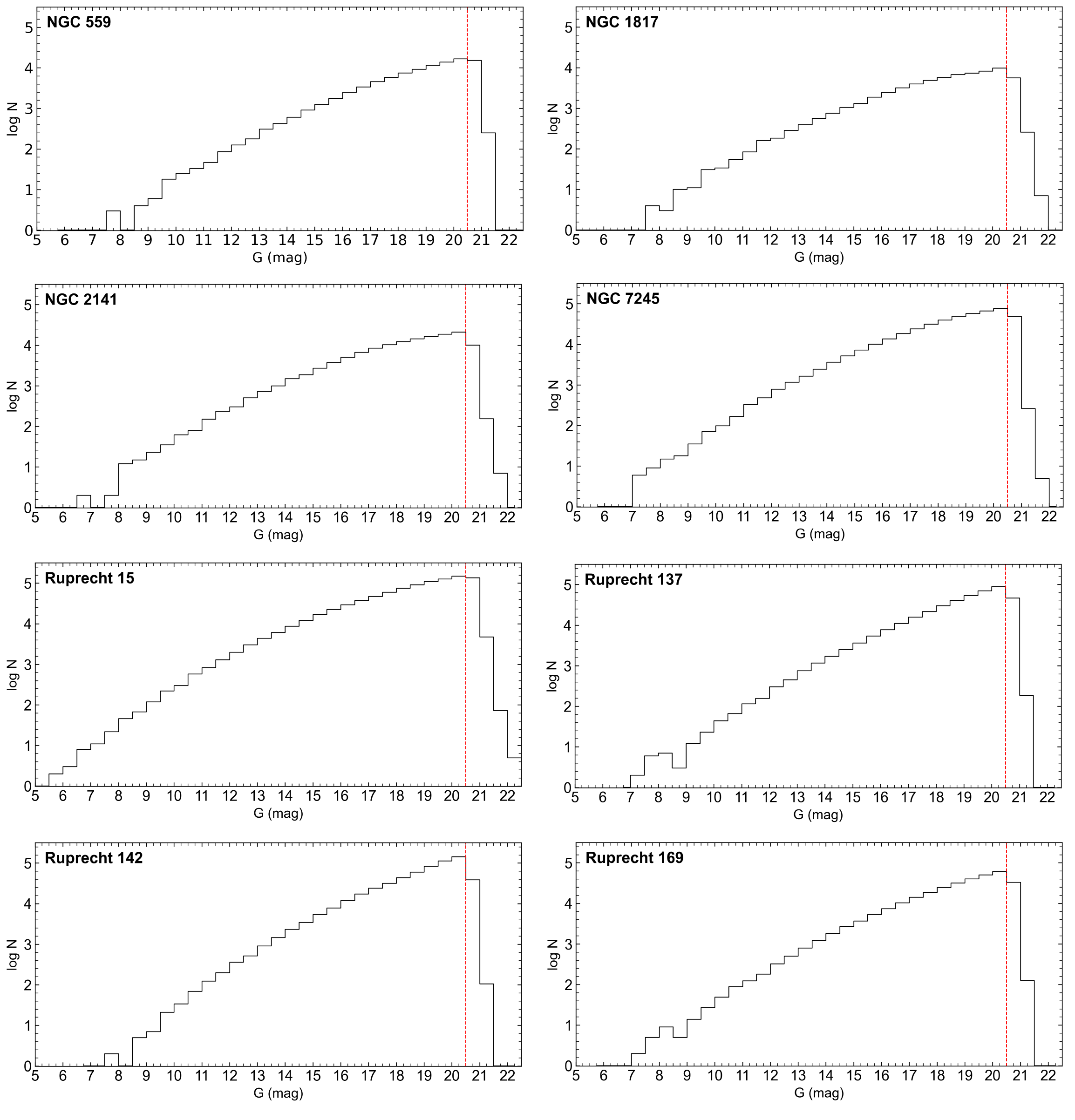}
    \caption{The distribution of $G$-apparent magnitudes for stars located in the fields of the eight OCs. Vertical red dashed lines denote the photometric completeness limits.}
    \label{fig:completeness}
\end{figure*}

The $Gaia$ DR3 database includes the five-parameter astrometry for approximately 1.8 billion sources along with their locations on the sky ($\alpha$, $\delta$), trigonometric parallaxes ($\varpi$) and the RA and DEC components of the proper motion ($\mu_{\alpha}\cos\delta$,~$\mu_{\delta}$) with a limiting magnitude of $G=21$ mag. The uncertainties in the respective proper motion components are up to $0.02-0.03$ mas $\rm yr^{-1}$ (at $G<15$ mag), 0.07 mas $\rm yr^{-1}$ (at $G\sim$ 17 mag), 0.50 mas $\rm yr^{-1}$ (at $G\sim$ 20 mag) and 1.40 mas $\rm yr^{-1}$ (at $G=21$). The trigonometric parallax values have uncertainties of $\sim$ 0.02–0.03 mas for sources with $G<15$ mag,$\sim$ 0.07 mas for sources with $G=17$ mag,$\sim$ 0.50 mas at $G=20$ mag and $\sim$ 1.30 mas at $G=21$ mag. As illustrated in Figure~\ref{fig:completeness}, the number of stars rises with increasing magnitude, reaching a peak at approximately $G$=20.5 mag.

The extraction was performed within circular regions centred on the cluster coordinates, adopting cluster-dependent search radii as summarised in Table~\ref{tab:photometric_uncertainies}. In all cases, the selected radii were chosen to be significantly larger than the literature cluster radii in order to include potential members in the outer regions and to adequately sample the surrounding field star population. The total number of sources retrieved from the $Gaia$ DR3 database for each cluster is listed in Table~\ref{tab:photometric_uncertainies}. The use of $Gaia$ DR3 is motivated by its substantial improvements over previous releases, including a $\sim$30\% enhancement in parallax precision and a twofold increase in proper-motion accuracy compared to $Gaia$ DR2.

The photometric quality of the $Gaia$ DR3 data for the eight target clusters is quantified through the mean uncertainties in the $G$ magnitude ($\sigma_G$) and colour index ($\sigma_{G_{\rm BP}-G_{\rm RP}}$), as a function of apparent magnitude. As summarised in Table~\ref{tab:photometric_uncertainies}, both photometric uncertainties show a clear and systematic increase toward fainter magnitudes in all cluster fields. For bright stars ($G \le 16$ mag), the typical uncertainties remain very small, with $\sigma_G \sim 0.003$ mag and $\sigma_{G_{\rm BP}-G_{\rm RP}} \le 0.01$ mag. Beyond $G \sim 18$ mag, colour uncertainties increase rapidly, reaching values of $\sigma_{G_{\rm BP}-G_{\rm RP}} \ge 0.1$ mag and exceeding 0.3–0.4 mag at the faintest magnitude bins ($G \ge 21$ mag). The overall mean photometric uncertainties vary slightly from cluster to cluster, reflecting differences in stellar density and crowding conditions, but follow consistent trends across all fields.

\section{Cluster Morphology and Density Profiles}

\subsection{Clusters' Centres}

\begin{figure*}[ht!]
\includegraphics[width=0.99\linewidth]{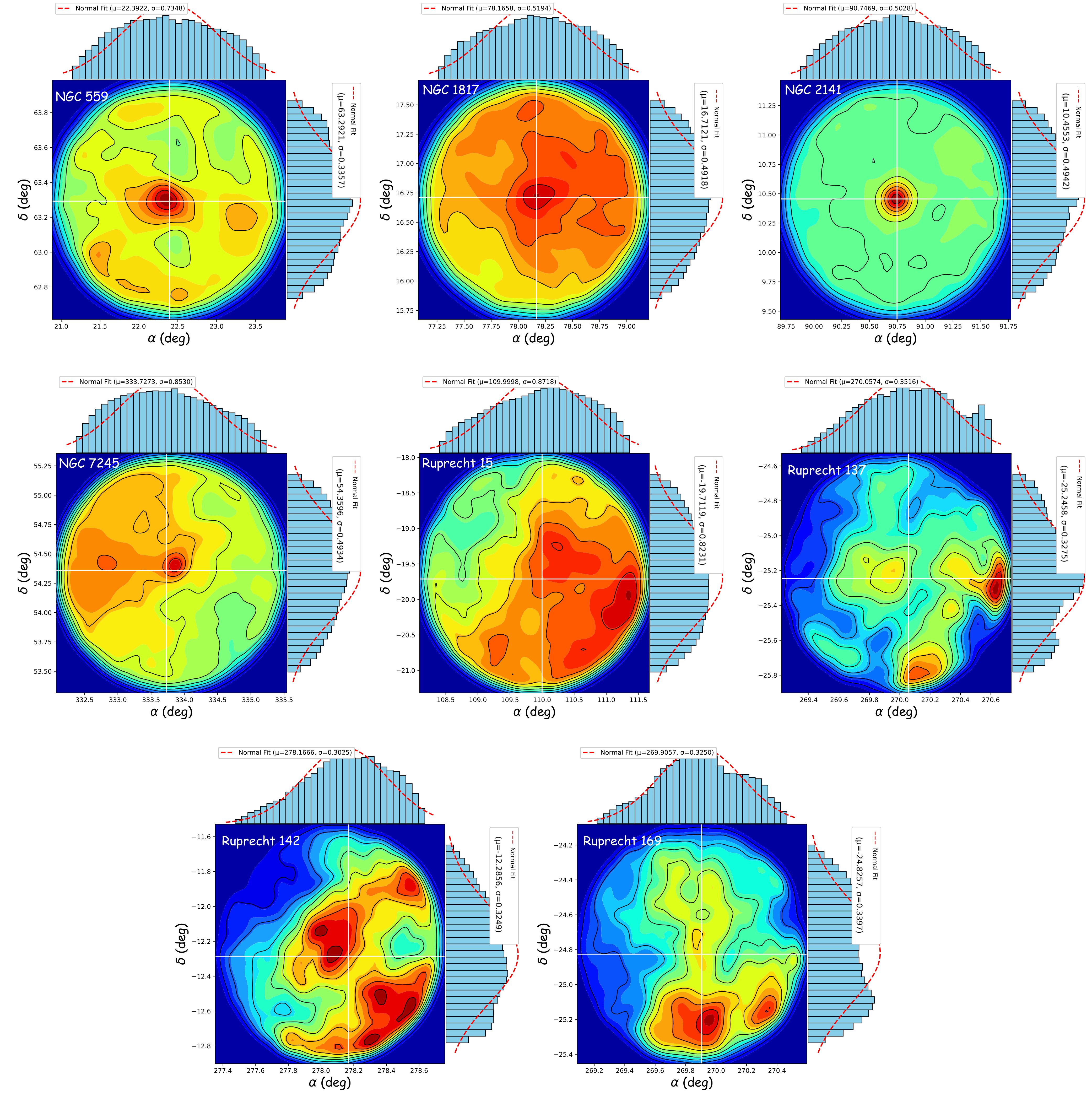}
\caption{Determination of the cluster centres for the eight target OCs. The peaks of the Gaussian fits (dashed red lines) are adopted as the new cluster centres, with the mean values and standard deviations noted in each panel.}
\label{New_center}
\end{figure*}

Cluster's centre is the region where the highest stellar aggregation is located when we plot the right ascension and declination of all stellar objects in the downloaded file from the $Gaia$ database. It can be located through the construction of $\alpha$ and $\delta$ histograms for each cluster. For each cluster, we divided the extracted region into bins of identical size of 0.05 deg in both $\alpha$ and $\delta$, and the stars were counted in each bin. Next, Gaussian fits were applied to histograms, and the peaks of fitting curves of the $\alpha$ and $\delta$ histograms of each cluster will mark the new cluster's centre, as shown in Figure \ref{New_center}.

\subsection{Radial Density Profile of OCs}

\begin{figure*}[ht!]
    \centering
        \includegraphics[width=0.47\linewidth]{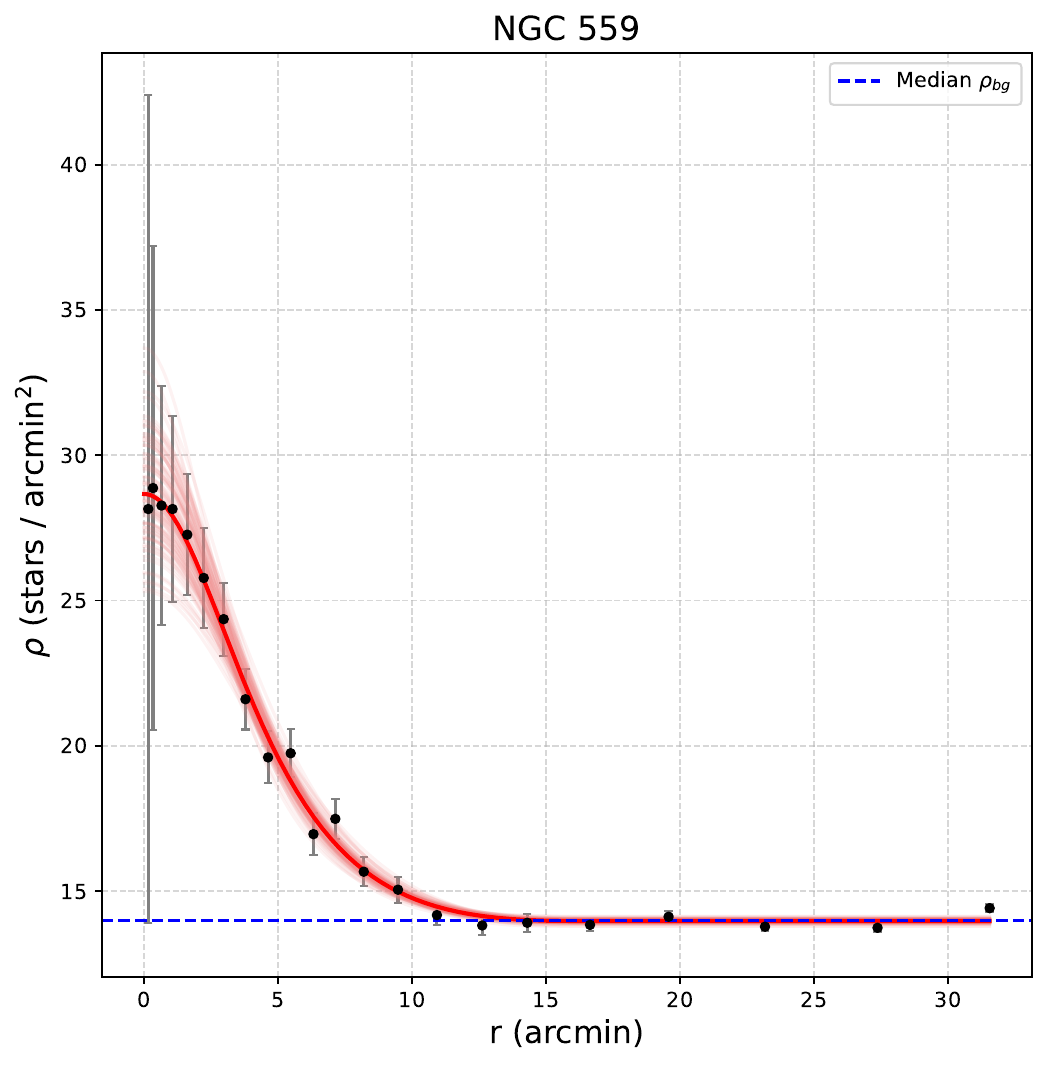}
    \includegraphics[width=0.47\linewidth]{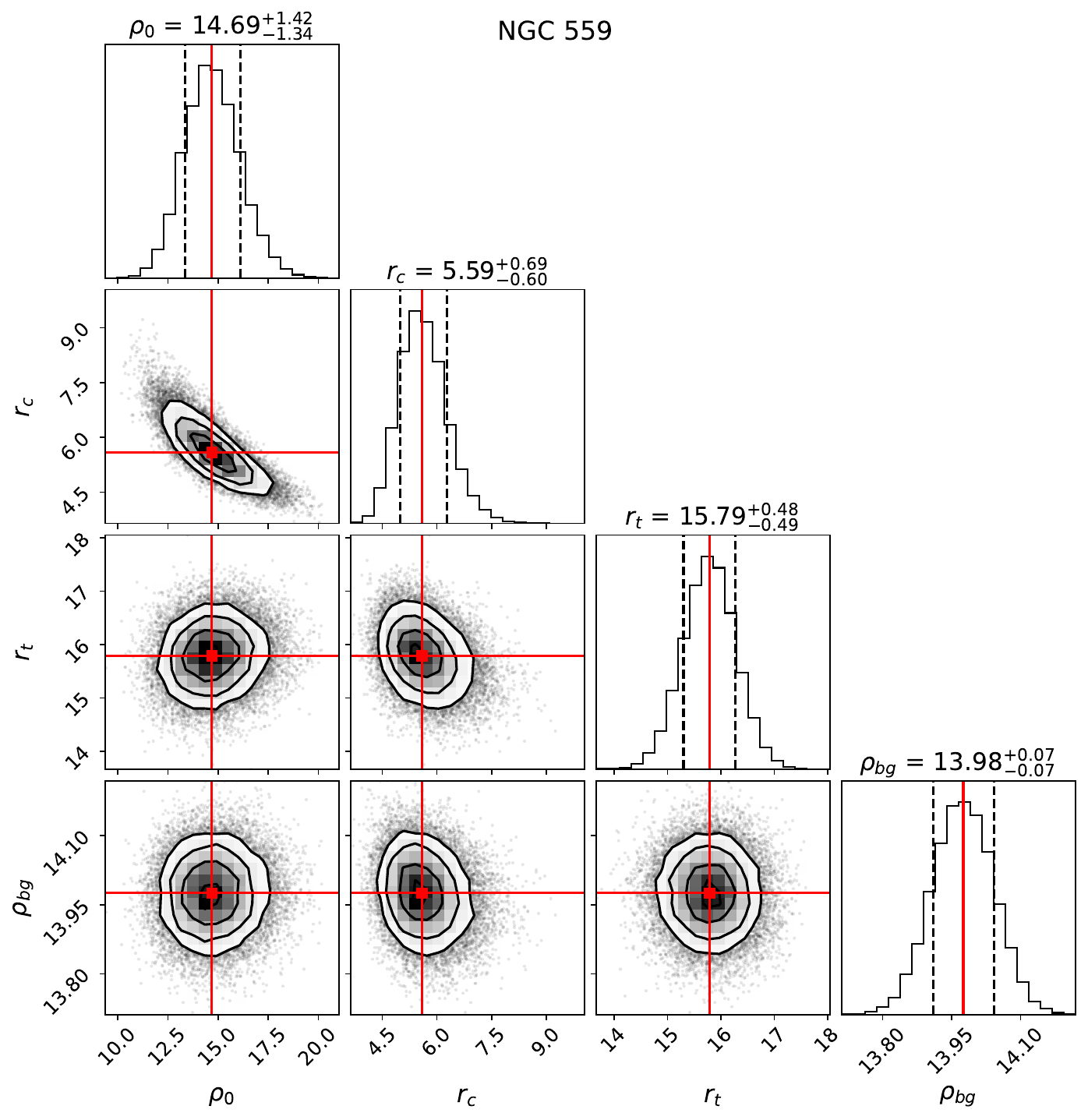}\\
        \includegraphics[width=0.47\linewidth]{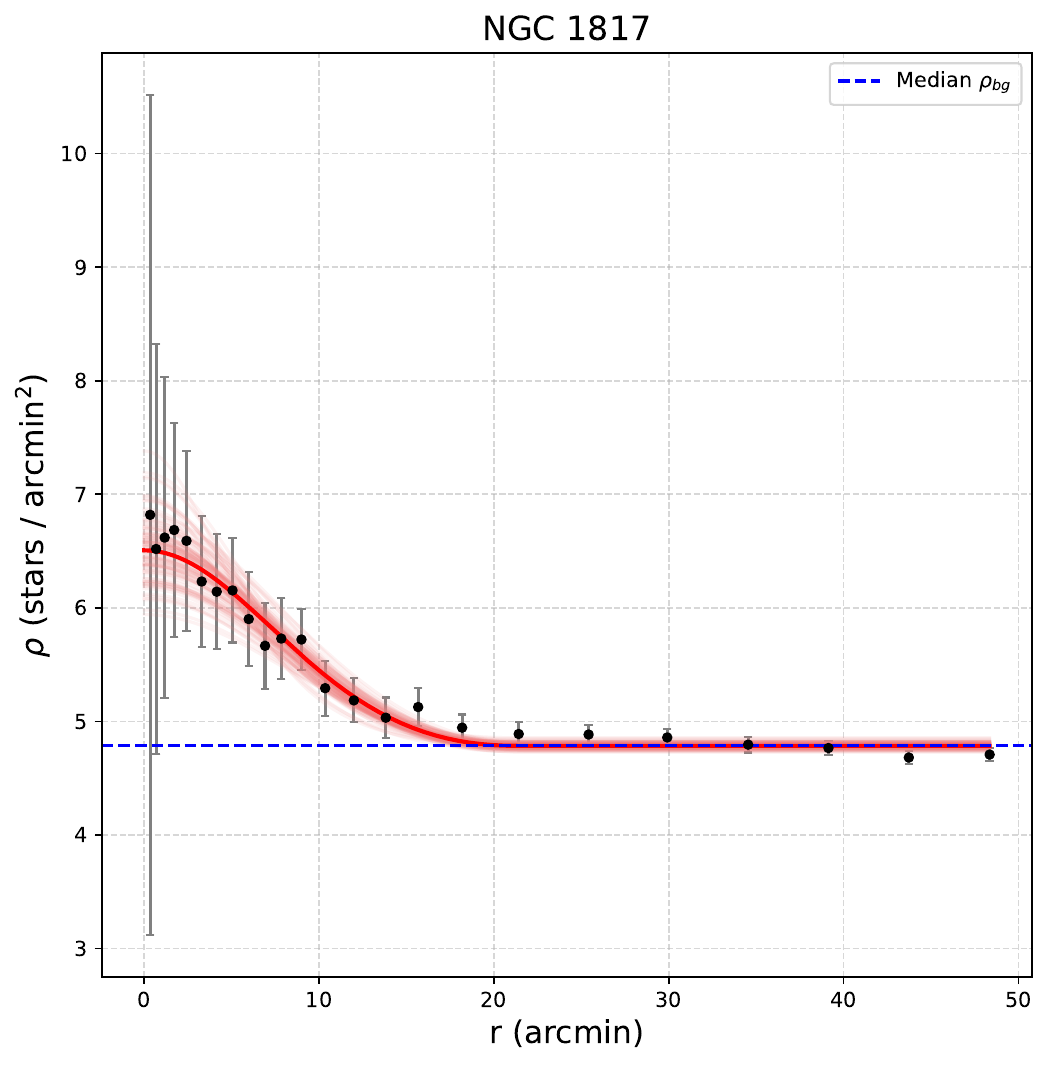}
    \includegraphics[width=0.47\linewidth]{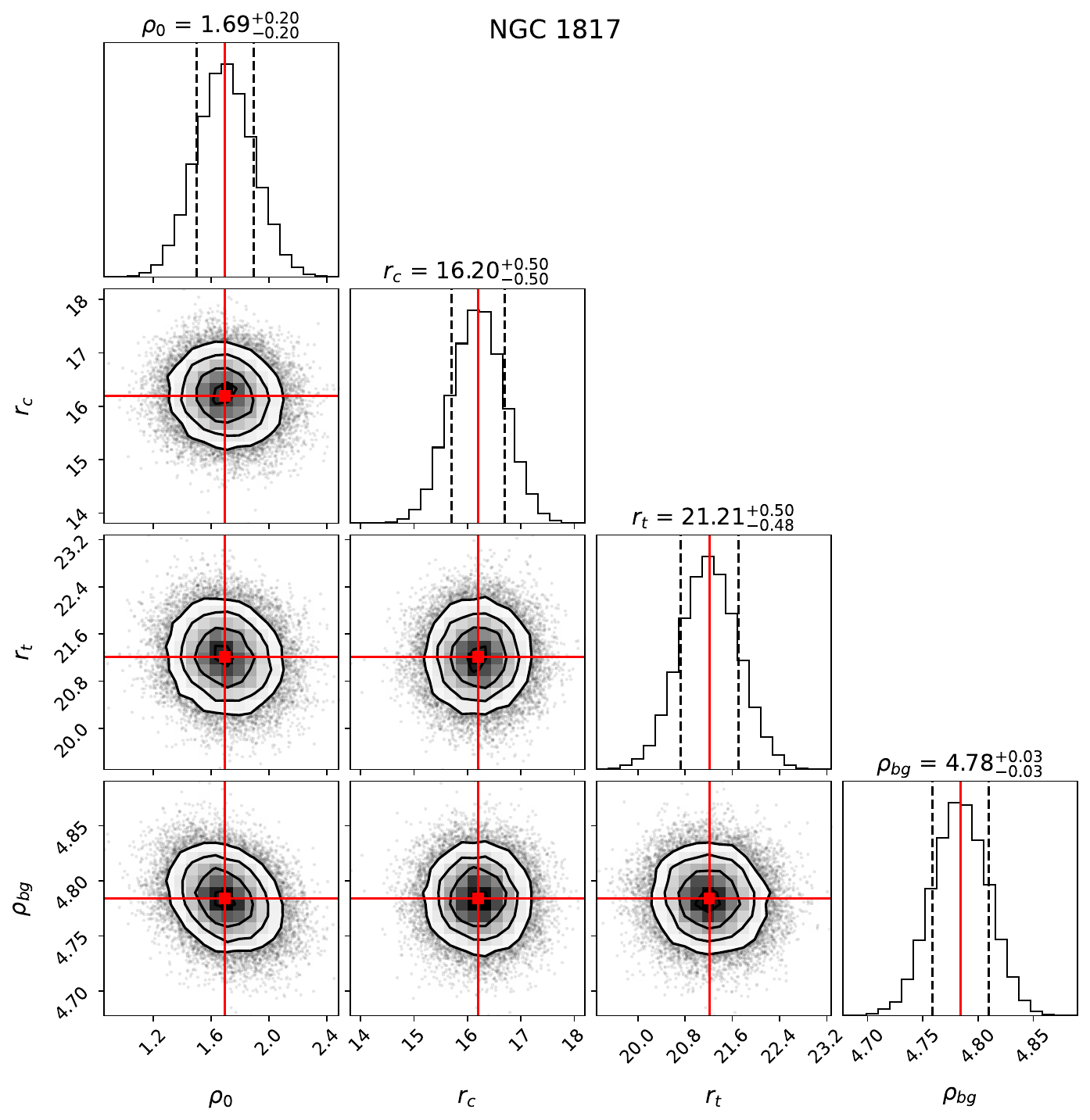}\\
    \caption{The RDPs of the NGC 559 \& NGC 1817 clusters represented by the black dots. The dashed red line represents the \cite{King1962} fitting while the blue dashed line marks the background density $f_{\rm bg}$.}
    \label{rdps}
\end{figure*}

By drawing the radial density profile (RDP) of a cluster, we can get information about its structure, such as whether this cluster is rich or poor, and whether the cluster has a very concentrated or elongated structure. The radial density profiles (RDPs) of the clusters were constructed using $Gaia$-based member stars. The sky region around each cluster was divided into concentric annuli, and the stellar surface density in each ring was computed as:
\begin{equation}
\rho(r_i)=\frac{N_i}{A_i},
\label{radial_density}
\end{equation}
with Poisson uncertainties estimated as $1/\sqrt{N_i}$. The resulting RDPs of NGC 559 and NGC 1817 clusters are presented in Figure~\ref{rdps}, together with the corresponding model fits, while the rest are presented in \ref{rdps-append}.

To derive structural parameters, we fitted the empirical \citet{King1962} surface-density model, a modern extension of the traditional King profile. In this study, we adopt the form:
\begin{equation}
\rho(r)=\rho_0 \left[ 
\frac{1}{\sqrt{1+(r/r_c)^2}} -
\frac{1}{\sqrt{1+(r_t/r_c)^2}}
\right]^2 + \rho_{\mathrm{bg}},
\label{eq:king_model_short}
\end{equation}
where $\rho_0$ is the central density above background, $r_c$ the core radius, $r_t$ the tidal radius, and $\rho_{\mathrm{bg}}$ the background density.

The parameters were estimated via maximum likelihood, using the log-likelihood expression:
\begin{equation}
\ln \mathcal{L} = -\sum_i 
\left( \frac{\rho_i - \rho_{i,\mathrm{Model}}}{\sigma_{\rho_i}} \right)^2 .
\end{equation}
Sampling was performed with the \texttt{emcee} MCMC algorithm \citep{Foreman-Mackey2013}, adopting uniform priors. Convergence was assessed through the Gelman--Rubin diagnostic \citep{gelman1992}. Final parameter estimates are given in Table~\ref{King_para}.

The concentration parameter was computed as $C = r_{\rm cl}/r_{\rm c}$ following the definition introduced by \citet{King1966}, and serves as a quantitative indicator of cluster compactness and dynamical evolution. The derived $C$ values for the analysed clusters span a broad range, reflecting diverse structural properties from relatively loose systems to more centrally concentrated clusters. As shown in Table~\ref{King_para}, our estimates are generally comparable to literature values, although noticeable differences are present for some clusters. These discrepancies are most likely attributable to differences in the adopted datasets, radial density profile construction, and fitting procedures, as well as to the improved depth and homogeneity of the $Gaia$ DR3 data used in this study.

\begin{table*}[ht]
\centering
\footnotesize
\caption{Derived structural parameters for the target OCs obtained from fitting \citet{King1962} model to the RDPs. Columns list the cluster radius ($r_{\rm cl}$), core radius ($r_{\rm c}$), central density ($\rho_0$), background density ($\rho_{\rm bg}$), and concentration parameter ($C$).
\label{King_para}}
\renewcommand{\arraystretch}{1.4}
\begin{tabular}{lcccccc} 
\hline\hline 
Cluster & $ r_{\rm cl}$ & $ r_{\rm c}$ & $\rho_0$ & $\rho_{\rm bg}$ & $C$ & $C_{\rm Literature}$ \\
& \multicolumn{2}{c}{\rm (arcmin)} & \multicolumn{2}{c}{(stars arcmin$^{-2}$)} & &  \\
\hline 
NGC 559 & $15.79^{+0.48}_{-0.49}$ & ~$5.59^{+0.69}_{-0.60}$ & $14.69^{+1.42}_{-1.34}$ & $13.98^{+0.07}_{-0.07}$ & 2.82 & 5.83$\rm ^a$\\
NGC 1817& $21.21^{+0.50}_{-0.48}$ & $16.20^{+0.50}_{-0.50}$ & ~$1.69^{+0.20}_{-0.20}$ & ~$4.78^{+0.03}_{-0.03}$ & 1.31 & 2.32$\rm ^a$ \\
NGC 2141& $21.33^{+0.49}_{-0.48}$ & ~$3.07^{+0.20}_{-0.18}$ & $39.87^{+2.86}_{-2.71}$ & $10.08^{+0.05}_{-0.05}$ & 6.95 & 3.87$\rm ^a$  \\
NGC 7245 & $25.43^{+0.50}_{-0.50}$ & $12.80^{+2.03}_{-1.56}$ & $10.85^{+1.00}_{-0.94}$ & $32.62^{+0.06}_{-0.06}$ & 1.99 & 3.00$\rm ^a$ \\
Ruprecht 15 & $16.96^{+0.10}_{-0.10}$ & ~$5.08^{+0.10}_{-0.10}$ & ~$6.20^{+0.79}_{-0.79}$ & $17.75^{+0.07}_{-0.07}$ & 3.33 & 6.88$\rm ^b$  \\
Ruprecht 137 & $13.58^{+0.10}_{-0.10}$ & ~$3.22^{+0.10}_{-0.10}$ & $30.51^{+2.33}_{-2.29}$ & $65.02^{+0.19}_{-0.19}$ & 4.22 & 1.26$\rm ^a$ \\
Ruprecht 142 & $25.97^{+0.10}_{-0.10}$ & ~$8.15^{+0.24}_{-0.24}$ & $65.26^{+1.72}_{-1.64}$ & $90.80^{+0.16}_{-0.16}$ & 3.19 & 10.00$\rm ^c$ \\
Ruprecht 169 & ~$9.97^{+0.10}_{-0.10}$ & ~$3.59^{+0.10}_{-0.10}$ & $21.79^{+2.13}_{-2.10}$ & $53.81^{+0.15}_{-0.16}$ & 2.78 & -- \\
\hline

\end{tabular}
\\
\noindent\footnotesize
\textbf{References.} (a) \citet{Hunt2024}; 
(b) \citet{Tadross2012}; 
(c) \citet{kharchenko2013global}.
\end{table*}

\section{Photometric Analysis}

\subsection{Membership Determination}\label{gaia-membership}

In order to estimate all the astrophysical parameters of the clusters, we need to distinguish true cluster members and exclude field stars. 
The present member selection procedures are based on the fact that all cluster members have the same space motion and spatial location, which distinguishes them from field stars. 
The use of proper motion is preferred over radial velocities, because proper motion provides a better description of the stellar motion in two dimensions, and it is also less affected by the orbital motion in unrecognized binaries \citep{Tian98}. 

In the present work, members of each cluster were selected following the statistical approach of \cite{Yad13} that is based on the procedure of \cite{Bal98}. 
\cite{Yad13} identifies the cluster zone on the vector point diagram (VPD), which is constructed by plotting the right ascension and declination components of the proper motion ($\mu_\alpha \cos\delta$, $\mu_\delta$) at different intervals of the \textit{G} magnitudes. 
Then, by inspecting the VPD at the brightest interval of \textit{G}, e.g., \textit{G} $<16$ mag, a clear assembly or concentration of stars can be found. This region is identified as the cluster zone, and stars outside are considered field stars. Then the differentiation of stars into those that lie within the selected cluster zone and the stars that lie outside is extended for the other intervals of \textit{G}.
\begin{figure}[htb!]
\scalebox{.5}{\includegraphics{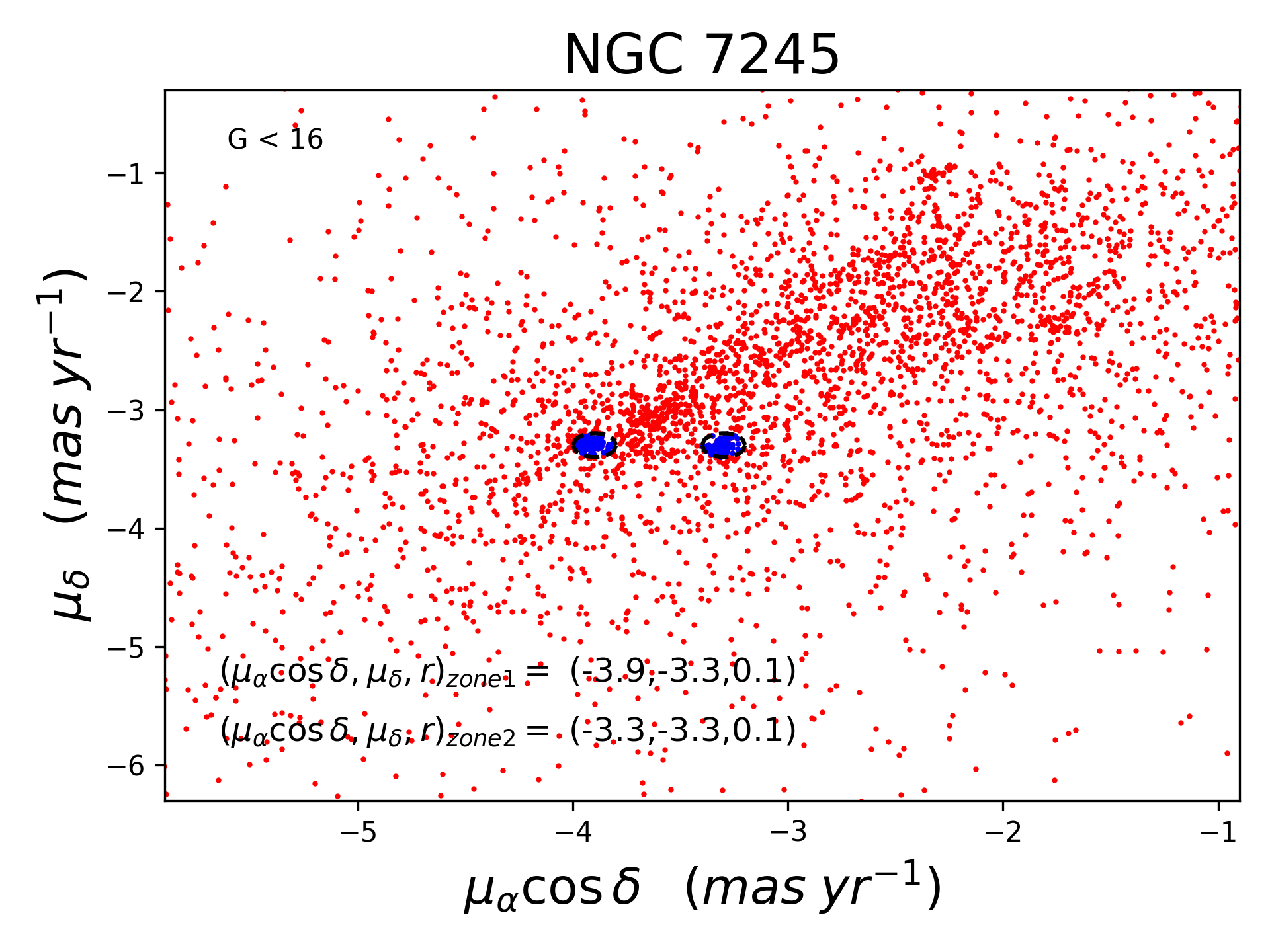}}
\caption{The VPD for stars in the field of NGC 7245 with $G<16$. The diagram clearly reveals two distinct kinematic sub-structures (NGC 7245a and NGC 7245b), whose selected proper motion centres and membership radii ($r_{zone}$) are indicated. 
\label{VPD_NGC7245}}
\end{figure}
It is interesting to note that the VPD of the NGC 7245 OC shows two clear concentrations of stars at the $G<15$ mag interval, i.e., there are two stellar sub-structures in this region, see Figure \ref{VPD_NGC7245}. 
Computing the membership probabilities for the stars that are located within the estimated radius of each cluster begins by calculating the frequency distribution for the cluster’s stars ($\phi_c^\nu$ ) and the frequency distribution for the field stars ($\phi_f^\nu$). The cluster’s frequency distribution is computed by
\begin{equation}
\phi_{c}^{\nu}= \frac{1}{2\pi\sqrt{\left(\sigma_{c}^{2}+ \epsilon_{xi}^{2}\right)\left(\sigma_{c}^{2}+\epsilon_{yi}^{2}\right)}} \times e^{-\frac{1}{2}\left[\frac{\left(\mu_{xi}-\mu_{xc}\right)^{2}}{\sigma_{c}^{2}+\epsilon_{xi}^{2}}+\frac{\left(\mu_{yi}-\mu_{yc}\right)^{2}}{\sigma_{c}^{2}+\epsilon _{yi}^{2}}\right]}\,, 
\end{equation}
where $\sigma_c$ is the intrinsic proper motion dispersion of cluster member stars that is assigned a value of 0.075 mas $\rm yr^{-1}$ \citep{Sin20} which is based on the assumption of \cite{Gir89} that radial velocity dispersions within OCs are equal to 1 km $\rm s^{-1}$ \citep{Sin20}. In the equation, $\mu_{xc}$ and $\mu_{yc}$ are the cluster's PM centre, $\mu_{xi}$ and $\mu_{yi}$ are  the RA and DEC components of the $\rm i^{th}$ star, and $\epsilon_{xi}$ and $\epsilon_{yi}$ are the corresponding observed errors.

Next, the frequency distribution of the field stars was computed using
\begin{align}
\phi_{f}^{\nu}=\frac {1}{2\pi\sqrt{\left(1-\gamma^{2}\right)}\sqrt{\left(\sigma_{xf}^{2}+\epsilon_{xi}^{2}\right)\left(\sigma_{yf}^{2}+\epsilon_{yi}^{2}\right)}} \times \nonumber\\
&\hspace{-5.0cm}e^{-\frac{1}{2\left(1-\gamma^{2}\right)}\left[\frac{\left(\mu_{xi}-\mu_{xf}\right)^{2}}{\sigma_{xf}^{2}+\epsilon _{xi}^{2}} - \frac{2\gamma\left(\mu_{xi}-\mu_{xf}\right)\left(\mu_{yi}-\mu_{yf}\right)}{\sqrt{\left(\sigma_{xf}^{2}+\epsilon _{xi}^{2}\right)\left(\sigma_{yf}^{2}+\epsilon _{yi}^{2}\right)}} + \frac{\left(\mu_{yi}-\mu_{yf}\right)^{2}}{\sigma_{yf}^{2}+\epsilon _{yi}^{2}}\right]}\, 
\end{align}
where $\mu_{xf}$ and $\mu_{yf}$ are the average value  of RA and DEC components of the proper motions of all field stars, while their corresponding dispersions are $\sigma_{xf}$ and $\sigma_{yf}$, and $\gamma$ is the correlation coefficient that is computed by:
\begin{equation} \gamma = \frac{\left(\mu_{xi}-\mu_{xf}\right)\left(\mu_{yi}-\mu_{yf}\right)}{\sigma_{xf}\sigma_{yf}} \,, \end{equation}

The distribution of all stars can be derived from
\begin{equation}
\Phi = \left(n_{c}\cdot\phi_{c}^{\nu}\right) + \left(n_{f}\cdot\phi_{f}^{\nu}\right)\,, 
\end{equation}
where $n_c$ and $n_f$ denote the normalized number of stars in the cluster and field regions, such that $n_c + n_f= 1$. For NGC 7245, the above equation is modified to include two stellar sub-structures as follows
\begin{equation}
\Phi = \left(n_{c1}\cdot\phi_{c1}^{\nu}\right)+\left(n_{c2}\cdot\phi_{c2}^{\nu}\right) + \left(n_{f}\cdot\phi_{f}^{\nu}\right)\,, 
\end{equation}
where $n_{c1}$ and $n_{c2}$ are the normalized number of stars for the two stellar sub-structures in NGC 7245, such that $n_{c1} +n_{c2}+ n_f= 1$.

Accordingly, the probability of $\rm i^{th}$ star’s membership is given by:
\begin{equation}
P_{\mu}(i) = \frac{\phi_{c}(i)}{\phi(i)}\,. \end{equation}

The search for members was limited to stars that lie within the estimated radii listed in Table \ref{King_para} with $G$ magnitudes equal to or greater than 20.5 to avoid the effect of the incompleteness in the $Gaia$ photometric data after that value, as discussed in section \ref{Data_info}. 
In addition, the $Gaia$ astrometric data includes a parameter called Renormalized Unit Weight Errors (RUWE) 
which is a crucial indicator that assesses the goodness-of-fit for a star's astrometric solution. Stars that behave as single point sources have RUWE values close to one, while stars whose motions suggest unresolved companions (binary/multiple system) have RUWE values equal to or greater than 1.4 \citep{2018A&A...616A...2L}.
Consequently, stars with renormalized unit weight errors (RUWE) exceeding 1.4 were excluded, which may influence the reliability of the analysis \citep{2024A&A...688A...1C}.

\begin{figure*}[ht!]
\centering
\includegraphics[width=0.85\linewidth]{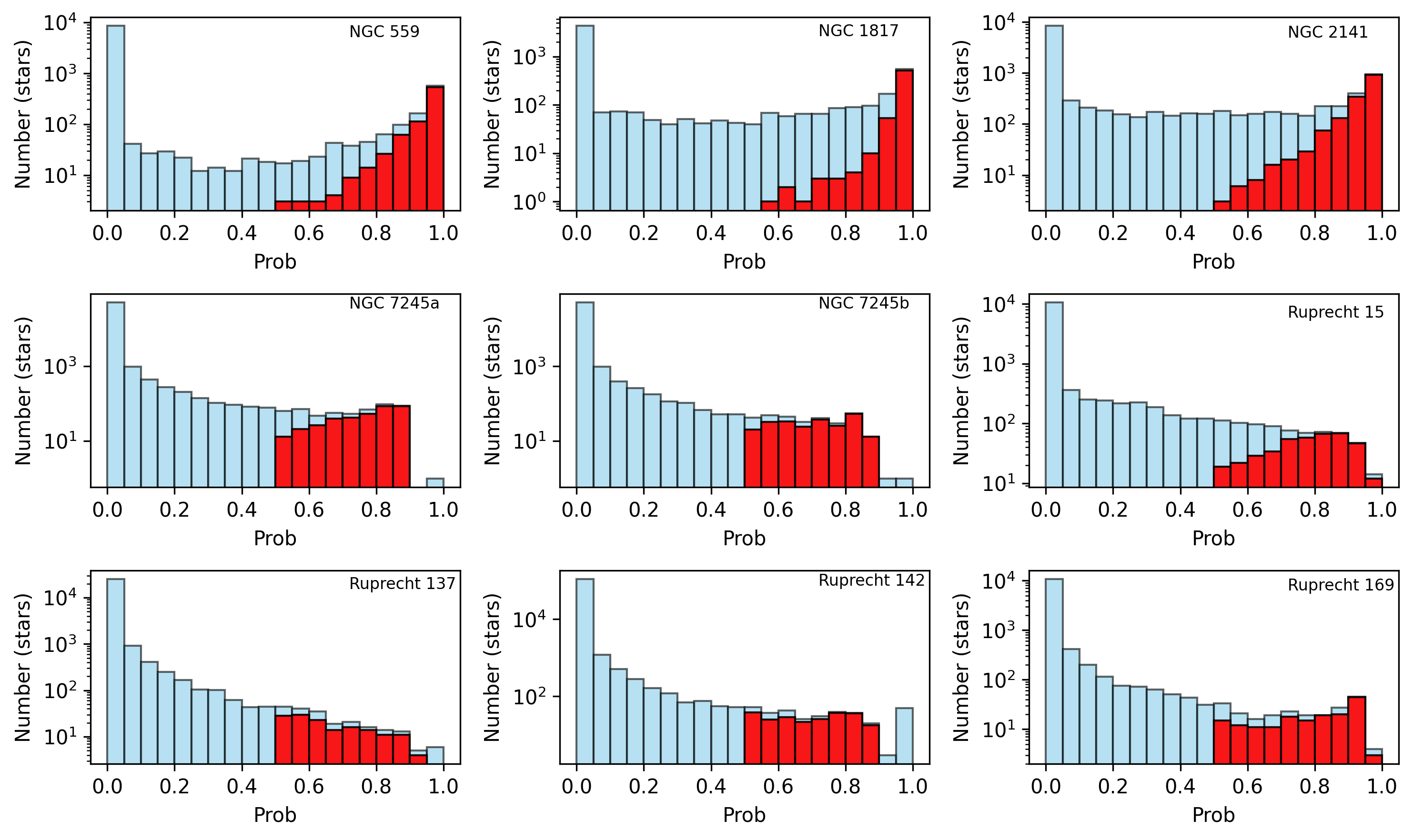}
\caption{Distributions of membership probabilities of all stars within the estimated radii of the eight open clusters (blue histograms), and those of the selected members only (red histograms). 
\label{PDists}}
\end{figure*}

Also, in order to put more constraints on our member selection process, members that have PMs differ from the mean values of the clusters by more than three times the standard deviation of stars in the selected cluster zones in the VPD were excluded. Next, the mean value and the standard deviation of the parallaxes for the new sample for each cluster, and stars that have trigonometric parallaxes values differing by more than one standard deviation from the mean value were rejected for each cluster.
Figure \ref{PDists} shows the distribution of the membership probabilities computed for all stars within the estimated radii of the eight open clusters (blue histograms), and those of only the selected members (red histograms). The figure shows that many stars with probabilities equal to or greater than 50\% were excluded, because their measured parallaxes significantly differ from the mean values of the clusters.

The values of parameters used in the membership test for the eight clusters and the radii of the selected zones ($r_{zone}$) are presented in Table \ref{member-parameters}. We got 779, 590, 1551, 365, 239, 410, 151, 231 and 168 members for NGC 559, NGC 1817, NGC 2141, NGC 7245a, NGC 7245b, Ruprecht 15, Ruprecht 137, Ruprecht 142, and Ruprecht 169, respectively, with probabilities equal to or greater than 50\%.

\begin{table*}[htb!]
\centering
\caption{Parameters adopted for the statistical membership analysis. Parameters include the mean proper motion ($\mu_{xf},~\mu_{yf}$) and dispersion ($\sigma_{xf},~\sigma_{yf}$) of field stars, the adopted cluster proper motion centre ($\mu_{xc},~\mu_{yc}$), the selection radius ($r_{zone}$), the normalized number of stars ($n_f, n_c$), and the resulting effectiveness ($E$) of the selection.}
\label{member-parameters}
\scalebox{0.7}{
\begin{tabular}{lrrrrrrrrr} 
\hline
Parameters & NGC 559 & NGC 1817 & NGC 2141 & \multicolumn{2}{c}{NGC 7245} & Ruprecht 15 & Ruprecht 137& Ruprecht 142 & Ruprecht 169 \\

\hline\hline

$\mu_{xf}$ $\rm (mas~yr^{-1}$) & $-$0.87 & 1.46 & 0.59 & \multicolumn{2}{c}{$-$2.47} & $-$1.13 & $-$0.73 & $-$1.24 & $-$0.61 \\ 

$\sigma_{xf}$ $\rm (mas~yr^{-1}$) & 4.13 & 4.15 & 2.82 & \multicolumn{2}{c}{3.59}& 2.45 & 2.75 & 2.56 & 2.69 \\

$\mu_{yf}$ $\rm (mas~yr^{-1}$) & $-$0.40 & $-$2.71 & $-$1.62 & \multicolumn{2}{c}{$-$2.34} & 1.40 & $-$2.71 & $-$3.52 & $-$2.60 \\

$\sigma_{yf}$ $\rm (mas~yr^{-1}$) & 2.45 & 5.91 &3.82 & \multicolumn{2}{c}{2.66} & 4.18 & 3.23 & 3.25 & 3.30\\

$n_f$ & 0.94 & 0.93 & 0.91 & \multicolumn{2}{c}{0.992} & 0.98 & 0.99 & 0.996 & 0.99 \\

\multicolumn{4}{c}{------------------------------------------------------------------} & NGC 7245a & NGC 7245b & \multicolumn{4}{c}{------------------------------------------------------------------------------} \\

$\mu_{xc}$ $\rm (mas~yr^{-1}$) & $-$4.28 & 0.39 & $-$0.07 & $-$3.90 & $-$3.30 & $-$0.72 & $-$0.02 & 0.76 & 0.30 \\

$\mu_{yc}$ $\rm (mas~yr^{-1}$) & 0.23 & $-$0.88 & $-$0.75 & $-$3.30 & $-$3.30 & 1.50 & $-$1.30 & $-$1.11 & $-$1.10 \\

$r_{zone}$ $\rm (mas~yr^{-1}$) & 0.30 & 0.20 & 0.25 & 0.10 & 0.10 & 0.15 & 0.20 & 0.20 & 0.20 \\

$n_c$ & 0.06 & 0.07 & 0.09 & 0.005 & 0.003 & 0.02 & 0.01 & 0.004 & 0.01 \\

$E$ & 0.80 & 0.70 & 0.64 & 0.39 & 0.30 & 0.40 & 0.29 & 0.33 & 0.40 \\

\hline 

\end{tabular}}
\end{table*}

The effectiveness of the member selection is computed using the formula of \cite{shao1996effectivity}
\begin{equation}
E=1-\frac{N \sum_{i=1}^{N} [P(i)(1-P(i))]}{\sum_{i=1}^{N} P(i) \sum_{i=1}^{N} (1-P(i))}
\, , \end{equation}
where higher values of $E$ reflect high effectiveness of the selection procedure. The values of $E$ in literature range between 0.2 to 0.9, and it has an optimum value at 0.55 \citep{shao1996effectivity}. The values of the effectiveness of membership determination for the investigated clusters are also listed in Table \ref{member-parameters}. OCs NGC 559, NGC 1817, and NGC 2141 have high $E$ values, which reflect the high effectiveness of member selection for these clusters, while the others have lower values due to the dispersed nature of these clusters, which makes the differentiation of member stars from field stars much harder.

\subsection{Astrometric Parameters}

\cite{lindegren2021gaia} show that the trigonometric parallax measurements of the $Gaia$ project are biased due to problems in the instrument and data processing. The authors found that the amount of parallax biases of stellar objects depends on their $G$ magnitudes, $(G_{\rm BP}-G_{\rm RP})$ colours, ecliptic latitudes, and the effective wavelength and the pseudo-colour of the observation. \cite{lindegren2021gaia} formulated two functions to compute the biases in the measured parallaxes of the sources included in the $Gaia$ DR3 with either five-parameter solutions or six-parameter solutions ($Z_5$ \& $Z_6$). The corrected parallaxes are computed by $\varpi_i^{corr}=\varpi_i - Z(x_i)$. They found that the parallax biases of the sources having five-parameter solutions vary between -94 and 36 $\rm \mu as$, and those of the sources having six-parameter solutions vary between -151 and 130 $\rm \mu as$. 

\begin{figure*}[htb!]
\centering
\includegraphics[width=0.9\linewidth]{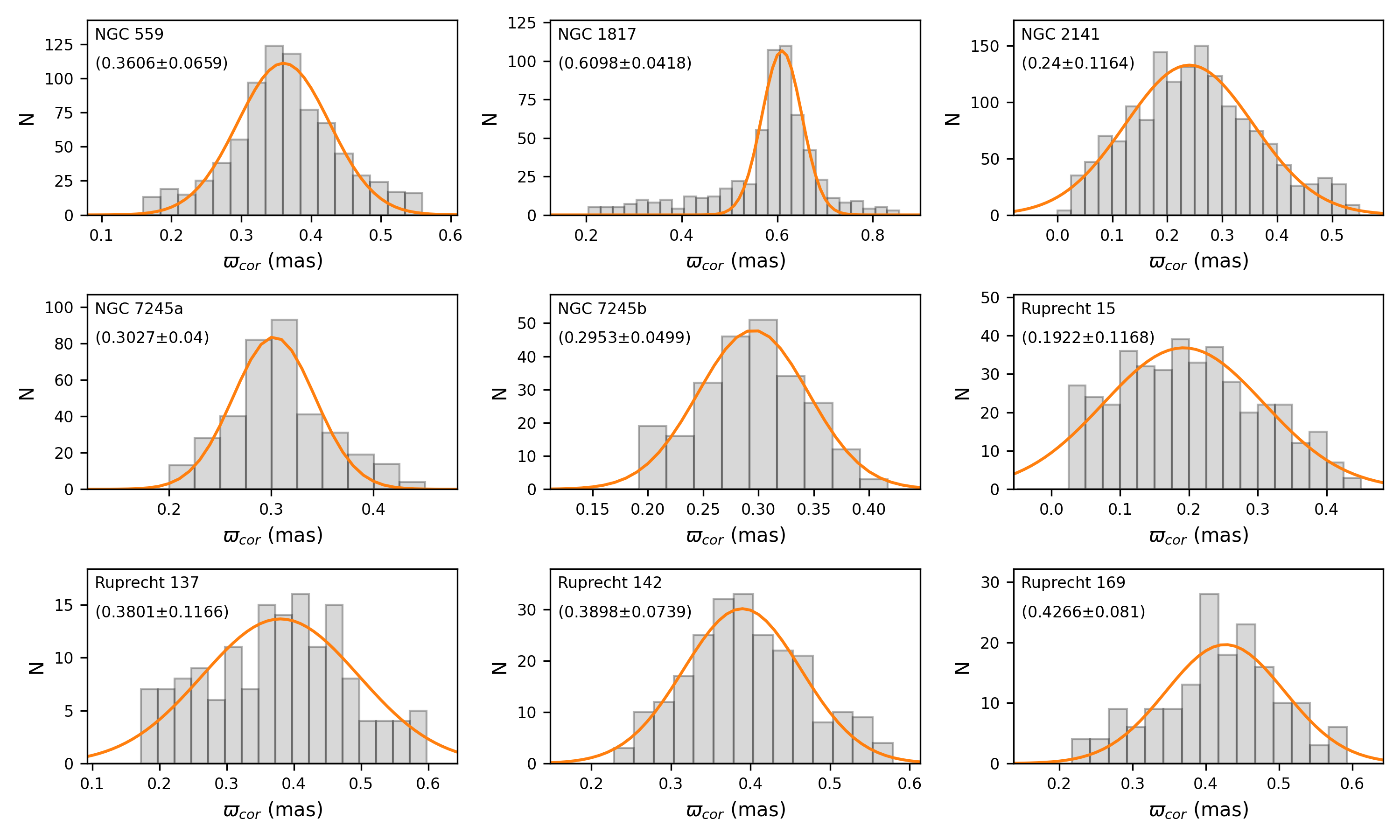}
\caption{Histograms of the corrected trigonometric parallaxes for all selected members of the clusters analysed in this study. The mean values and standard deviations are indicated in each panel, while the red curves represent the Gaussian fits to the trigonometric parallax distributions.
\label{PlxHists}}
\end{figure*}

Python implementations of the two functions (\textit{Gaia DR3-zero point python package}) available in the $Gaia$ web pages \footnote{\url{https://www.cosmos.esa.int/web/gaia/edr3-code}} \citep{lindegren2021gaia}, which were used in this study to calculate the parallax biases of the selected probable members and shown in Figure~\ref{PlxHists}. The mean parallax zero-points of the probable cluster members range between 23.0 and 37.0 $\mu$as. The mean trigonometric parallaxes, parallax-based distances, and the mean proper-motion components $(\mu_{\alpha}\cos\delta,~\mu_{\delta})$ of the clusters are summarised in Table~\ref{all_results}.

\begin{figure*}[htb!]
\captionsetup[subfigure]{labelformat=empty}
\centering
\subfloat[]{\scalebox{.4}{\includegraphics{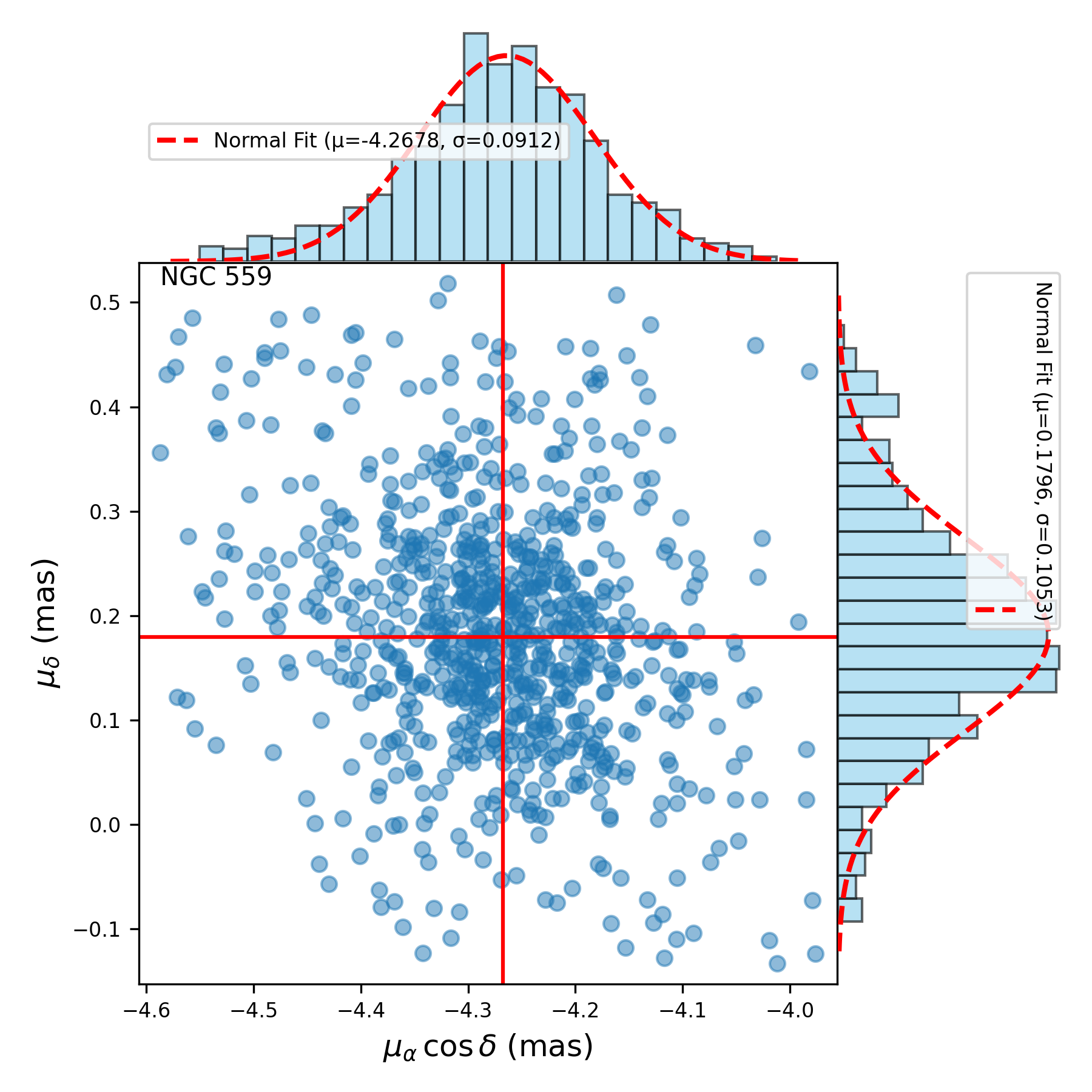}}}
\subfloat[]{\scalebox{.4}{\includegraphics{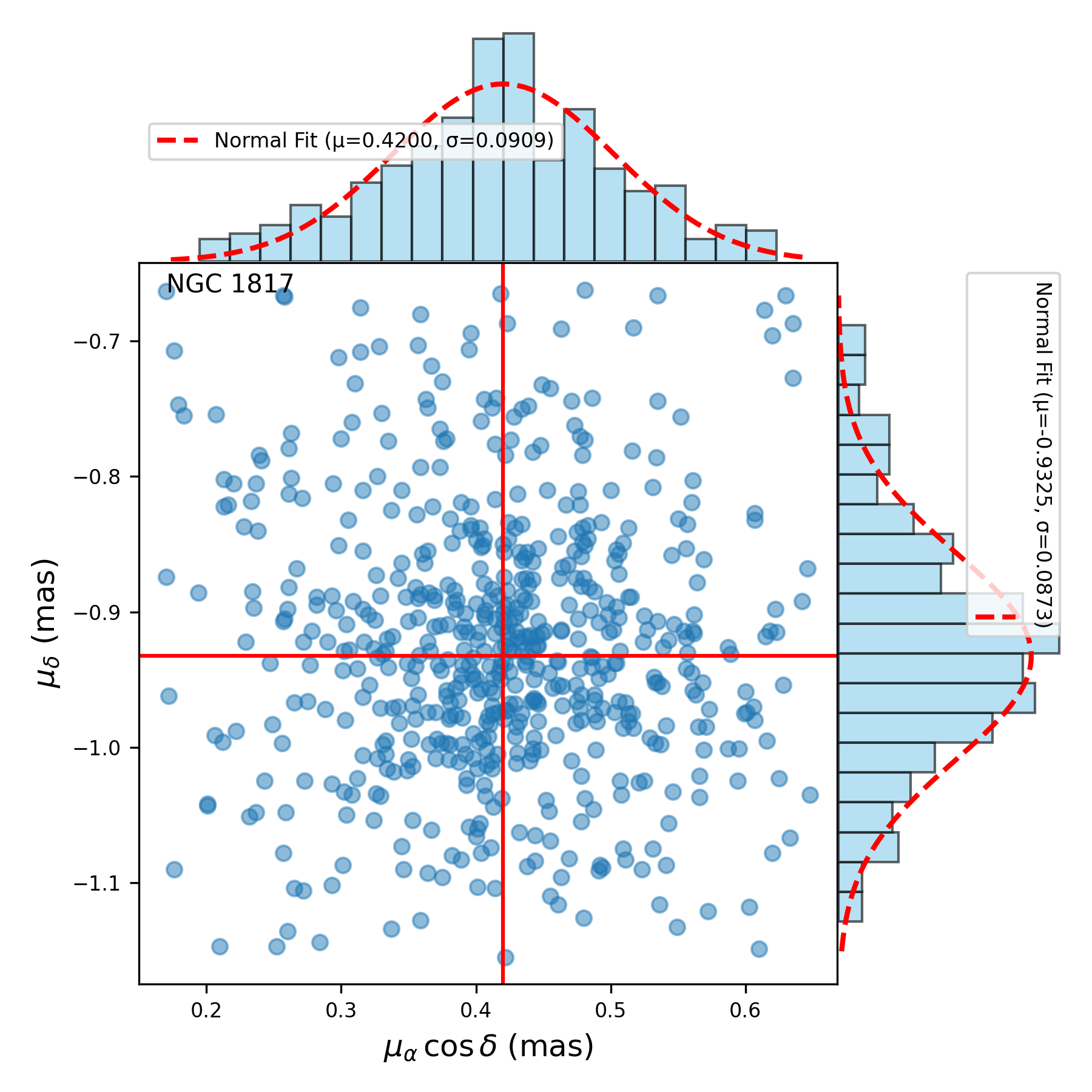}}}
\subfloat[]{\scalebox{.4}{\includegraphics{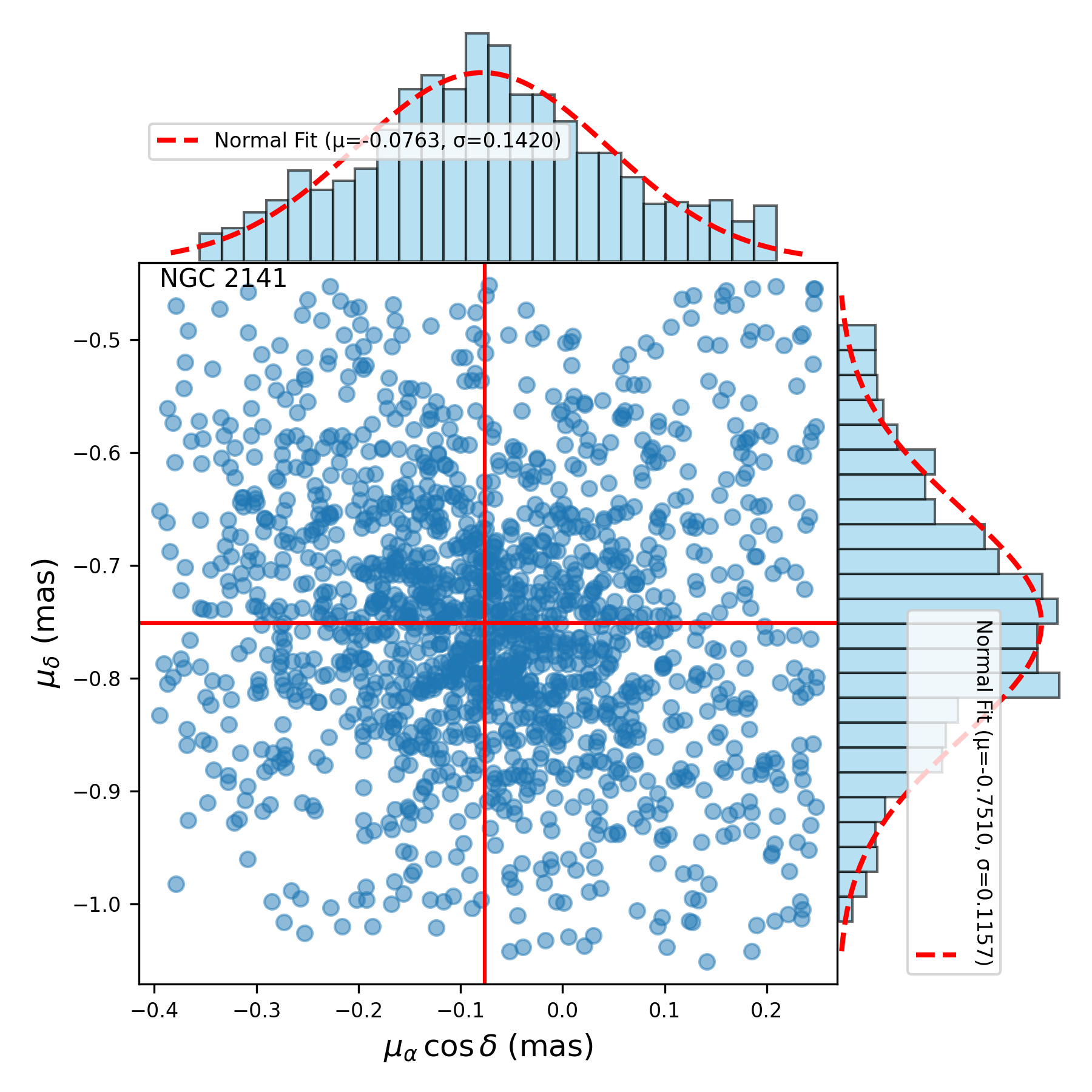}}}\\
\subfloat[]{\scalebox{.4}{\includegraphics{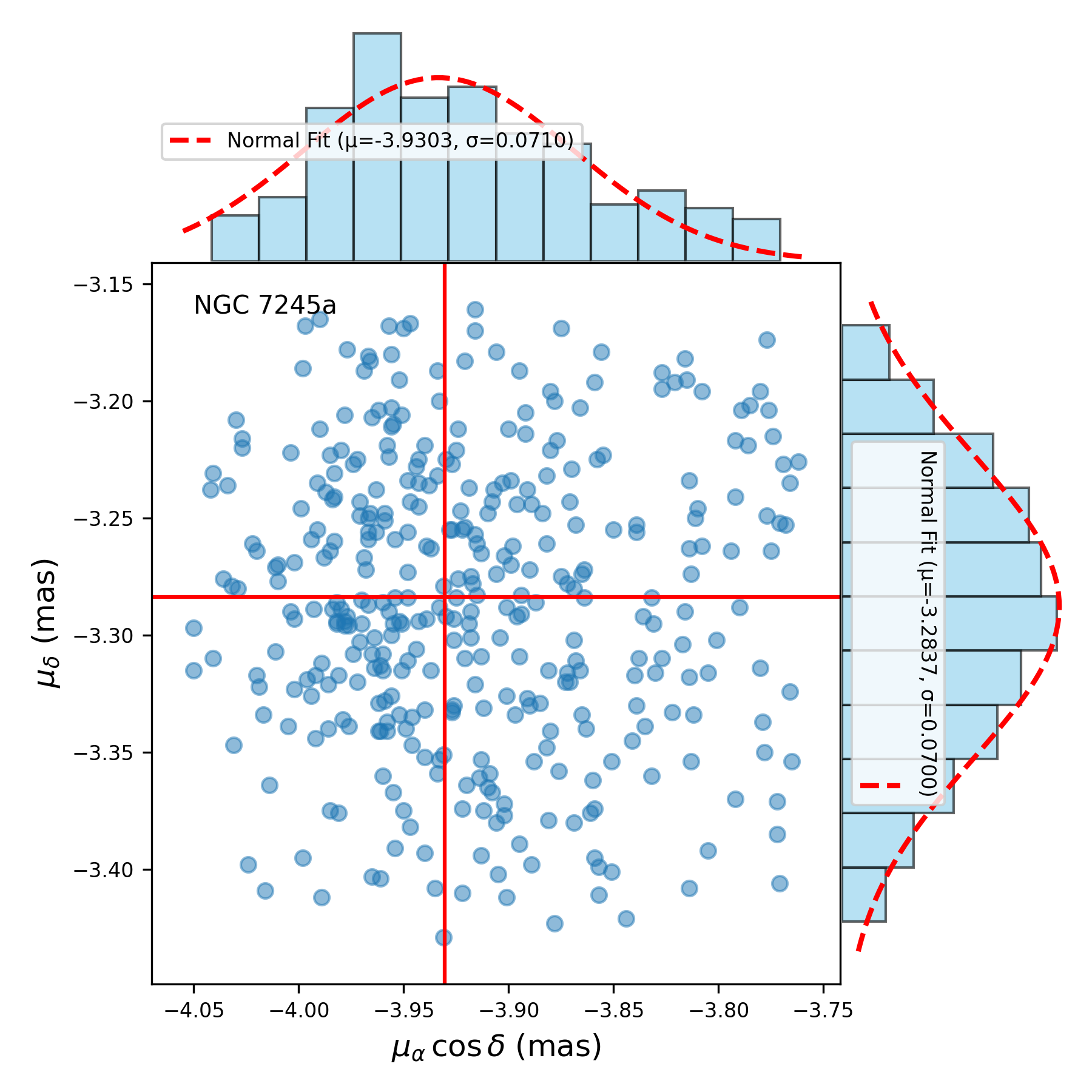}}}
\subfloat[]{\scalebox{.4}{\includegraphics{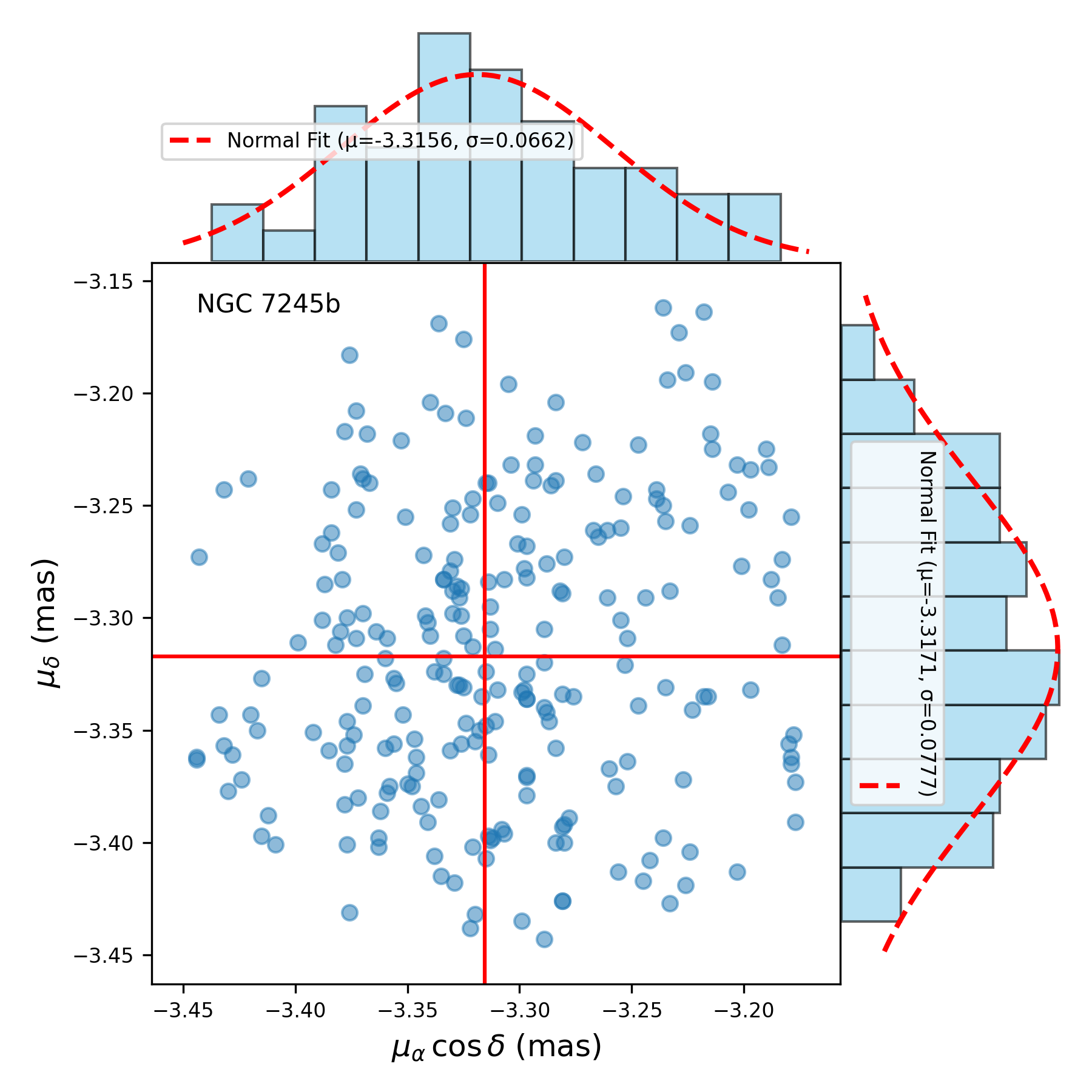}}}
\subfloat[]{\scalebox{.4}{\includegraphics{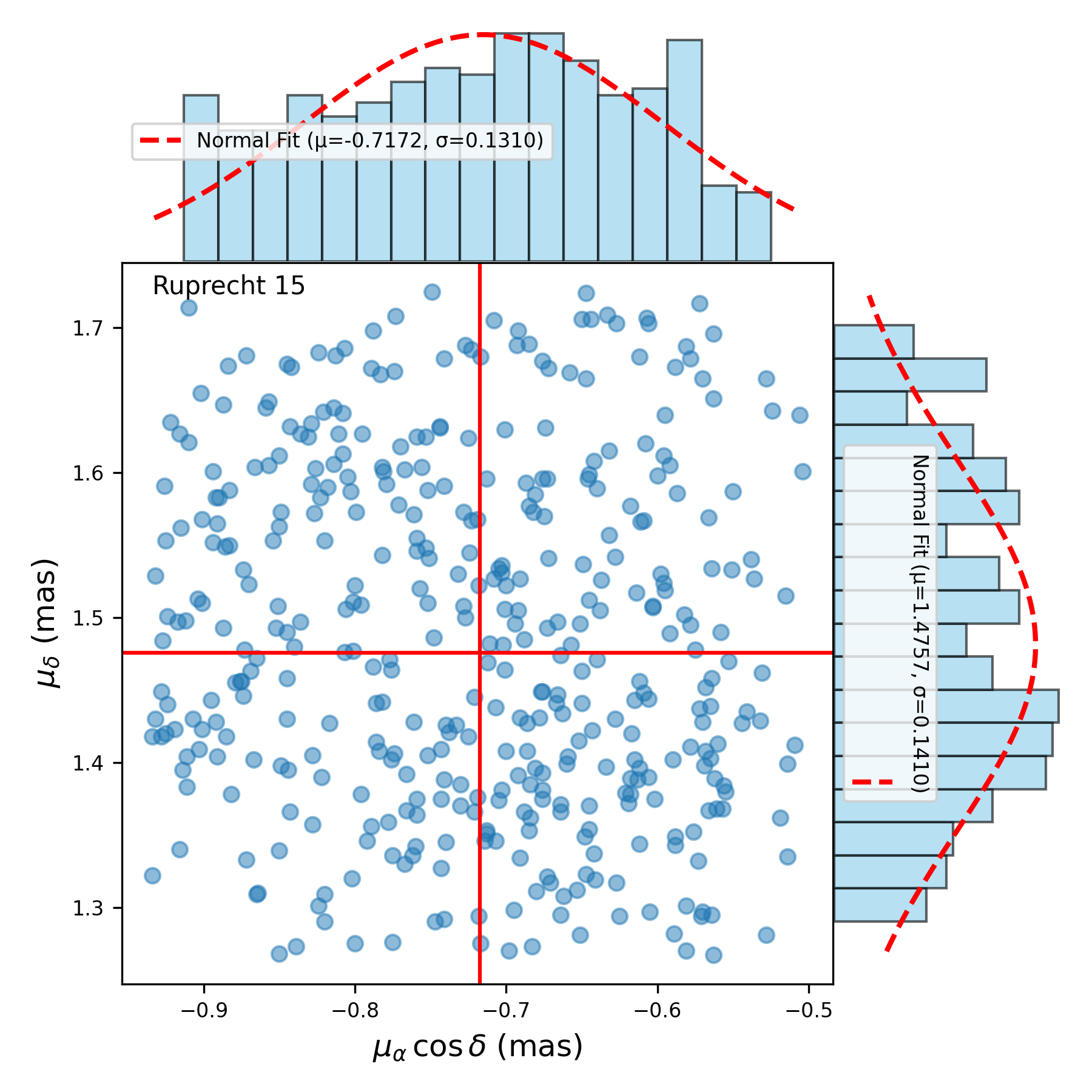}}}\\
\subfloat[]{\scalebox{.4}{\includegraphics{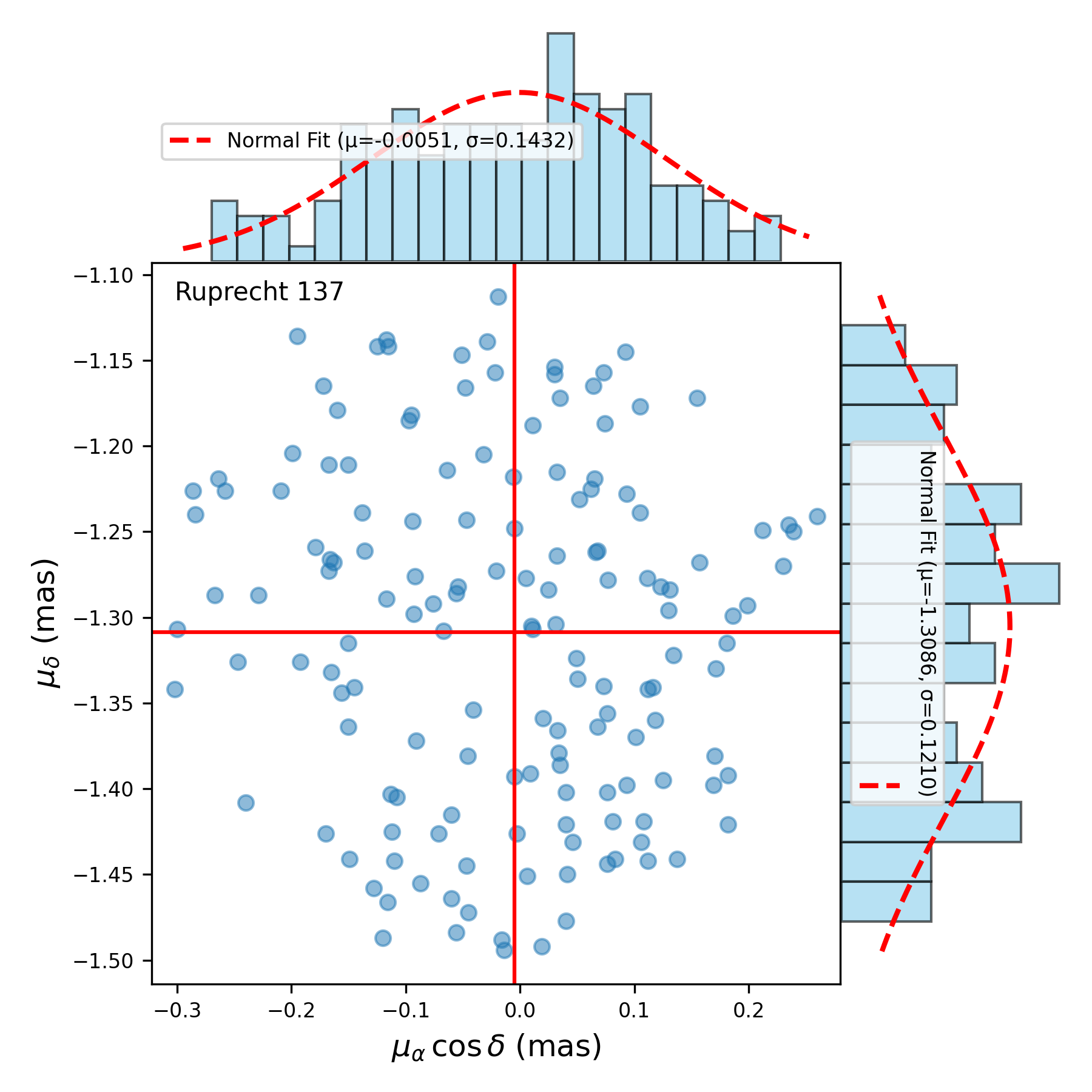}}}
\subfloat[]{\scalebox{.4}{\includegraphics{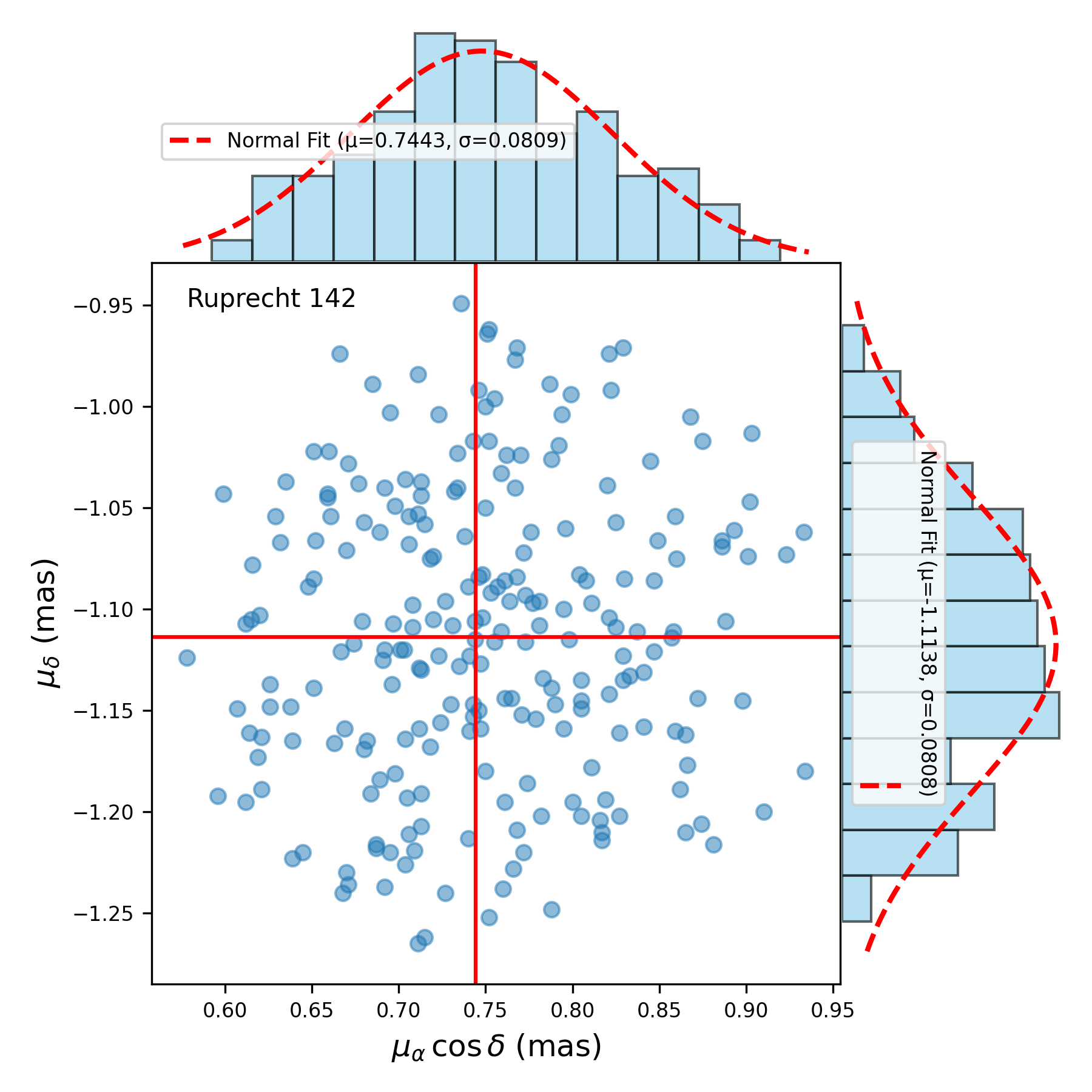}}}
\subfloat[]{\scalebox{.4}{\includegraphics{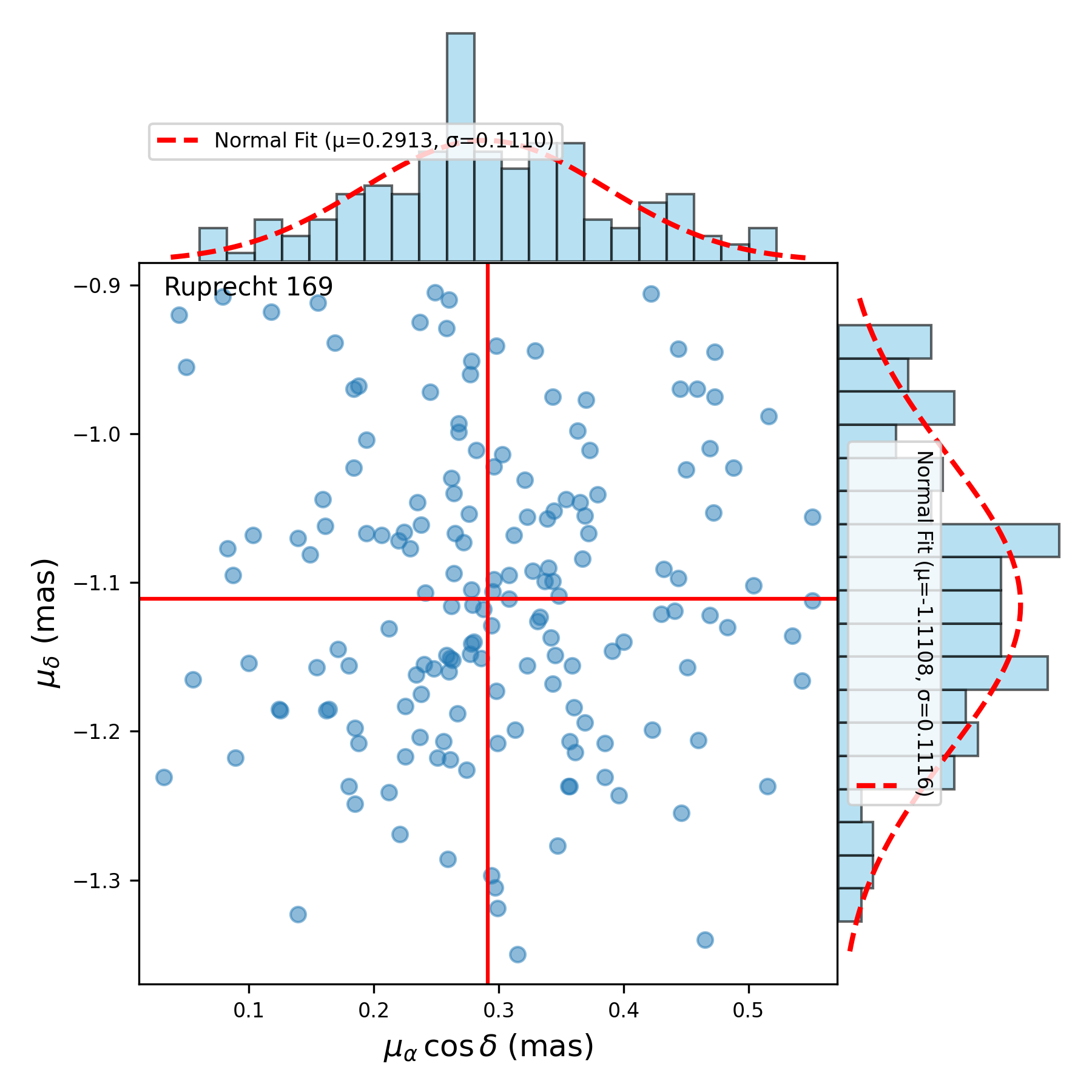}}}
\caption{Histograms of the proper motion components ($\mu_{\alpha}\cos\delta$,~$\mu_{\delta}$) for the most probable members of each target cluster. The mean value and standard deviation for each distribution are noted in the panels.
\label{fig: PM}}
\end{figure*}

Estimates of the mean proper-motion components in both directions, and the mean trigonometric parallax for every cluster were obtained by constructing histograms for the proper-motion components and parallaxes of all member stars, and then we searched for the best-fit Gaussian profiles of the constructed histograms. Next, the mean values and standard deviations of the best fit profiles were assigned to the clusters as shown in Figures \ref{PlxHists} and \ref{fig: PM}. The estimated mean proper-motion components in both directions, and parallaxes of the studied clusters are listed in Table \ref{all_results}.

\begin{figure}[htb!]
\scalebox{.5}{\includegraphics{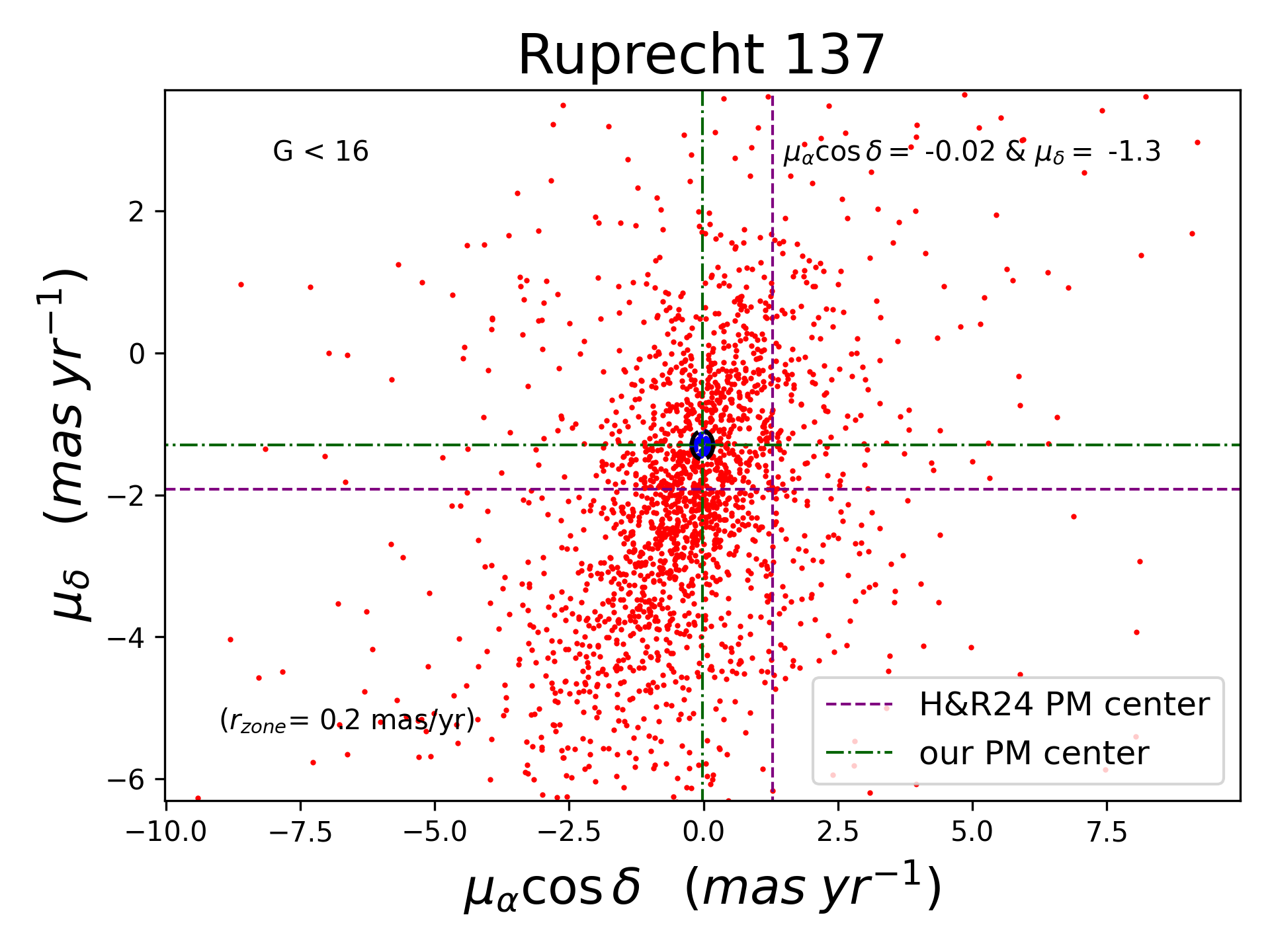}}
\caption{The VPD for bright stars ($G<16$) in the field of Ruprecht 137. The proper motion centre adopted in this study is marked by the green dashed-dotted lines, which differ significantly from the literature value reported by \cite{Hunt2024} (H\&R24), marked by the purple dashed lines.
\label{VPD_Ruprecht137}}
\end{figure}

Comparison with the mean proper motion components and parallaxes obtained from literature listed in Table \ref{tab:literature} shows that our mean proper-motion components and mean parallaxes estimates agree well with the corresponding values in the studies that used either $Gaia$ DR2 or DR3 databases, e.g. \cite{Hunt2024}, \cite{Pog21} and\cite{Cant20a}, while our results differ significantly from those of works that used other older databases, eg. \cite{dias2014proper} \& \cite{kharchenko2013global}. In addition, we present the measured values of the newly discovered substructure in the field of the NGC 7245 OC, i.e. NGC 7245b, see Table \ref{all_results}.

We note that our mean proper-motion estimates of the OC Ruprecht 137 differ significantly from those of \cite{Hunt2024} (1.286 \& $-$1.918). However, the VPD of the bright stars of this cluster, $G<16$, existing within its estimated radius, shows that there is no clear stellar assembly in their assigned mean proper-motion components of this cluster, see Figure \ref{VPD_Ruprecht137}. On the contrary, there is a clear stellar aggregation in our estimated mean proper-motion components.

Also, mean distance estimates of the clusters, $d_\varpi$,  were computed using their mean trigonometric parallaxes, which are also listed in Table \ref{all_results}. Similarly, our computed distances of NGC 559, NGC 1817, NGC 2141, NGC 7245, and Ruprecht 137 agree fairly well with the distances reported in the studies that used the $Gaia$ database in their works, as shown in Table \ref{tab:literature}. Our results of Ruprecht 15, Ruprecht 137, Ruprecht 142, and Ruprecht 169 differ significantly from the reported values in literature that were prior to the $Gaia$ era. However, our estimations are based on the more precise proper motion and trigonometric parallax measurements of the $Gaia$ DR3 dataset, unlike the other works.

\begin{figure*}[ht!]
\centering
\includegraphics[width=0.85\linewidth]{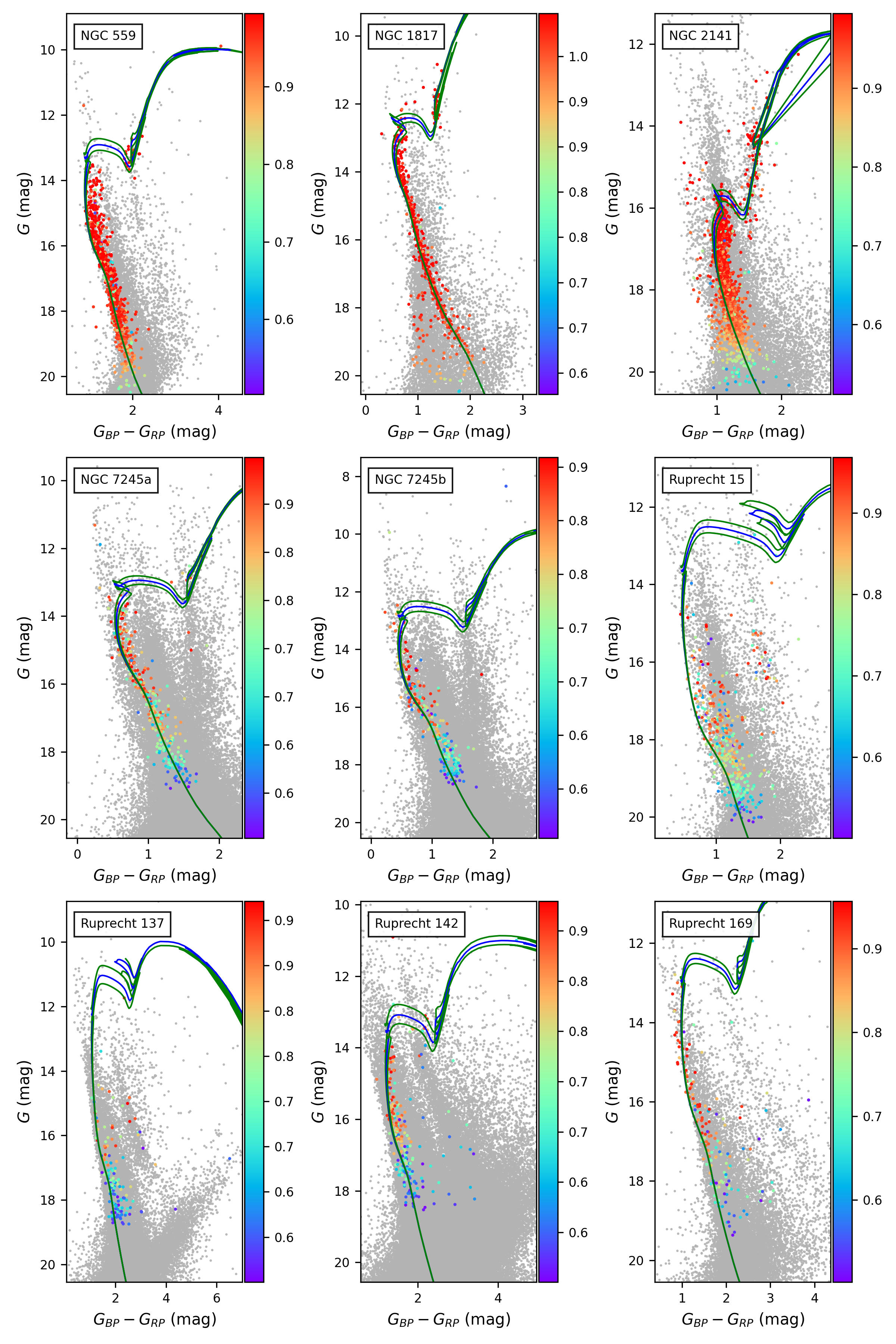}
\caption{CMDs for the clusters analysed in this study. Candidate members with membership probability is \(0.5 \leq P\) are shown using individual colour schemes, while probable field stars are plotted as grey points. Overlaid are the adopted {\sc PARSEC} stellar isochrones in blue; their confidence envelopes are drawn in green to indicate the model uncertainties.
\label{CMDs}}
\end{figure*}

\subsection{Astrophysical Parameters}\label{CMD-Age-Dist}

In this study, the fundamental astrophysical parameters of the clusters were obtained by fitting their observed colour–magnitude diagrams (CMDs) with theoretical stellar evolutionary models. We adopted the {\sc PARSEC} isochrones \citep{Bressan2012}, choosing the metallicity values that best represent each system. Isochrones with $Z=0.015$ were adopted for NGC~2141, Ruprecht~15, Ruprecht~137, Ruprecht~142, and Ruprecht~169, whereas models with $Z=0.009$, $Z=0.007$, and $Z=0.011$ were used for NGC~559, NGC~1817, and the two components of NGC~7245, respectively.

For clusters with available iron-abundance estimates, the [Fe/H] values were transformed into the corresponding heavy-element mass fraction $Z$ to remain consistent with the metallicity scale of the {\sc PARSEC} models. This conversion was performed using analytical prescriptions calibrated for the PARSEC framework \footnote{\url{https://github.com/jobovy/isodist/blob/master/isodist/Isochrone.py}}:
\begin{equation}
z_{\rm x} = 10^{{\rm [Fe/H]} + \log\left(\frac{Z_{\odot}}{1 - 0.248 - 2.78~Z_{\odot}}\right)},
\end{equation}
\citep{2023AJ....165...79Y,2023AJ....166..263G} where $Z$ is related to $z_{\rm x}$ using the following formula
\begin{equation}
Z = \frac{z_{\rm x} (1 - 0.2485)}{1 + 2.78~z_{\rm x}}.
\end{equation}

Here, the adopted solar heavy-element abundance is $Z_{\odot}=0.0152$ \citep{Bressan2012}. These relations allow a seamless conversion between iron abundance and mass fraction, ensuring that the isochrone fitting procedure remains consistent across all clusters in the sample \citep{YontanCanbay2023, Tasdemir2023, Yontan2022, Tasdemir2025a}. Based on this transformation, the metallicity values adopted in the analysis were [Fe/H] = 0.00 dex for NGC~2141, Ruprecht~15, Ruprecht~137, Ruprecht~142, and Ruprecht~169, while NGC~559, NGC~1817, and the two components of NGC~7245 were assigned metallicities of $-$0.20 dex, $-$0.30 dex, and $-$0.10 dex, respectively.

Figure \ref{CMDs} shows the {\it G } versus $(G_{\rm BP} - G_{\rm RP})$ CMDs for the studied clusters and their best fit isochrones obtained through visual inspection, i.e. comparing the observed CMD of each cluster with isochrones of different ages, where each isochrone was shifted vertically and horizontally until the best-fit isochrone was obtained. Then, the selected clusters' ages were those of the best-fit isochrones.

The isochrone data includes absolute magnitudes, $M_G$, and intrinsic colours, $(G_{\rm BP} - G_{\rm RP})$, computed for different stellar masses at the same age, which can be used to determine the cluster's observed distance modulus and reddening. Cluster's observed distance modulus is equal to the difference between the observed stellar magnitudes and the absolute magnitudes of the best-fit isochrone, $m_G-M_G$, while its colour excess is the difference between the observed stellar colours and the corresponding values of the isochrone, $E(G_{\rm BP} - G_{\rm RP})$. 

As shown in Figure \ref{CMDs}, the best-fit isochrone ages of the analysed clusters span a wide range, from very young systems of about $\sim$90–100 Myr to intermediate- and old-age clusters reaching up to $\sim$2.2 Gyr. Similarly, the derived distance moduli vary between approximately 11.6 and 15.8 mag, while the colour excesses cover a broad interval, reflecting differing extinction conditions along the lines of sight. Rather than listing individual values here, all adopted isochrone parameters, including ages, distance moduli, and colour excesses for each cluster, are summarised in detail in Table~\ref{all_results}.

The corresponding $E(B-V)$ colour excesses are derived from the $Gaia$ colours using $E(B-V)=0.775 \times E(G_{\rm BP}-G_{\rm RP})$ \citep{cardelli1989relationship} and are found to vary between approximately 0.25 and 1.0 mag. The line-of-sight extinction in the $Gaia$ $G$ band is then estimated using $A_G = 2.74 \times E(B-V)$ \citep{casagrande2018use,zhong2019substructure,2023AJ....165..163C}, yielding $A_G$ values in the range $\sim$0.7–2.8 mag. All individual reddening and extinction parameters are listed in Table~\ref{all_results}. The photometric distances of the clusters can be obtained using their estimated distance moduli by the following relation:
\begin{equation}
d=10^{((m-M)_{obs}-A_{G}+5)/5}\,.
\end{equation}

The photometric distances of NGC 559, NGC 1817, NGC 2141, NGC 7245 (a\&b), Ruprecht 15, Ruprecht 137, Ruprecht 142 and Ruprecht 169 are 2273 $\pm$ 411, 1548 $\pm$ 182, 4267 $\pm$ 405, 2677 $\pm$ 311, 2935 $\pm$ 340, 7290 $\pm$ 1325, 2228 $\pm$ 557, 2333 $\pm$ 437, and 2360 $\pm$ 340 pc, respectively, which agree within uncertainties with their computed astrometric distances. 
 
Our estimated age of NGC 559 is lower than that of \cite{kharchenko2013global}, and it is higher than the rest of the values mentioned in Table \ref{tab:literature}. However, we did not assume a solar metallicity like the rest of the other works.
The selected ages of NGC 1817 and NGC 2141 agree within uncertainties with those of \cite{Pog21} and \cite{Cant20b}, while they are lower than those of \cite{Hunt2024}. 
The selected age of NGC 7245a agrees within uncertainties with those of \cite{Hunt2024}, \cite{Pog21} and \cite{Cant20b}, while the other substructure that was not identified by these studies, NGC 7245b, has a lower age, which agrees with the reported age of \cite{sampedro2017multimembership}. 
The age of Ruprecht 142 agrees within uncertainties with the reported values in literature, while the ages of Ruprecht 15, Ruprecht 137, and Ruprecht 169 are younger than the reported values in literature, see Table \ref{tab:literature}.
Our estimated photometric distances mostly agree within uncertainties with the corresponding trigonometric distances, except Ruprecht 15, which has a higher value. 

Using the measured heliocentric distances, the distances of the clusters from the Galactic plane, $Z_\odot$, and their projected distances from the Sun, $X_\odot$ and $Y_\odot$, and their distances from the Galactic centre ($R_{\rm GC}$) were computed by:
\begin{align}
X_{\odot} = d \cos b \cos l\,, Y_{\odot} = d \cos b \sin l \,,Z_{\odot} = d \sin b \,
\end{align}
\begin{equation}
R_{GC} = \sqrt {R_{\odot}^{2}+\textcolor{red}{d^{2}}-2R_{\odot} d \cos b \cos l}\,, 
\end{equation}
where $d$ is the distance to the Sun in parsec, $R_\odot $ is the distance between the Sun and the Galactic centre \citep{2019AdSpR..63.1360T}. $R_\odot $ was assigned a value of 8.20 $\pm$ 0.10 kpc as reported by \cite{bland2019galah}. The updated Galactocentric distances ($R_{GC}$) of the analysed clusters span a wide range, from $\sim$5.9 kpc for the innermost systems to $\sim$13.8 kpc for the most distant outer-disk cluster. These revised $R_{GC}$ values are derived consistently for all targets and are summarised in Table~\ref{all_results}.

\begin{table*}[!ht]
\centering

\caption{Summary of derived astrophysical, astrometric, and spatial parameters for the target OCs. Parameters include coordinates ($\alpha,~\delta$), Galactic position ($l,~b$), number of members ($N_{stars}$), proper motions ($\mu_{\alpha}\cos\delta,~\mu_{\delta}$), parallax ($\varpi_{cor}$), distance ($d_\varpi$), photometric results (distance modulus, reddening, distance $d_{DM}$, age), Galactic coordinates ($X,~Y,~Z,~R_{GC}$), and metallicities ($Z$, [Fe/H]).}
\label{all_results}
\resizebox{\textwidth}{!}{
\begin{tabular}{lcccc c cccc c} 
\hline\hline
Cluster & $\alpha$ & $\delta$ & $\ell $ & $b$ & $N_{stars}$ & $\mu_{\alpha}\cos\delta$ & $\mu_{\delta}$ & $\varpi_{cor}$ & $\Delta\varpi_{cor}$ & $d_\varpi$ \\
& (hh:mm:ss) & (dd:mm:ss) & (deg) & (deg) & & $\rm (mas\;yr^{-1})$ & $\rm (mas\;yr^{-1})$ & (mas) & (mas) & (kpc) \\
\hline
NGC 559      & 01:29:34 & +63:17:32 & 127.2009 &    0.7368 & 779  & $-$4.268 $\pm$ 0.003 & 0.180 $\pm$ 0.004   & 0.361 $\pm$ 0.002 & 0.027 &  2.773 $\pm$ 0.018 \\
NGC 1817     & 05:12:40 & +16:42:44 & 186.1984 & $-$13.0167 & 590  & 0.420 $\pm$ 0.004    & $-$0.932 $\pm$ 0.004 & 0.610 $\pm$ 0.002  & 0.030 & 1.640 $\pm$ 0.005\\
NGC 2141     & 06:02:59 & +10:27:19 & 198.0454 &  $-$5.7970 & 1551 & $-$0.076 $\pm$ 0.004 & $-$0.751 $\pm$ 0.003 & 0.240 $\pm$ 0.003  & 0.025 &  4.167 $\pm$ 0.052\\
NGC 7245a    & 22:14:55 & +54:21:35 & 101.3539 &  $-$1.8339  & 365  & $-$3.930 $\pm$ 0.004 & $-$3.284 $\pm$ 0.004 &  0.303 $\pm$ 0.002 & 0.026 &  3.304 $\pm$ 0.023 \\
NGC 7245b    & 22:14:55 & +54:21:35 & 101.3539 &  $-$1.8339 & 239  & $-$3.316 $\pm$ 0.004 & $-$3.317 $\pm$ 0.005 & 0.295 $\pm$ 0.003  & 0.026 &  3.386 $\pm$ 0.037 \\
Ruprecht 15  & 07:20:00 & $-$19:42:43 & 233.6628  &  $-$2.8463  & 410  & $-$0.717 $\pm$ 0.006 & 1.476 $\pm$ 0.007    & 0.192 $\pm$ 0.006  & 0.023 &  5.203 $\pm$ 0.156 \\
Ruprecht 137 & 18:00:14 & $-$25:14:45 & 4.8306 &  $-$0.9229 & 151  & $-$0.005 $\pm$ 0.012 & $-$1.309 $\pm$ 0.010 & 0.380 $\pm$ 0.010  & 0.035 &  2.631 $\pm$ 0.066 \\
Ruprecht 142 & 18:32:40 & $-$12:17:08 & 19.8759 &  $-$1.4830 & 231  & 0.744 $\pm$ 0.005   & $-$1.114 $\pm$ 0.005  & 0.390 $\pm$ 0.005  & 0.037 & 2.565 $\pm$ 0.032 \\
Ruprecht 169 & 17:59:37 & $-$24:49:33 & 5.1269 &  $-$0.5948 & 168  & 0.291 $\pm$ 0.009    & $-$1.111 $\pm$ 0.009 & 0.427 $\pm$ 0.006  & 0.034 &  2.344 $\pm$ 0.034 \\
\hline
\end{tabular}}
\vspace{5pt} 
\resizebox{\textwidth}{!}{
\begin{tabular}{lcccccccccccc} 
\hline\hline
\noalign{\smallskip}
Cluster & $(m-M)_{obs}$ & $E(G_{\rm BP}-G_{\rm RP})$ & $E(B-V)$ & $A_{G}$ & $d_{DM}$ & Age & $X_{\odot}$ & $Y_{\odot}$ & $Z_{\odot}$ & $R_{GC}$ & $Z$ & [Fe/H] \\
 & (mag) & (mag) & (mag) & (mag) & (kpc) & (Myr) & (kpc) & (kpc) & (kpc) & (kpc) & & (dex) \\
\noalign{\smallskip}
\hline
\noalign{\smallskip}
NGC 559      & $13.80 \pm 0.20$ & $0.95 \pm 0.05$ & 0.74 & 2.02 & $2.273 \pm 0.411$ & $450 \pm 50$ 
 & $-$1.374 & 1.810 & 0.029  & 9.74 $\pm$ 0.31 & $0.009$ & $-0.20$ \\
NGC 1817     & $11.65 \pm 0.10$ & $0.33 \pm 0.05$ & 0.26 & 0.70 & $1.548 \pm 0.182$ & $1275 \pm 100$ & $-$1.499 & $-$0.163 & $-$0.349 & 9.70 $\pm$ 0.18 & $0.007$ & $-0.30$ \\
NGC 2141     & $14.00 \pm 0.10$ & $0.40 \pm 0.05$ & 0.31 & 0.85 & $4.267 \pm 0.405$ & $2200 \pm 200$ & $-$4.036 & $-$1.315 & $-$0.431 & 12.31 $\pm$0.77 & $0.015$ & $0.00$  \\
NGC 7245a    & $13.20 \pm 0.10$ & $0.50 \pm 0.05$ & 0.39 & 1.06 & $2.677 \pm 0.311$ & $660 \pm 50$  & $-$0.527 & 2.623  & $-$0.086 & 9.11 $\pm$ 0.15 & $0.011$ & $-0.10$ \\
NGC 7245b    & $13.40 \pm 0.10$ & $0.50 \pm 0.05$ & 0.39 & 1.06 & $2.935 \pm 0.340$ & $450 \pm 50$  & $-$0.578 & 2.876  &$-$0.094 & 9.24 $\pm$0.17 & $0.011$ & $-0.10$ \\
Ruprecht 15  & $15.80 \pm 0.20$ & $0.70 \pm 0.05$ & 0.55 & 1.49 & $7.290 \pm 1.325$ & 
$100 \pm 10$  & $-$4.314 & $-$5.865 & $-$0.362 & 13.82 $\pm$1.00 & $0.015$ & $0.00$  \\
Ruprecht 137 & $14.50 \pm 0.30$ & $1.30 \pm 0.05$ & 1.01 & 2.76 & $2.228 \pm 0.557$ & $90 \pm 15$   & 2.220  & 0.188  & $-$0.036 & 5.98 $\pm$0.55 & $0.015$ & $0.00$  \\
Ruprecht 142 & $14.60 \pm 0.30$ & $1.30 \pm 0.05$ & 1.01 & 2.77 & $2.333 \pm 0.437$ & $300 \pm 50$  & 2.193  & 0.793  & $-$0.060 & 6.06 $\pm$0.27 & $0.015$ & $0.00$  
\\
Ruprecht 169 & $14.20 \pm 0.15$ & $1.10 \pm 0.05$ & 0.85 & 2.34 & $2.360 \pm 0.340$ & $250 \pm 20$  &2.350  & 0.211  & $-$0.024 & 5.85 $\pm$0.34 & $0.015$ & $0.00$  \\
\noalign{\smallskip}
\hline
\end{tabular}
}
\end{table*}

\section{Luminosity and Mass Functions}

The luminosity function (LF) of a cluster is a histogram of the luminosities or absolute magnitudes of all its members, and to construct it, we divide the full range of magnitudes into a number of equally spaced bins, and then we count stars in each bin \citep{2015Ap&SS.355..267Y, 2015MNRAS.453.1095B, Bostanci2018}. The absolute magnitudes of member stars in a cluster can be computed using their observed apparent magnitudes and the cluster's observed distance modulus \citep{Alzahrani2025a, Alzahrani2025b}.
Clusters' LFs can help us to get further information about the clusters, such as the different stellar populations that exist within them.

Next, by studying the distribution of stellar masses within the cluster, we can get valuable insight into its structure and evolutionary state. Significant variations are observed among the clusters, not only in terms of their total mass but also in the number of stars and the mean stellar mass per OC. These differences reflect the heterogeneity in their stellar populations and the various stages of evolution they may be undergoing. $\langle M_C \rangle$ of the stars in each OC offers a further understanding of the distribution of stellar masses within the population. 

As mentioned earlier, the theoretical isochrone of \cite{marigo2017new} computed at a certain age contains information about the luminosities of stars with different masses.

An approximate relation between stellar mass and luminosity can be obtained by fitting isochronous mass and luminosity data with a fourth-degree polynomial over an absolute magnitude range of interest, which is known as the mass-luminosity relationship (MLR), and is represented by
\begin{equation}
M_{\rm c} = a_0 + a_1 \times M_{\rm G} + a_2 \times M_{\rm G}^2 + a_3 \times M_{\rm G}^3 + a_4 \times M_{\rm G}^4,
\label{Eq: ML}
\end{equation}
where $a_0,~a_1,~a_2,~a_3$, and $a_4$ are polynomial coefficients that we get from the fitting \citep{2017SerAJ.194...59A}. 
The values of the absolute magnitudes of the members of the study clusters vary between $-$4.0 and 7.0 mag (for NGC 559), $-$1.0 and 9.0 mag (for NGC 1817), $-$2.0 and 6.5 mag (for NGC 2141), $-$2.0 and 6.0 mag (for NGC 7245a), $-$5.0 and 5.5 mag (for NGC 7245b), $-$3.0 and 4.5 mag (for Ruprecht 15), $-$2.5 and 4.5 mag (for Ruprecht 137), $-$4.0 and 4.0 mag (for Ruprecht 142) and $-$2.5 and 5.5 mag (for Ruprecht 169).

The luminosity function can be converted into the mass function through MLRs, which are used to compute the masses of the individual members of the cluster \citep{eker2024, 2019Ap&SS.364..152Y, 2016Ap&SS.361..126A}. 
Next, the cluster's total mass $(M_{\text{C}};~M_{\odot})$, can be obtained by summing the masses of all members, and its mean mass $(\langle M_C \rangle;~M_{\odot})$ is obtained by dividing the total mass by the number of members. Fitting polynomial coefficients, total mass, and average mass of each cluster are listed in Table \ref{MLR_results}.

\begin{table*}[htbp]
\centering
\caption{MF and MLR results for the target OCs. Listed are the total derived cluster mass ($M_{\text{C}}$), the mean stellar mass ($\langle M_C \rangle$), the derived MF slope ($\alpha$), the coefficients ($a_0...a_4$) for the polynomial MLR fit, and the correlation coefficient ($R^2$).}
\label{MLR_results}
\scalebox{0.8}{
\begin{tabular}{lcccccccccc} 
\hline\hline
Cluster & $M_{\text{C}}\;(M_\odot)$ & $\langle M_C \rangle \;(M_\odot)$ & $\alpha$ & $a _0$ & $a _1$ & $a _2$ & $a _3$ & $a _4$ & $R^2$\\ 
\hline
NGC 559& 1089 $\pm$ 87  &  1.40  &  1.96 $\pm$ 0.12  &  2.59  &  $-$0.32  &  $-$0.11  &  0.04  &  $-$0.003  &  0.9987 \\ 
NGC 1817& 632 $\pm$ 33  &  1.07  &  2.33 $\pm$ 0.21  &  1.93  &  $-$0.05  &  $-$0.10  &  0.02  & $-$0.001  &  0.9939 \\
NGC 2141& 1916 $\pm$ 233  &  1.24  &  2.13 $\pm$ 0.22  &  1.33  &  0.65  &  $-$0.38  &  0.07  &  $-$0.004  &  0.9966\\
NGC 7245a& 515 $\pm$ 23  &  1.41  &  2.12 $\pm$ 0.31  &  2.42 &  $-$0.28  & $-$0.08  &  0.02  &  $-$0.002 &  0.9979 \\
NGC 7245b& 355 $\pm$ 20  &  1.48  &  2.72 $\pm$ 0.41  &  2.62  &  $-$0.35  &  $-$0.11  &  0.04  &  $-$0.003  &  0.9978 \\
Ruprecht 15& 807 $\pm$ 78 &  1.97  &  2.29 $\pm$ 0.25  &  3.44  &  $-$0.92  &  0.02  &  0.04  &  $-$0.004  &  0.9997 \\
Ruprecht 137 & 257 $\pm$ 35  &  1.70  &  3.07 $\pm$ 0.39  &  3.50  &  $-$1.01  &  0.02  &  0.05  &  $-$0.007  &  0.9999 \\
Ruprecht 142& 430 $\pm$ 41  &  1.86  &  1.97 $\pm$ 0.33  &  2.81  &  $-$0.46  & $-$0.09  &  0.04  &  $-$0.0004  &  0.9970\\
Ruprecht 169 & 296 $\pm$ 19  & 1.76  &  2.04 $\pm$ 0.38  &  2.98  & $-$0.57  &  $-$0.06  &  0.04  &  $-$0.003  &  0.9984\\
\hline 
\end{tabular}}
\end{table*}
Study of \cite{salpeter1955} shows that the number of stars decreases with the increase of stellar mass following a power-law relation of a power law index ($\alpha$) equal to 2.35 written as:
\begin{equation}
\frac{dN}{dM} \propto M^{-\alpha}\,.
\label{sec:MF}
\end{equation}

\begin{figure*}[htb!]
\includegraphics[width=0.99\linewidth]{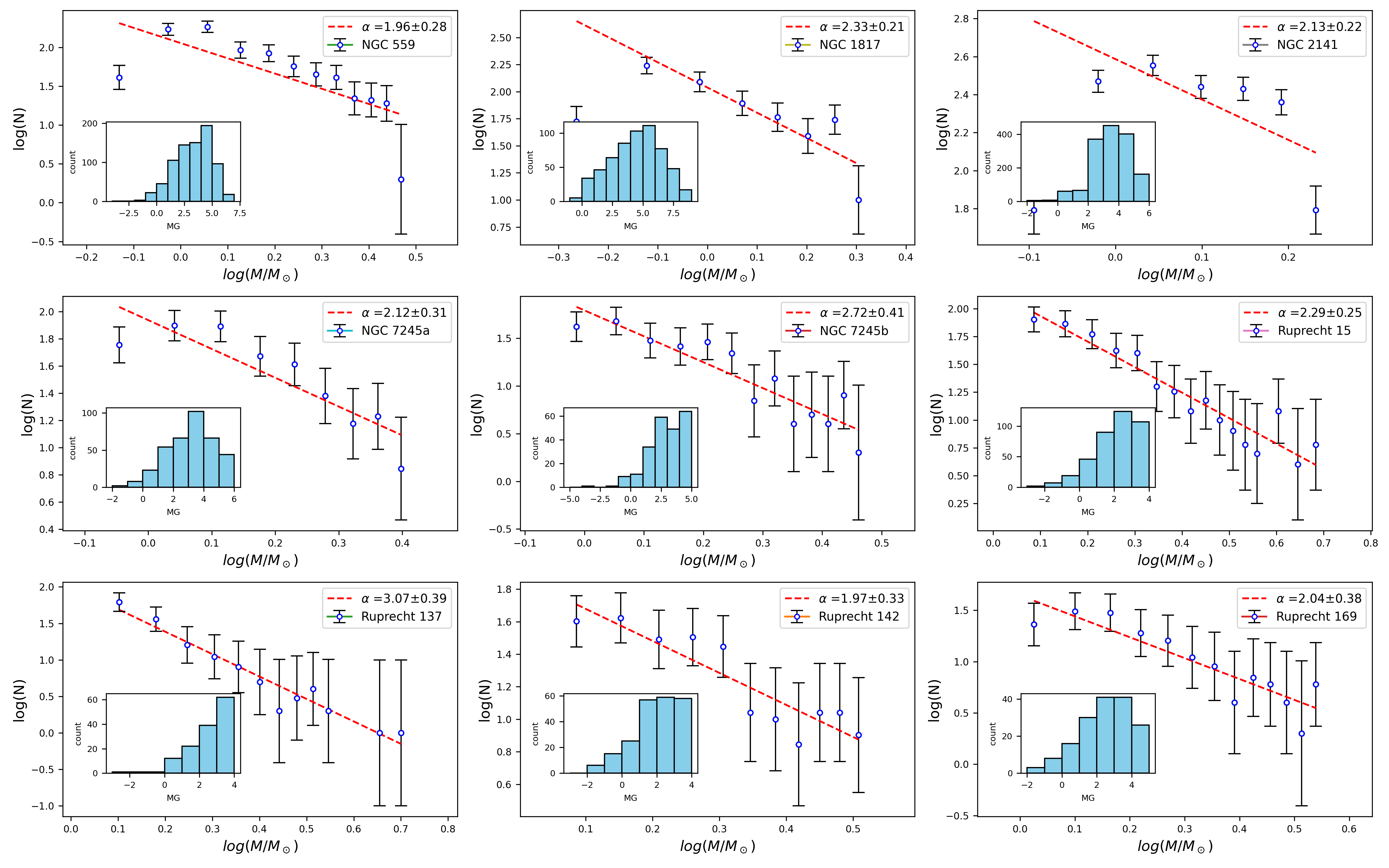}
\caption{The MFs and LFs of the investigated OCs, where the slope of the linear fits, $\alpha$, of the MFs are shown in the legends.
\label{IMFs}}
\end{figure*}

Figure \ref{IMFs} shows the LF and MF of the investigated clusters. Linear relations were fitted for the observed MFs where their slopes are the $\alpha$ values of these clusters, which are listed in Table \ref{MLR_results} as well. The incompleteness of data for $G<6$ and $G<5$ for NGC 1817 and NGC 2141, respectively, shown in Figure \ref{IMFs} leads to a significant underestimation of mass in the lowest mass bin in each cluster, which may cause a bias in the results. Consequently, the first bin count in each cluster was excluded from the linear fit. Results reflect good agreement with \citet{salpeter1955} value within uncertainties.

\section{Dynamical State of the OCs}

Most of the observed star clusters have almost Gaussian star velocity distributions with small internal dispersions (0.1–0.5 km s$^{-1}$), which are typical of virial equilibrium systems \citep{Soubiran_2018}. However, several young clusters show streaming movements and kinematic clumps in their structure, indicating that many star clusters are born with complex velocity structures \citep{Kounkel2019}. Some OCs have also been observed to rotate. According to \citet{Galli2021}, the Hyades and Praesepe exhibit indications of light internal rotation, which may be a result of tidal interactions or a reminder of their formation process.

Even though star clusters may be born with random internal structure, gravitational interactions between the members help them to relax and reach an equilibrium state, and to have a uniform Gaussian velocity distribution. 
However, not all clusters are allowed to reach such an equilibrium state, especially clusters that lie in or close to the Galactic disc. The disruptive, gravitational effects of the giant molecular clouds and the galactic spiral arms can cause the breakdown of most of OCs in this region \citep{Spitzer1971, Kruijssen2020}.

\subsection{Dynamical Relaxation Time}

The interval of time cluster needs to reach equilibrium or equipartition state via gravitational interactions between the members is known as the dynamical relaxation time ($T_{\rm relax}$).The cluster's dynamical relaxation time is a function of cluster's size, mass distribution, and number of stars. Most of OCs relax over time intervals ranging from tens to hundreds of Myr \citep{binney08}.

During this period, the cluster evolves from a beginning, potentially asymmetrical condition to a dynamically relaxed configuration. Following the prescription of \citet{Spitzer1971}, $T_{relax}$ is calculated using the relation:
\begin{equation}
\label{Eq: t_relax}
T_\text{relax} = \frac{8.9 \times 10^5 N^{1/2} R_\text{h}^{3/2}}{\langle M_C \rangle^{1/2} \log(0.4N)},
\end{equation}
where $N$ represents the cluster's total number of stars, $R_{\rm h}$ is the half-mass radius (pc), and $\langle M_C \rangle$ represents the mean cluster mass. \citet{Larsen2006} provided a formula that was used to derive the half-mass radius:
\begin{equation}
R_{\rm h} = 0.547 \times r_{\rm c} \times \left( \frac{r_{\rm t}}{r_{\rm c}} \right)^{0.486},
\end{equation}
where $r_{\rm cl}$ (pc) is the cluster radius derived from the RDP analysis, as reported in Table \ref{King_para}. Therefore, the obtained results of $T_{relax}$ (Myr) are 70 $\pm$ 3 (NGC 559), 93 $\pm$ 2 (NGC 1817), 135 $\pm$ 9 (NGC 2141), 190 $\pm$ 5 (NGC 7245a), 170 $\pm$ 6 (NGC 7245b), 121 $\pm$ 8 (Ruprecht 15), 21 $\pm$ 2 (Ruprecht 137), 71 $\pm$ 4 (Ruprecht 142) and 16 $\pm$ 1 (Ruprecht 169). The calculated $R_{\rm h}$ (pc) and $T_{\rm relax}$ are listed here with Table \ref{tab:evol_kin_dyn_parameters}.

The dynamical evolution parameter, $\tau$ = age/$T_{relax}$, may be utilized to characterize the dynamical state of clusters. This parameter is significantly greater than unity ($\tau \gg 1$) for relaxed clusters and vice versa for non-relaxed ones. Values of $\tau$ are also listed here with Table \ref{tab:evol_kin_dyn_parameters}, which shows that most of these open clusters are dynamically relaxed ones, except NGC 7245a and NGC 7245b are recent relaxed clusters, while Ruprecht 15 is a non-relaxed one.

The period during which all member stars are expelled as a result of internal stellar interactions is represented by the evaporation time, which is $ \tau_{\text{ev}} \simeq 10^2 T_{\text{relax}}$ (in Myr) \citep{Adams2001}. Therefore, low-mass stars usually escape through the Lagrange points in this mechanism at low velocities \citep{koposov2008}. Our obtained results are summarized in Table \ref{tab:evol_kin_dyn_parameters}, which reflects that these clusters are relaxed ones except Ruprecht 15.

\begin{figure*}[htb!]
\includegraphics[width=0.99\linewidth]{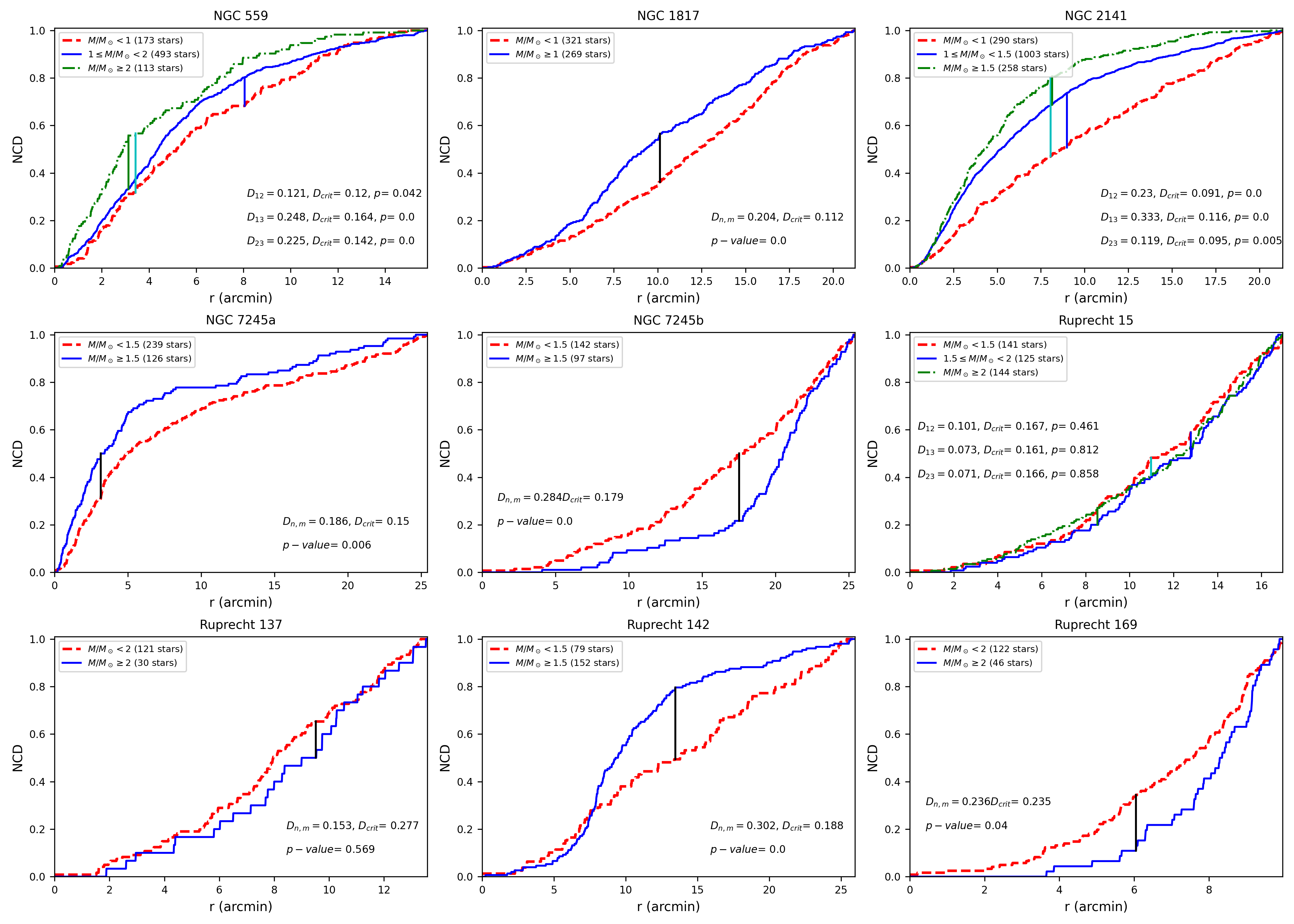}
\caption{The cumulative distributions of the estimated masses of the members of each cluster. 
\label{KS_tests}}
\end{figure*}

\subsection{Mass Segregation} 

As star clusters evolve to the equilibrium state or energy equipartition state, massive stars will have smaller velocities and will sink toward the centre, while the less massive stars will have larger velocities and move closer to the cluster's external regions. Consequently, clusters are expected to show mass segregation in their structures. However, it is still a matter of debate whether this observed mass segregation results from the dynamical interactions between the member stars within the clusters as discussed above, or whether the clusters were formed and born that way. 

The cumulative radial distribution of member stars within the cluster for two or three mass ranges is used to look for the presence of mass segregation. By knowing the presence or absence of mass segregation in the clusters and their dynamical relaxation time calculated above, we can look for an answer to this debate. Such that if the cluster shows a clear mass segregation and its age is less than the relaxation time, this will indicate that the cluster was born this way. On the other hand, if the cluster shows no mass segregation and its age is less than the relaxation time, this will indicate that mass segregation results from dynamical interactions. 

The Kolmogorov–Smirnov test was used to look for signs of mass segregation in the investigated cluster. As it searches for the maximum difference between the cumulative radial distributions of two samples with sizes n and m, $D_{n,m}$, which is computed using the following relation:
\begin{equation}
D_{n,m}=sup_x |F(x)-G(x)|
\end{equation}

The null hypothesis, which is that the two distributions are similar, is rejected if $D_{n,m}$ is greater than the critical value $D_{crit}$ computed for the same sizes as those of the two samples and for a selected level of significance $\alpha$.

Figure \ref{KS_tests} presents the cumulative radial distribution of stars in two or three mass ranges for each cluster, where the difference between the distribution and the probability of resemblance, p-value, is shown in Figure \ref{KS_tests}. Also, only Ruprecht 15, Ruprecht 137, the youngest two clusters in our sample, show no mass segregation signs, while the other older clusters show clear mass segregation for $\alpha$ = 0.05 (or 5\% resemblance). Such results may support the opinion that mass segregation results from dynamical interactions.

\section{Convergent Point}
On the celestial sphere, the proper motion (PM) vectors of stars belonging to a co-moving system, such as an OC appear to converge, due to projection effects, toward a specific location known as the convergent point (CP) \citep{Elsanhoury2018, Bisht2022a, Bisht2022b, 2024AJ....167..188B, Elsanhoury2025, Elsanhoury2025b}. CP method was originally employed in early 20th-century astrometry to estimate distances to nearby clusters, such as the Hyades. With the advent of high-precision astrometric data from missions like Gaia, the technique has since undergone substantial refinement and enhancement. When a cluster moves nearly parallel through space, the CP corresponds to the point on the sky toward which the observed PM vectors of its member stars seem to converge. Mathematically, this phenomenon is governed by vector motion relations and spherical trigonometry \citep{Galli_2012}. 

\begin{figure*}[!ht]
\centering
\includegraphics[width=0.8\linewidth]{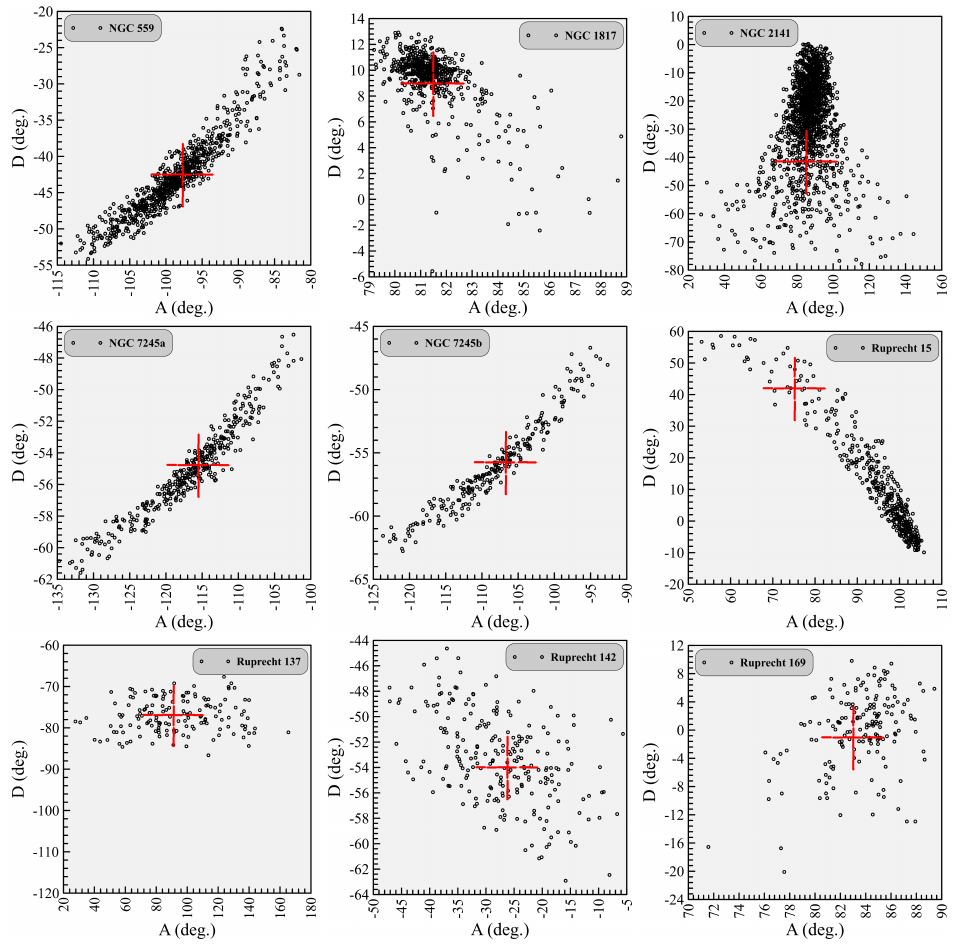}
\caption{AD-diagrams of NGC 559, NGC 1817, NGC 2141, NGC 7245a, NGC 7245b, Ruprecht 15, Ruprecht 137, Ruprecht 142, and Ruprecht 169. The cross symbols mark the positions of the computed apex coordinates $(A_0,~D_0)$.}
\label{fig: CP}
\end{figure*}

The advent of $Gaia$ DR3 has substantially advanced the application of the CP technique. For instance, \citet{2023A&A...678A..75Z} refined cluster membership lists and derived LFs across different age ranges by employing an enhanced CP method for approximately 50 $\%$ OCs within 0.5 kpc. Their analysis revealed that older clusters, likely influenced by dynamical evolution, exhibit greater diversity, whereas younger clusters ($\approx$ 50 Myr) display more uniform luminosity profiles. In another study, \citet{2025A&A...694A.258R} detected tidal tails in 19 of 21 nearby OCs using the CP method. These extended stellar features form when stars escape the gravitational potential of their parent clusters. To improve the detection of such structures, they introduced the self-compact convergent point (SCCP) technique. Interestingly, some tidal tails were found to extend outward from the Galactic centre-contrary to traditional expectations regarding their orientation.

Determining the apex position provides essential insights into the cluster’s dynamical coherence and overall kinematic structure. An improvement on the traditional CP method is the AD-diagram method, also known as the apex diagram method. This is typically achieved using the AD-diagram method, which is an improvement originally proposed by \citet{Chupina2001, Chupina2006}. As illustrated in Figure \ref{fig: CP}, the apex coordinates ($A_0,~D_0$) are obtained from the mean velocity components of individual cluster members. Here, $A_0$ and $D_0$ represent the right ascension and declination of the convergent point, respectively, and are computed as follows:

\begin{equation}
\label{Eq: A_o and D_o}
A_0 = \tan^{-1}\left(\dfrac{\overline{V}_y}{\overline{V}x}\right),
\end{equation}
\begin{equation}
\label{Eq: 8}
D_0 = \tan^{-1} \left(\dfrac{\overline{V}z}{\sqrt{\overline{V}^{2}{x} + \overline{V}^{2}{y}}}\right),
\end{equation}
where $\overline{V}_x$, $\overline{V}_y$, and $\overline{V}_z$ denote the mean space velocity components along the respective Cartesian axes. These components are derived from the observed parameters using the transformation equations given by \citet{Melchior1958}:

\begin{equation}
\label{Eq. 9-10-11}
\begin{pmatrix}
V_x \\\\
V_y \\\\
V_z
\end{pmatrix} 
= 
\begin{pmatrix}
-4.74~d~\mu_\alpha\cos{\delta}\sin{\alpha} - 4.74~d~\mu_\delta\sin{\delta}\cos{\alpha} \\
+ V_{\rm r} \cos\delta \cos\alpha \\
+4.74~d~\mu_\alpha\cos{\delta}\sin{\alpha} - 4.74~d~\mu_\delta\sin{\delta}\cos{\alpha} \\
+ V_{\rm r} \cos\delta \cos\alpha \\
+4.74~d~\mu_\delta\cos\delta + V_{\rm r}\sin\delta
\end{pmatrix}
\end{equation}

We derived radial velocities ($V_r$) estimates for all target clusters utilizing spectroscopic data from $Gaia$ DR3. This approach ensures a homogeneous dataset, avoiding the potential biases introduced by adopting heterogeneous literature values.

We employed a two-tiered approach based on the availability of $V_r$ measurements for member stars. For clusters with a statistically significant sample size ($N \geq 19$), the data often exhibited large scatter. To mitigate dispersion and obtain a coherent set of estimates, we applied an iterative $1\sigma$-clipping algorithm. In each iteration, stars deviating from the mean by more than one standard deviation were excluded until convergence. This method was applied to NGC\,559, NGC\,1817, NGC\,2141, and Ruprecht\,142. Conversely, for clusters with sparse data ($N < 19$), such as NGC\,7245 (a \& b) and the majority of the Ruprecht objects, we adopted the mean and standard deviation of the entire available sample without clipping.

The final derived radial velocities, the number of stars used ($N$), and comparisons with literature values are summarized in Table~\ref{tab:rv_results}. Our estimates for NGC\,559, NGC\,1817, NGC\,2141, and NGC\,7245 show excellent agreement with those reported by \citet{Hunt2024}, and our velocity estimate of Ruprecht\,137 has a fair agreement with that of \cite{dias2014proper} within the assigned uncertainties. To our knowledge, we present the first radial velocity determinations for Ruprecht\,15, Ruprecht\,142, and Ruprecht\,169.

\begin{table}
\caption{Derived radial velocities for the studied open clusters using $Gaia$ DR3 data.}
\footnotesize
\label{tab:rv_results}
\centering
\begin{tabular}{l c c c}
\hline\hline
\noalign{\smallskip}
Cluster & $N$ & This Work & Literature \\
 & & ($V_r$; km s$^{-1}$) & ($V_r$; km s$^{-1}$) \\
\noalign{\smallskip}
\hline
\noalign{\smallskip}
NGC 559      & 14 & $-77.43 \pm 0.53$  & $-77.17 \pm 2.13^a$ \\
NGC 1817     & 38 & $65.72 \pm 1.86$   & $66.74 \pm 22.88^a$ \\
NGC 2141     & 47 & $25.22 \pm 1.83$   & $26.14 \pm 11.54^a$ \\
NGC 7245a    & 11 & $-77.73 \pm 14.47$ & $-74.44 \pm 1.22^a$ \\
NGC 7245b    & 3  & $-67.10 \pm 11.84$ & -- \\
Ruprecht 15  & 7  & $86.48 \pm 26.99$   & -- \\
Ruprecht 137 & 7  & $4.38 \pm 30.36$ & $19.00 \pm 10.00^b$ \\
Ruprecht 142 & 9  & $10.57 \pm 0.47$   & -- \\
Ruprecht 169 & 3  & $-30.76 \pm 25.50$ & -- \\
\noalign{\smallskip}
\hline
\multicolumn{4}{l}{\footnotesize References: (a) \citet{Hunt2024}; (b) \cite{dias2014proper}.}
\end{tabular}
\end{table}

The transformation equations given in Eqs. (\ref{Eq. 12}) were used to translate the space velocity components from the equatorial coordinate system to the Galactic frame. The velocity components ($V_x,~V_y,~V_z$; km s$^{-1}$) can be transformed into their Galactic equivalents (i.e., $U,~V,~W$; km s$^{-1}$), which represent the velocity vector components with respect to the Galactic centre, using these equations. These components were ascertained using the following formulas:
\begin{equation}
\label{Eq. 12}
\begin{pmatrix}
U \\\\
V \\\\
W
\end{pmatrix} 
= 
\begin{pmatrix}
-0.0518807421 \; V_{x} -
0.872222642 \; V_{y} -\\
0.4863497200 \; V_{z} \\
+0.4846922369 \; V_{x} -
0.4477920852 \; V_{y} +\\
0.7513692061 \; V_{z}\\
-0.873144899 \; V_{x} -
0.196748341 \; V_{y} +\\
0.4459913295 \; V_{z} \\
\end{pmatrix}
\end{equation}

Thus, the individual velocity components of each star in the cluster are averaged to determine the mean Galactic space velocity components $\overline{U}$,~$\overline{V}$,~and~$\overline{W}$. The following formulas are used to express these average velocities:

\begin{equation} 
\label{eq: 15}
\overline{U} = \dfrac{1}{N}\sum ^{N}_{i=1}U_i,\;\;
\overline{V} = \dfrac{1}{N}\sum ^{N}_{i=1}V_i, \;\; \text{and} \; \; \overline{W} = \dfrac{1}{N}\sum ^{N}_{i=1}W_i,
\end{equation}
where $N$ is the cluster's total number of stars. Moreover, Table \ref{tab:evol_kin_dyn_parameters} displays the average space velocity components $(U,~V,~W)$ for the clusters. 

\subsection{Solar Motion Elements}

The Sun's velocity in relation to the Local Standard of Rest (LSR), a reference frame that depicts the typical motion of stars in the solar neighbourhood, is referred to as the "solar motion". Because of their widespread distribution over the Galactic disc and coherent kinematics, which are made up of stars with a shared origin, age, and velocity, are ideal tracers for figuring out solar motion. The mean spatial velocity components ($\overline{U},~\overline{V},~\overline{W}$) of a given stellar cluster in Galactic coordinates can be used to calculate the solar space velocity components, which are expressed in km s$^{-1}$.

The following relationships are used to evaluate these components \citep{Bisht2021, Elsanhoury2022, elsanhoury2026, Haroon2025, Bisht2025}:
\begin{equation}
\label{Eq.16}
U_{\odot} = -\overline{U}, \; V_{\odot} = -\overline{V}, \; \text{and} \; W_{\odot} = -\overline{W}.
\end{equation}

The magnitude of the solar space velocity ($S_{\odot}$) relative to the observed objects is then calculated using the following expression. Conversely, the Galactic coordinates of the solar apex ($l_\text{A}$, $b_\text{A}$) are determined from the relations:
\begin{equation}
\label{Eq. 17}
S_{\odot}=\sqrt{(\overline{U})^2+(\overline{V})^2+(\overline{W})^2}.
\end{equation}
and
\begin{equation}
\label{Eq. 18}
l_\text{A} = \tan^{-1}\left(\frac{-\overline{V}}{\overline{U}}\right) \;\; \text{and} \;\;
b_\text{A} = \sin^{-1} \left(\frac{-\overline{W}}{S_\odot}\right),
\end{equation}

The heliocentric velocity components ($\overline{U}$, $\overline{V}$, $\overline{W}$) were transformed to the Local Standard of Rest (LSR) after adopting the solar motion with respect to the LSR, $(U, V, W)_{\rm LSR} = (8.83 \pm 0.24,, 14.19 \pm 0.34,, 6.57 \pm 0.21)$ km s$^{-1}$ \citep{Coskunoglu2011}. In addition, first–order Galactic differential–rotation corrections were applied using the Oort constants following \citet{MihalasBinney1981} and \citet{Johnson1987}, thereby removing velocity gradients induced by the non–rigid rotation of the Galactic disc \citep{2025AJ....169...87C,2012MNRAS.421.3362B,2012MNRAS.419.2844C}. After these adjustments, the resulting kinematic parameters represent the clusters’ motion relative to the Galactic reference frame rather than the Sun. The final LSR– and rotation–corrected quantities are listed in Table~\ref{tab:evol_kin_dyn_parameters}.

\section{Characterization of the Clusters' Galactic Orbits}

To constrain the Galactic dynamical histories of the sample OCs, a comprehensive suite of kinematic and astrometric data was utilized as input for the \textsc{galpy} Python library \citep{Bovy2015}. These inputs encompassed the clusters' mean astrometric parameters ($\alpha$,~$\delta$), $\langle\mu_{\alpha}\cos\delta$, and $\mu_{\rm \delta}\rangle$, the mean radial velocities $\langle V_R\rangle$, and their photometrically derived isochrone distances ($d_{\rm DM}$). Orbital integrations were performed employing the Galactic potential model, \textsc{MWPotential2014}.

\begin{figure*}[!ht]
    \includegraphics[width=0.33\linewidth]{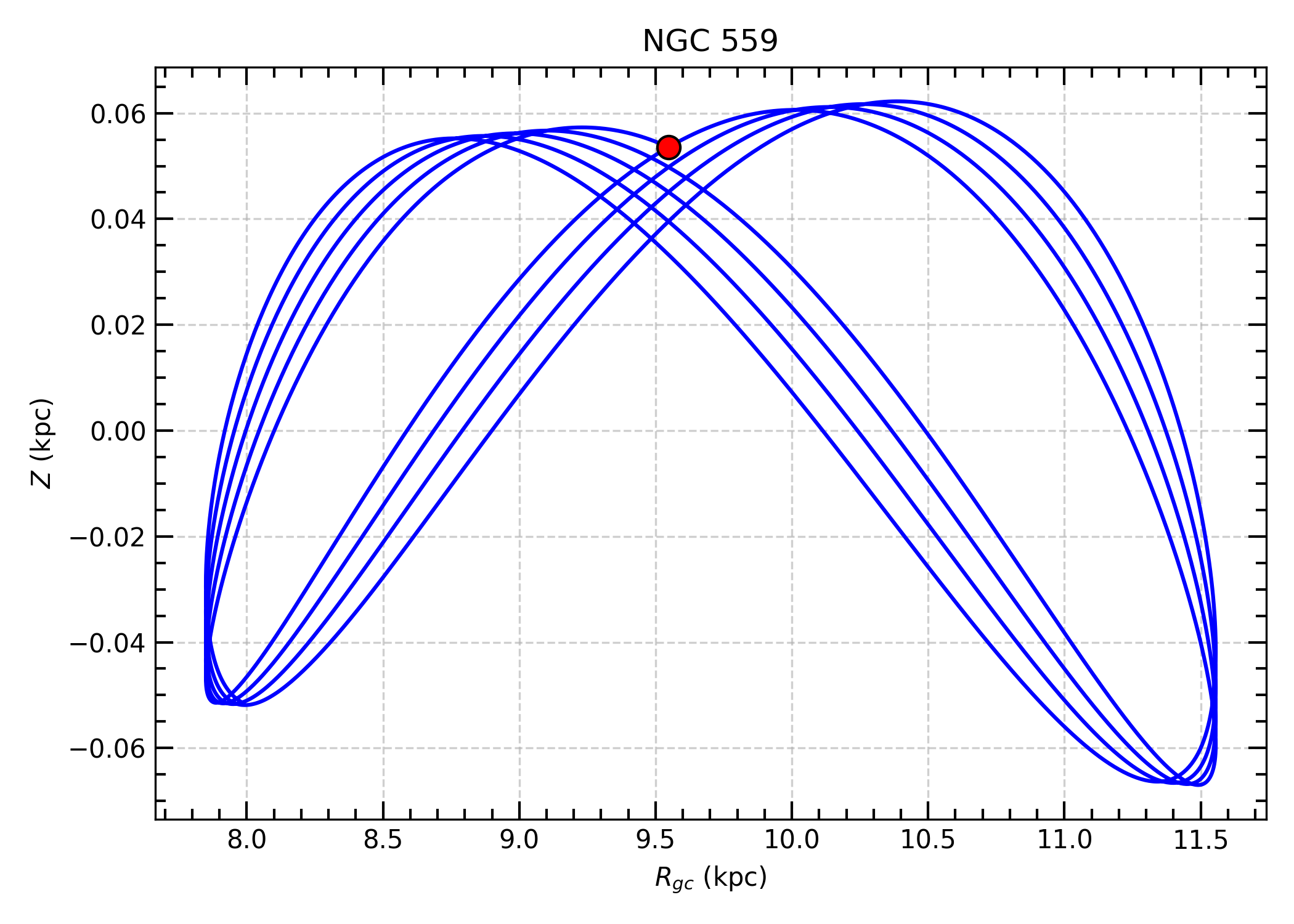}
    \includegraphics[width=0.33\linewidth]{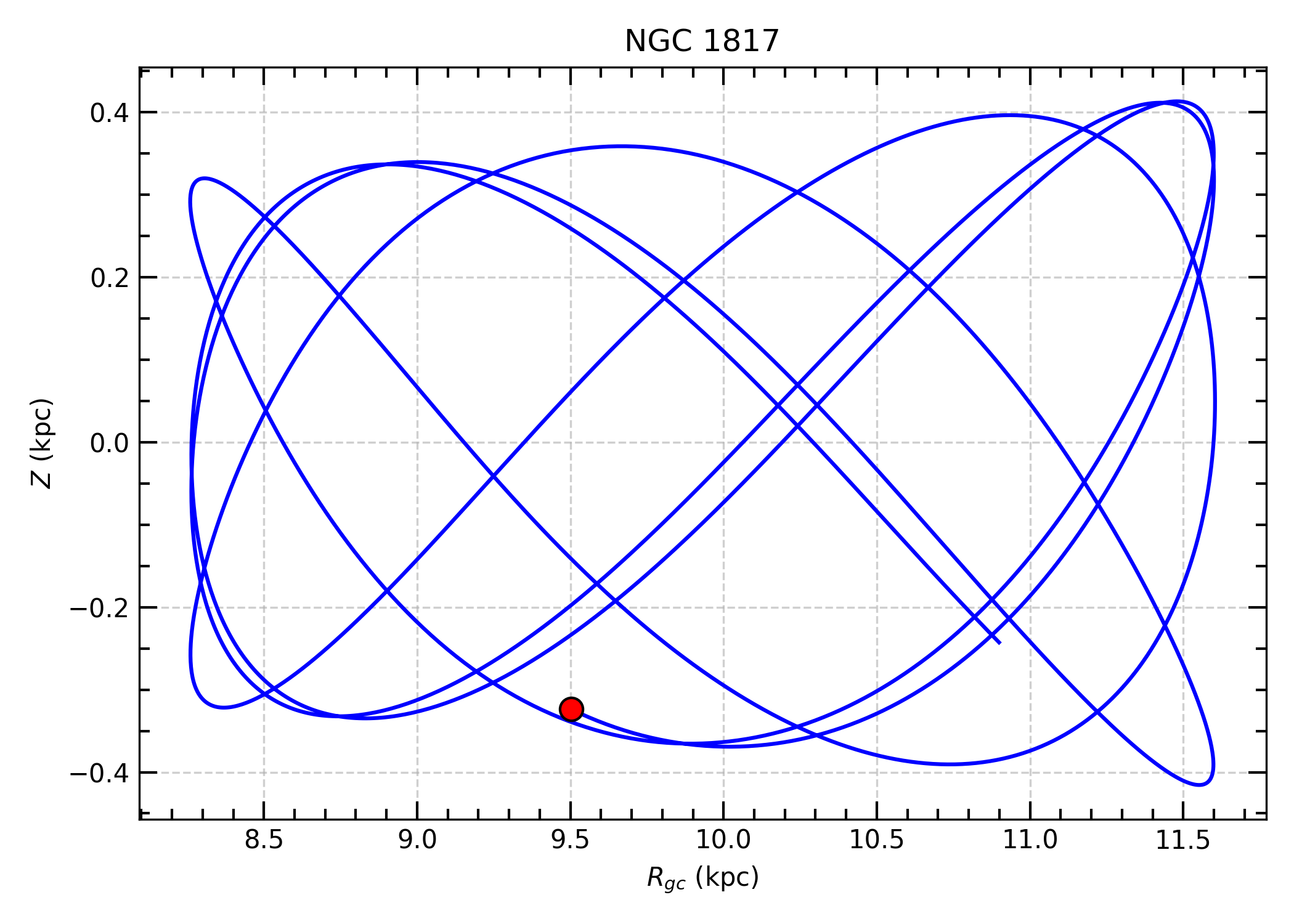}
    \includegraphics[width=0.33\linewidth]{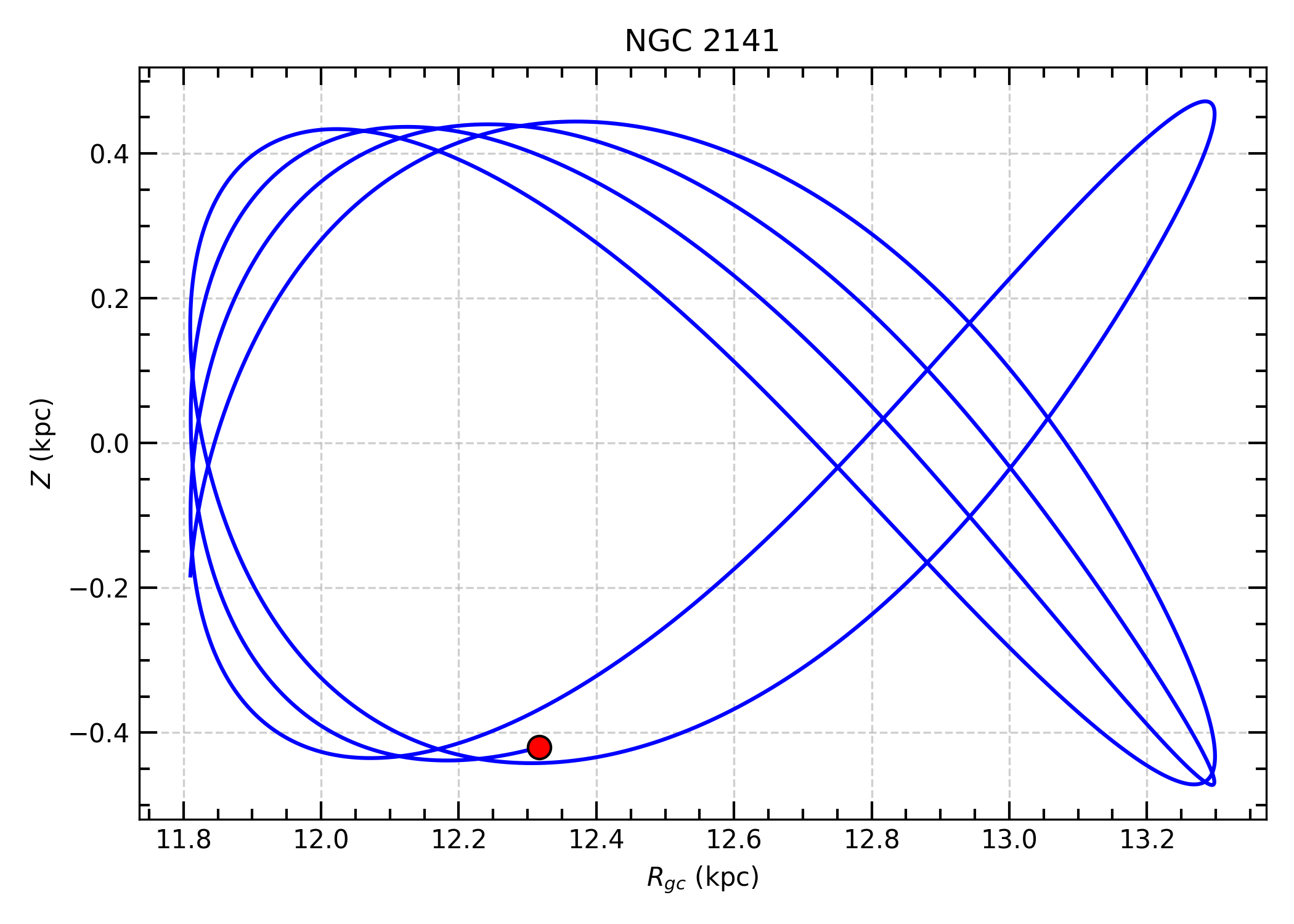}\\
    \includegraphics[width=0.33\linewidth]{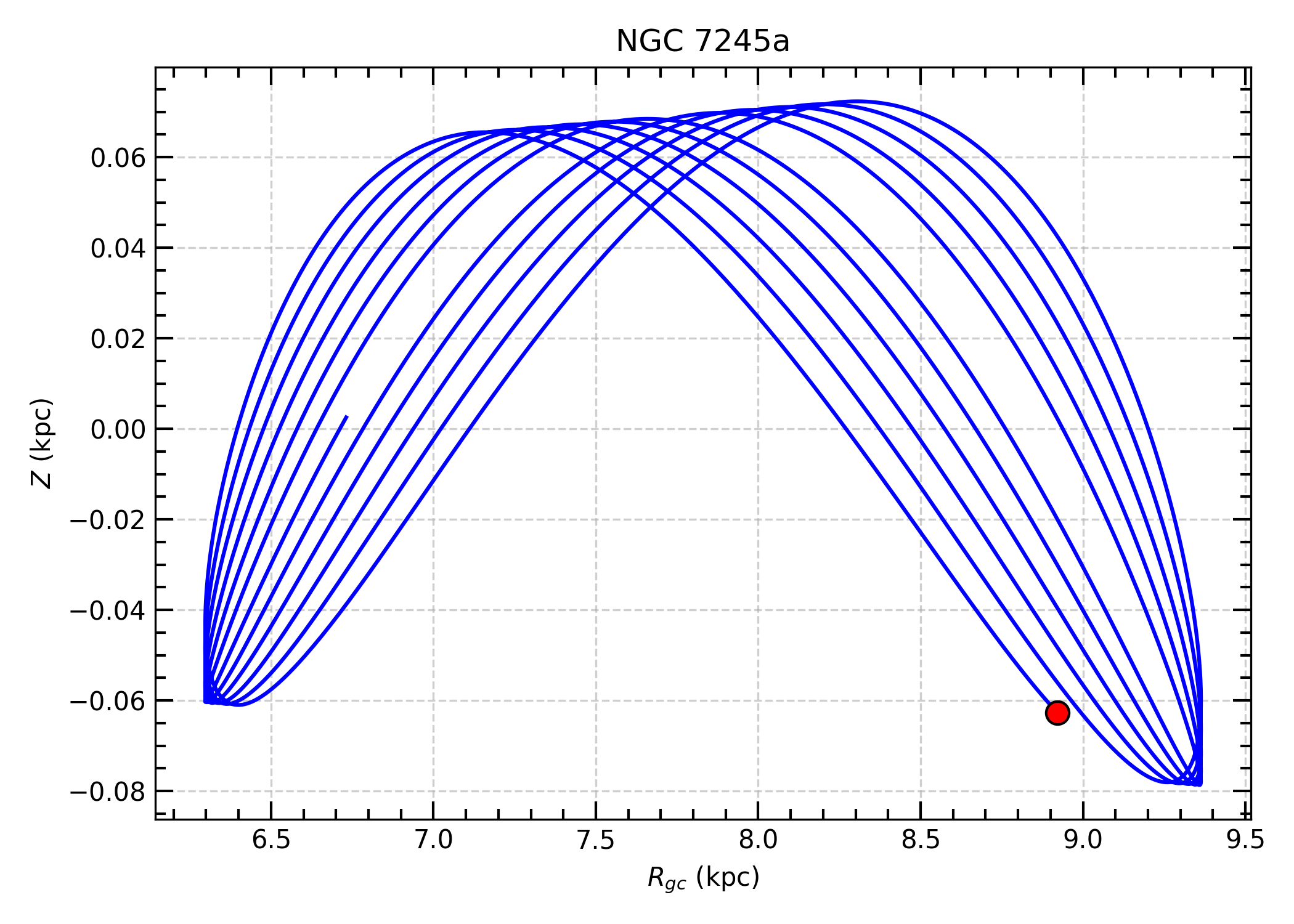}
    \includegraphics[width=0.33\linewidth]{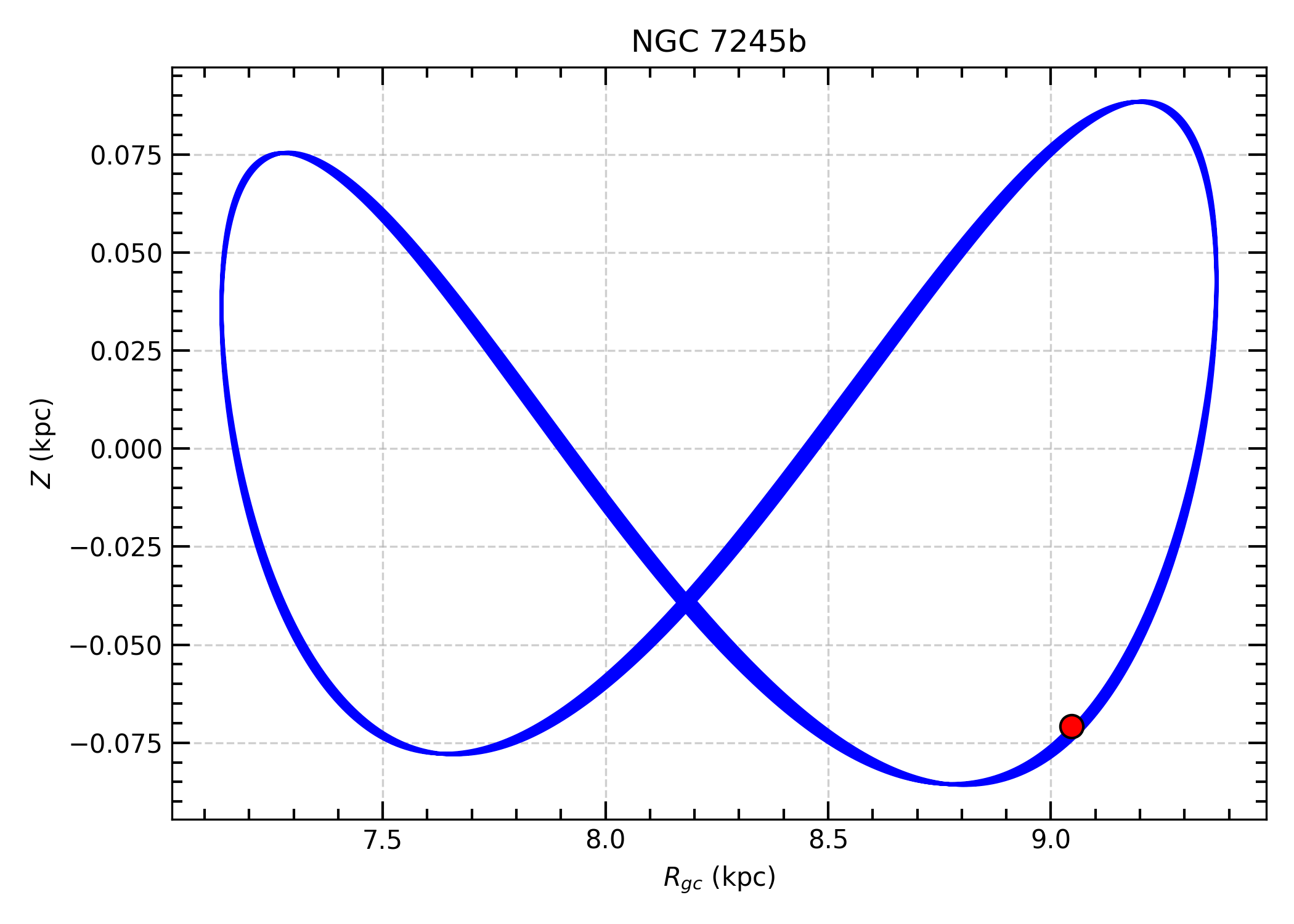}
    \includegraphics[width=0.33\linewidth]{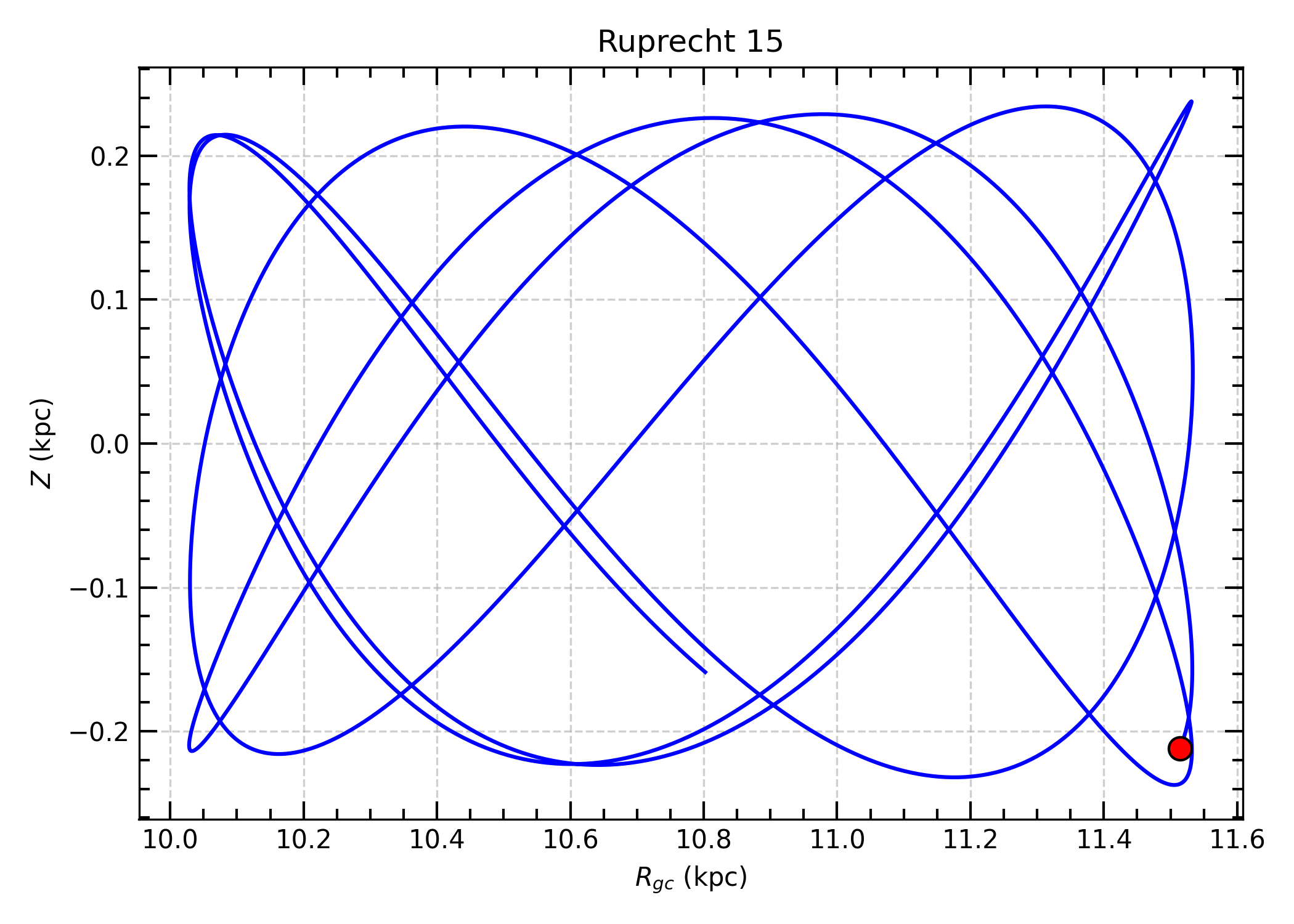}\\
    \includegraphics[width=0.33\linewidth]{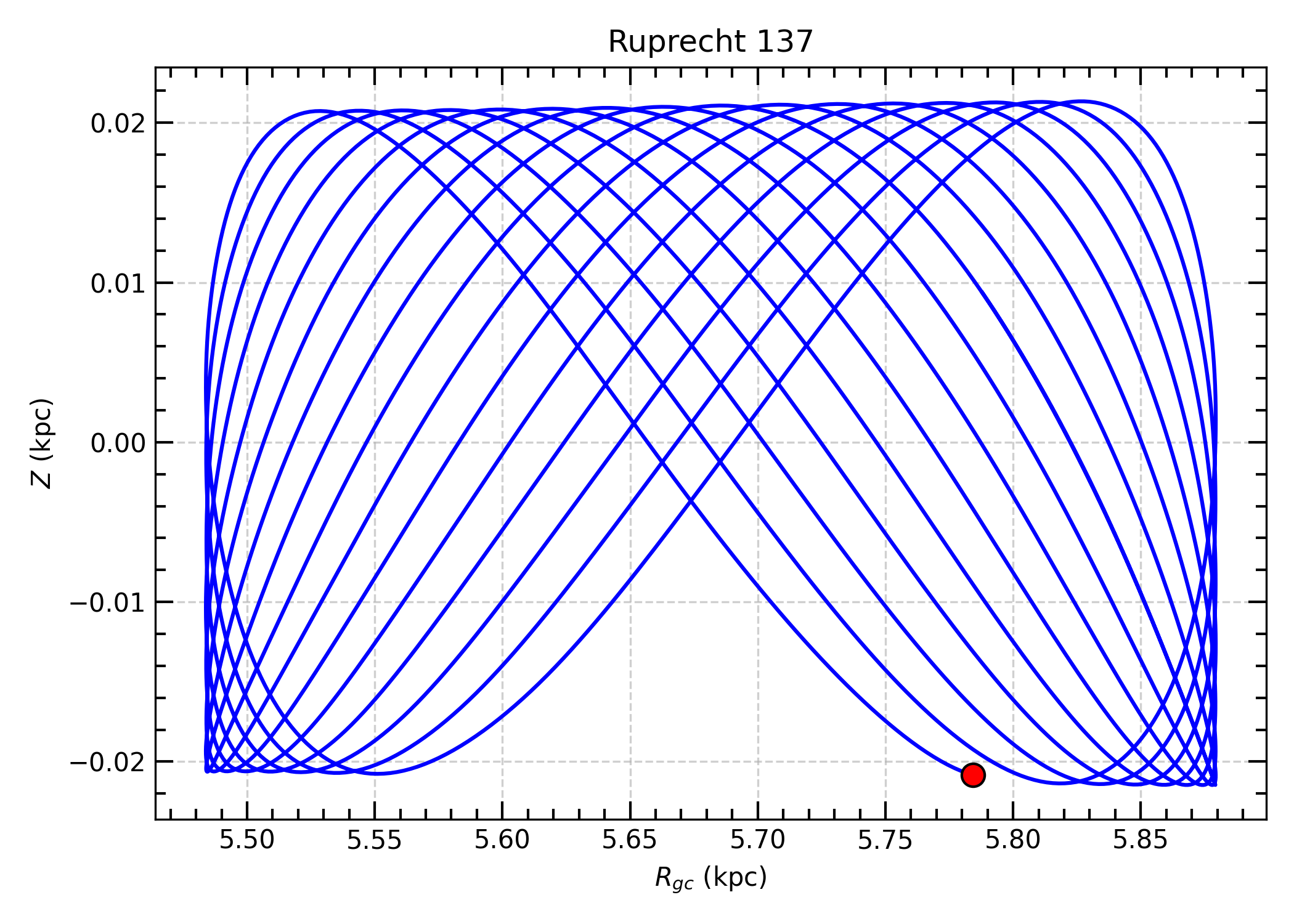}
    \includegraphics[width=0.33\linewidth]{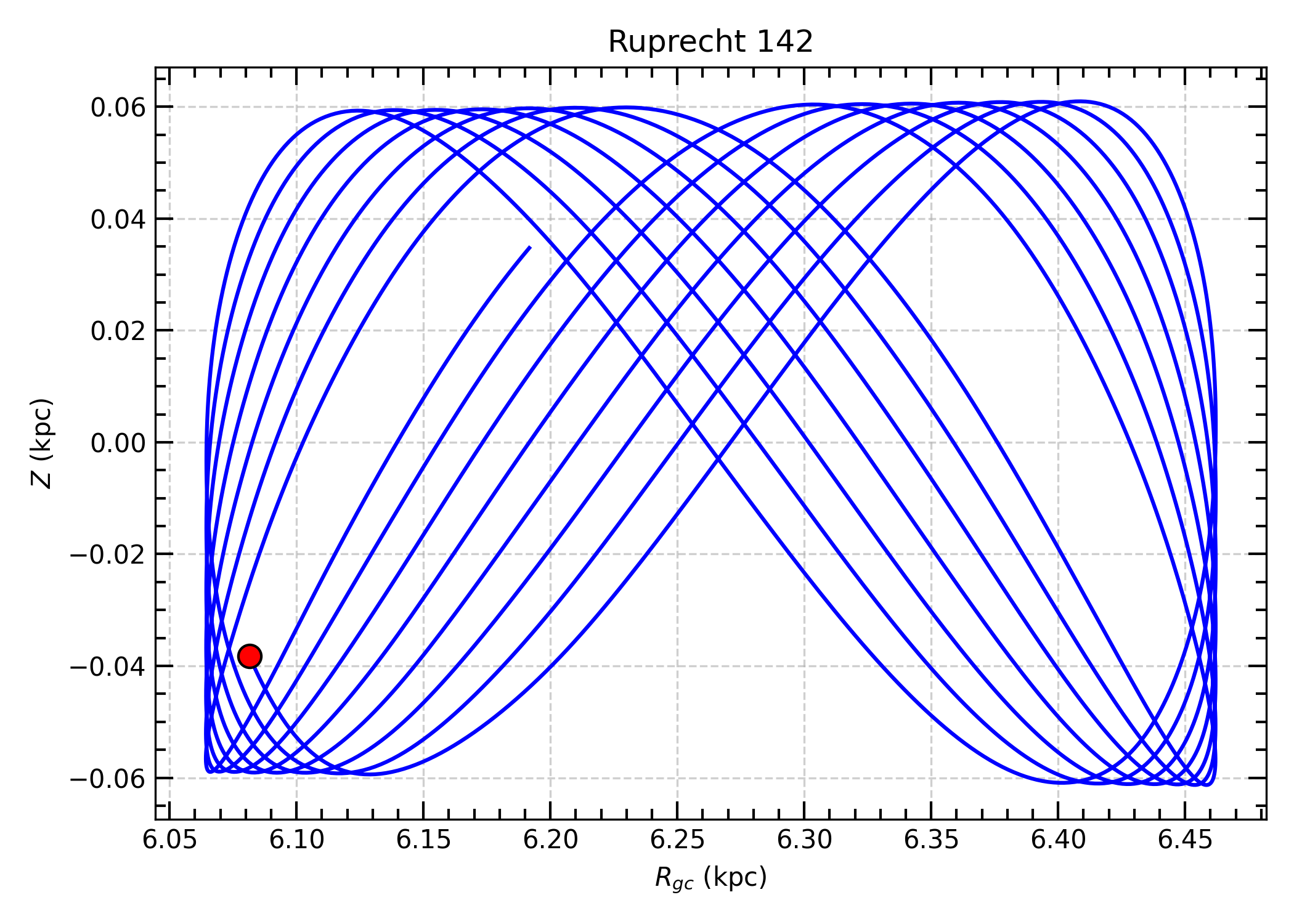}
    \includegraphics[width=0.33\linewidth]{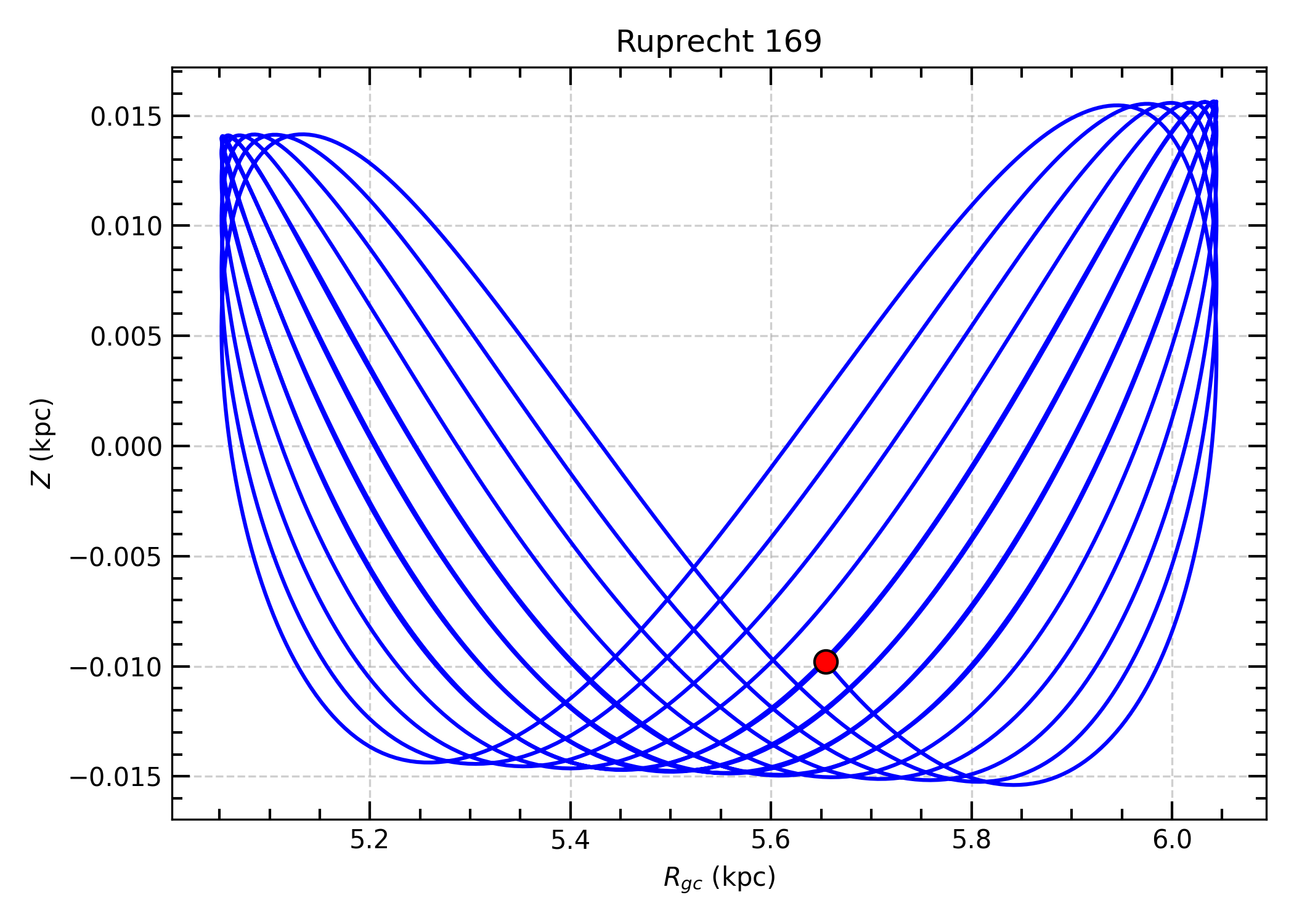}
    \caption{The orbits for the studied OCs, showing their motion in the meridional plane ($R_{\mathrm{gc}}$ vs. $Z$). The red dots mark the present-day positions of the clusters.}
    \label{Orbits}
\end{figure*}

To rigorously constrain the dynamical evolution of each cluster, their orbits were integrated backward in time for $5 \mbox{ Gyr}$ with a $1.5 \mbox{ Myr}$ time step. This integration window, spanning multiple Galactic rotations, ensures a robust determination of the key orbital parameters. This procedure yielded the fundamental dynamical properties for each cluster: the perigalactic ($R_{\rm p}$) and apogalactic ($R_{\rm a}$) distances, orbital eccentricity ($e$), maximum vertical excursion from the Galactic plane ($Z_{\rm max}$), and the orbital period ($T_{\rm P}$). The complete set of parameters derived from this orbital analysis is compiled in Table~\ref{tab:evol_kin_dyn_parameters}.

As visually confirmed by the orbital projections in the $R_{\rm gc}~\times Z$ plane (Figure ~\ref{Orbits})], this homogeneity is striking \citep[e.g.,][]{Tasdemir2025b, Yucel2024,Dursun2024}. All OCs in the sample trace near-circular paths around the Galactic centre. Their eccentricities are tightly constrained, ranging from the near-perfectly circular orbit of Ruprecht 142 ($e = 0.03 \pm 0.01$) to the modest value of $e = 0.19 \pm 0.01$ for NGC 559, which represents the most eccentric path in our sample.

These kinematic properties are fundamental in diagnosing the Galactic population membership of the clusters. The maximum vertical excursion from the Galactic plane ($Z_{\rm max}$) serves as the primary diagnostic for this classification. All clusters in this study are dynamically confined to the immediate vicinity of the Galactic plane. While some members, such as Ruprecht 137 and Ruprecht 169 ($Z_{\max} \approx 0.02 \mbox{ kpc}$), exhibit negligible vertical motion, even the cluster with the largest excursion, NGC 2141 ($0.47\pm0.09 \mbox{ kpc}$), remains well within thin-disk boundaries ($Z_{\max} \leq 0.5 \mbox{ kpc}$). We find no kinematic evidence for any cluster being associated with the thick disc (which typically requires $Z_{\max} \geq 1.0 \mbox{ kpc}$) \citep{2015Ap&SS.357...72A}. This combination of low eccentricities and small $Z_{\max}$ excursions, in complete agreement with dynamical classification schemes \citep[e.g.,][]{Schuster12}, unequivocally identifies all clusters in this study as bona fide members of the Galactic thin disc population \citep[e.g.,][]{Cinar2025}.

\begin{table*}[h!]
\centering
\caption{Summary of evolving, kinematical and orbital parameters for the target OCs. Parameters include heliocentric space velocity components ($V_x,~V_y,~V_z$), Galactic space velocity components ($\overline{U},~\overline{V},~\overline{W}$), solar motion elements ($S_{\rm \odot}$), LSR-corrected velocities ($U_{\rm LSR},~V_{\rm LSR},~W_{\rm LSR}$), apex coordinates ($A_o,~D_o,~l_A,~b_A$), and orbital parameters ($Z_{\rm max},~R_a,~R_p,~e,~T_p$).}
\resizebox{\textwidth}{!}{
\begin{tabular}{llccccccccc}
\hline
& \textbf{Parameter} & \textbf{NGC 559} & \textbf{NGC 1817} & \textbf{NGC 2141} & \textbf{NGC 7245a} & \textbf{NGC 7245b} & \textbf{Ruprecht 15} & \textbf{Ruprecht 137} & \textbf{Ruprecht 142} & \textbf{Ruprecht 169} \\
\hline
\hline
\multicolumn{11}{c}{\textbf{Evolving Parameters}} \\
\midrule
& $R_h$ (pc) & 4.09 $\pm$ 0.37 & 4.82 $\pm$ 0.80 & 5.22 $\pm$ 0.26 & 9.39 $\pm$ 1.00 & 9.63 $\pm$ 1.02 & 7.56 $\pm$ 0.12 & 2.71 $\pm$ 0.07 & 5.84 $\pm$ 0.13 & 2.20 $\pm$ 0.05 \\
& $T_{\rm relax}$ (Myr)& 70 $\pm$ 3 & 93 $\pm$ 2 & 135 $\pm$ 9 & 190 $\pm$ 5 & 170 $\pm$ 6.00 & 121 $\pm$ 8 & 21 $\pm$ 2 & 71 $\pm$ 4 & 16 $\pm$ 1 \\
& $\tau$ & 6.47 & 13.68 & 16.33 & 3.47 & 2.64 & 0.83 & 4.28 & 4.21 & 16.12 \\
& $\tau_{\rm ev}$ (Myr)& 7000 & 9300 & 13500 & 19000 & 17000 & 12100 & 2100 & 7100 & 1600\\
\hline
\multicolumn{11}{c}{\textbf{Kinematical Parameters}}\\
\midrule
& $\overline{V_x}$ (km s$^{-1}$) & $-$9.96 $\pm$ 3.16 & 9.70 $\pm$ 3.11 & 2.39 $\pm$ 0.65 & $-$29.18 $\pm$ 5.40 & $-$17.70 $\pm$ 4.21 & 43.79 $\pm$ 6.62 & $-$0.25 $\pm$ 0.01 & 11.23 $\pm$ 3.35 & 3.91 $\pm$ 0.51\\
& $\overline{V_y}$ (km s$^{-1}$) & $-$73.72 $\pm$ 8.59 & 64.77 $\pm$ 8.05 & 30.80 $\pm$ 5.55 & $-$60.94 $\pm$ 7.81 & $-$58.77 $\pm$ 7.67 & 174.66 $\pm$ 13.22 & 4.72 $\pm$ 0.46 & $-$5.55 $\pm$ 0.43 & 34.00 $\pm$ 5.83\\
& $\overline{V_z}$ (km s$^{-1}$) & $-$67.92 $\pm$ 8.24 & 10.77 $\pm$ 3.28 & $-$27.76 $\pm$ 5.27 & $-$96.20 $\pm$ 9.81 & $-$89.90 $\pm$ 9.48 & 162.27 $\pm$ 12.74 & $-$20.32 $\pm$ 4.51 & $-$17.26 $\pm$ 4.15 & $-$0.28 $\pm$ 0.01\\

& $\overline{U}$ (km s$^{-1}$)  & 97.85 $\pm$ 9.90 & $-$62.23 $\pm$ 7.89 & $-$13.49 $\pm$ 3.67 & 101.45 $\pm$ 10.07 & 95.90 $\pm$ 9.80 & $-$81.39 $\pm$ 15.28 & 5.78 $\pm$ 0.42 & 12.65 $\pm$ 3.56 & $-$29.72 $\pm$ 5.45\\

& $\overline{V}$ (km s$^{-1}$)  & $-$22.85 $\pm$ 4.78 & $-$16.21 $\pm$ 4.03 & $-$33.50 $\pm$ 5.79 & $-$59.14 $\pm$ 7.69 & $-$49.81 $\pm$ 7.06 & $-$46.87 $\pm$ 8.06 & $-$17.50 $\pm$ 4.18 & $-$5.04 $\pm$ 0.45 & $-$13.54 $\pm$ 3.68\\

& $\overline{W}$ (km s$^{-1}$)  & $-$7.09 $\pm$ 2.66 & $-$16.41 $\pm$ 4.05 & $-$20.53 $\pm$ 4.53 & $-$5.44 $\pm$ 0.43 & $-$13.08 $\pm$ 3.62 & $-$3.05 $\pm$ 0.01 & $-$9.78 $\pm$ 3.13 & $-$16.41 $\pm$ 4.05 & $-$10.22 $\pm$ 3.20\\

& $S_{\rm \odot}$ (km s$^{-1}$) & 100.73 $\pm$ 10.04 & 66.37 $\pm$ 8.15 & 41.54 $\pm$ 6.45 & 117.55 $\pm$ 10.84 & 108.85 $\pm$ 10.43 & 142.40 $\pm$ 15.57 & 20.86 $\pm$ 4.57 & 21.33 $\pm$ 4.62 & 34.22 $\pm$ 5.85\\

& $U_{\rm LSR}$ (km s$^{-1}$) & 65.67 $\pm$ 9.90 & $-$47.24 $\pm$ 7.89 & 32.46 $\pm$ 3.68 & 88.46 $\pm$ 10.07 & 78.50 $\pm$ 9.80 & $-$9.04 $\pm$ 15.28 & 11.31 $\pm$ 0.43 & 32.25 $\pm$ 3.56 & $-$31.39 $\pm$ 5.45\\
& $V_{\rm LSR}$ (km s$^{-1}$) &$-$51.52 $\pm$ 4.79 & 2.28 $\pm$ 4.04 & $-$18.87 $\pm$ 5.80 &$-$136.24 $\pm$ 7.69 & $-$134.43 $\pm$ 7.06 & $-$130.90 $\pm$ 8.06 & $-$12.97 $\pm$ 4.18 & $-$13.27 $\pm$ 0.45 & $-$16.67 $\pm$ 3.68\\
& $W_{\rm LSR}$ (km s$^{-1}$) & $-$0.52 $\pm$ 2.67 & $-$9.84 $\pm$ 4.06 & $-$13.96 $\pm$ 4.54 & $-$0.10 $\pm$ 0.48 & $-$11.13 $\pm$ 3.62 & $-$4.49 $\pm$ 0.11 & $-$14.61 $\pm$ 3.13 &$-$24.63 $\pm$ 4.05 & $-$21.83 $\pm$ 3.20\\
& $S_{\rm LSR}$ (km s$^{-1}$) & 83.47 $\pm$ 11.32 & 48.31 $\pm$ 9.75 & 40.05 $\pm$ 8.23 & 162.44 $\pm$ 12.68 & 156.07 $\pm$ 12.61 & 131.29 $\pm$ 17.28 & 22.57 $\pm$ 5.24 & 42.69 $\pm$ 5.41 & 41.71 $\pm$ 7.31\\
\hline
\multicolumn{11}{c}{\textbf{Dynamical Parameters}}\\
\midrule
& $A_o~(^\circ)$ & $-$97.70 $\pm$ 0.10 & 81.48 $\pm$ 0.11 & 85.56 $\pm$ 0.11 & $-$115.59 $\pm$ 0.09 & $-$106.76 $\pm$ 0.10 & 75.93 $\pm$ 0.11 & 93.00 $\pm$ 0.18 & $-$26.32 $\pm$ 0.21 & 83.45 $\pm$ 0.11\\

& $D_o~(^\circ)$ & $-$42.40 $\pm$ 0.18 & 9.33 $\pm$ 0.01 & $-$41.94 $\pm$ 0.14 &$-$54.92 $\pm$ 0.14 & $-$55.68 $\pm$ 0.14 & 42.03 $\pm$ 0.15 & -76.90 $\pm$ 0.11 & $-$54.02 $\pm$ 0.13 & $-$0.47 $\pm$ 0.01\\

& $l_A~(^\circ)$ & 13.14 & $-$14.60 & $-$68.07 & 30.24 & 27.45 & 15.54 & 71.74 & 21.70 & $-$24.50\\

& $b_A~(^\circ)$ & 4.04 & 14.32 & 29.62 & 2.65 & 6.90 & 0.06 & 27.94 & 50.30 & 17.39\\
\hline
& $Z_{\rm max}$ (kpc) & 0.07 $\pm$ 0.01 & 0.42 $\pm$ 0.05 & 0.47 $\pm$ 0.11 & 0.08 $\pm$ 0.01 & 0.09 $\pm$ 0.03 & 0.24 $\pm$ 0.06 & 0.02 $\pm$ 0.01 & 0.06 $\pm$ 0.02 & 0.02 $\pm$ 0.01\\
& $R_{\rm a}$ (kpc) & 11.56 $\pm$ 0.85 & 11.61 $\pm$ 0.20 & 13.31 $\pm$ 0.92 & 9.36 $\pm$ 0.23 & 9.37 $\pm$ 0.24 & 11.53 $\pm$ 1.05 & 5.88 $\pm$ 0.45 & 6.46 $\pm$ 0.32 & 6.04 $\pm$ 0.62\\
& $R_{\rm p}$ (kpc) & 7.85 $\pm$ 0.63 & 8.26 $\pm$ 0.15 & 11.81 $\pm$ 0.58 & 6.30 $\pm$ 0.63 & 7.14 $\pm$ 0.88 & 10.03 $\pm$ 1.35 & 5.48 $\pm$ 0.72 & 6.06 $\pm$ 0.31 & 5.05 $\pm$ 0.10\\
& $e$ & 0.19 $\pm$ 0.01 & 0.17 $\pm$ 0.01 & 0.06 $\pm$ 0.01 & 0.20 $\pm$ 0.04 & 0.14 $\pm$ 0.05 & 0.07 $\pm$ 0.02 & 0.03 $\pm$ 0.03 & 0.03 $\pm$ 0.01 & 0.09 $\pm$ 0.04\\
& $T_{\rm p}$ (Gyr) & 0.28 $\pm$ 0.02 & 0.28 $\pm$ 0.01 & 0.37 $\pm$ 0.02 & 0.22 $\pm$ 0.01 & 0.23 $\pm$ 0.02 & 0.31 $\pm$ 0.04 & 0.15 $\pm$ 0.02 & 0.17 $\pm$ 0.01 & 0.15 $\pm$ 0.01\\
\hline
\end{tabular}}
\label{tab:evol_kin_dyn_parameters}
\end{table*}

\section{Summary and Conclusions}
\label{sec:summary}

In this study, we have presented a comprehensive structural, astrophysical, and dynamical analysis of eight OCs: NGC 559, NGC 1817, NGC 2141, NGC 7245, Ruprecht 15, Ruprecht 137, Ruprecht 142, and Ruprecht 169. Our study is based on the high-precision astrometric and photometric data from the $Gaia$ DR3 database, which provides an unprecedented opportunity to investigate their properties homogeneously.

An important first step was the identification of probable cluster members using a statistical approach based on $Gaia$ DR3 proper motions and trigonometric parallaxes. This method successfully filtered field star contamination, resulting in 779, 590, 1551, 410, 151, 231, and 168 members for NGC 559, NGC 1817, NGC 2141, Ruprecht 15, Ruprecht 137, Ruprecht 142, and Ruprecht 169, respectively. A significant finding emerged during the analysis of NGC 7245, where the VPD revealed two distinct kinematic sub-structures. We designate these components NGC 7245a (365 members) and NGC 7245b (239 members), providing strong evidence that NGC 7245 is a binary OCs candidate.

We determined the structural parameters for all clusters by fitting the empirical \citet{King1962} model to their RDPs. After converting the angular radii to physical scales using our derived astrometric distances, we find that the core radii ($r_c$) of the clusters span a range from 2.49 pc (Ruprecht 137) to 12.67 pc (NGC 7245b). The cluster limiting radii ($r_{cl}$) were found to be in the range of 6.96 pc (Ruprecht 169) to 25.17 pc (NGC 7245b).

The fundamental astrophysical parameters were derived by fitting {\sc PARSEC} stellar isochrones to the member-cleaned CMDs. We determined logarithmic ages, $\log(t)$, spanning from 7.95 (Ruprecht 137; 90 Myr) to 9.34 (NGC 2141; 2.2 Gyr). The two components of NGC 7245 were confirmed to have distinct ages of 660 Myr (NGC 7245a) and 450 Myr (NGC 7245b), further supporting their binary nature. We corrected the $Gaia$ trigonometric parallaxes for the known zero-point bias and derived mean astrometric distances ($d_\varw$), placing the clusters between 1.640 $\pm$ 0.006 kpc (NGC 1817) and 5.203 $\pm$ 0.153 kpc (Ruprecht 15). These astrometric distances show excellent agreement with our photometrically-derived distances ($d_{DM}$). We also determined the line-of-sight reddening, finding $E(B-V)$ values from 0.26 mag (NGC 1817) to a significant 1.01 mag (Ruprecht 137 \& Ruprecht 142).

Our analysis of the cluster LFs and MFs provided insights into their stellar content. By integrating the MF, we calculated total cluster masses ($M_{\text{C}}$) ranging from 257 $\pm$ 35 $M_{\odot}$ (Ruprecht 137) to a substantial 1916 $\pm$ 233 $M_{\odot}$ (NGC 2141). The slopes of the mass functions ($\alpha$) for all nine cluster components were found to be in the range of 1.96 $\pm$ 0.12 to 3.07 $\pm$ 0.39. This result demonstrates that the stellar mass distributions in all our sample clusters are consistent with the canonical initial mass function slope derived by \citet{salpeter1955} ($\alpha = 2.35$).

A key component of our work was the investigation of the clusters' dynamical state. We calculated the dynamical relaxation time ($T_{relax}$) for each cluster and compared it to its isochrone age to find the dynamical evolution parameter, $\tau = \text{age} / T_{\text{relax}}$. We find that all clusters in our sample except Ruprecht 15 ($\tau=0.83$) are dynamically relaxed. This finding correlates perfectly with our mass segregation analysis. The Kolmogorov-Smirnov test revealed clear statistical evidence of mass segregation in six of the seven older, dynamically relaxed clusters. This segregation was, however, demonstrably absent in the youngest two open clusters in our sample, Ruprecht 15 and Ruprecht 137. This result provides strong support for the conclusion that mass segregation in this cluster is an evolutionary feature driven by internal two-body dynamical interactions, rather than a primordial condition of their formation.

Finally, we performed a detailed kinematic analysis and orbital integration for all clusters using the \textsc{galpy} Python library with the \textsc{MWPotential2014} Galactic potential. We determined mean radial velocities, calculated the CPs (i.e., $A_0,~D_0$), and derived the full $(\overline{U},~\overline{V},~\overline{W}$; km s$^{-1})$ space velocity components $(\overline{V_x},~\overline{V_y},\overline{V_z}$; km s$^{-1})$. The orbital integrations, traced back for 5 Gyr, show that all clusters trace near-circular paths around the Galactic centre. The orbital eccentricities ($e$) are all low, ranging from 0.03 $\pm$ 0.01 (Ruprecht 137 \& Ruprecht 142) to 0.20 $\pm$ 0.04 (NGC 7245a).

This combination of low orbital eccentricities and small maximum vertical excursions from the Galactic plane ($Z_{max}$), which are all confined well within thin-disk boundaries ($Z_{max} \leq 0.5$ kpc), unequivocally identifies all eight clusters in this study as bona fide members of the Galactic thin disc population. Their properties and Galactic orbits confirm their status as reliable tracers for the ongoing dynamical and chemical evolution of the MW's disc.

While this study has presented a detailed picture of eight OCs based on $Gaia$ DR3 data, it also opens new doors for future research. The kinematic and orbital parameters we have obtained, particularly for the less-studied Ruprecht clusters, should be supported by high-resolution spectroscopic follow-up studies. Such observations will not only improve the radial velocity measurements, thereby increasing the precision of orbital calculations, but will also reveal the clusters' chemical abundance patterns (e.g., [Fe/H] and $\alpha$-element ratios). This is critical for testing the metallicity assumptions used in isochrone fitting and for chemically tagging the clusters to their birthplaces in the Galactic disc. Furthermore, the binary cluster candidate we identified in NGC 7245 merits dedicated photometric and spectroscopic monitoring campaigns to confirm the physical association of its two components, model their dynamical interactions, and understand their common evolution.

\section*{Acknowledgements}
We sincerely thank the anonymous referee for their thoughtful comments and helpful suggestions, which greatly enhanced both the clarity and overall quality of this manuscript. This study presents results derived from the European Space Agency (ESA) space mission Gaia. The data from $Gaia$ are processed by the $Gaia$ Data Processing and Analysis Consortium (DPAC). Financial support for DPAC is provided by national institutions, primarily those participating in the $Gaia$ Multi-Lateral Agreement (MLA). For additional information, the official $Gaia$ mission website can be accessed at \url{https://www.cosmos.esa.int/gaia}, and the $Gaia$ archive is available at \url{https://archives.esac.esa.int/gaia}.The authors would like to express their gratitude to the Deanship of Scientific Research at Northern Border University, Arar, KSA, for funding this research under project number "NBU-FFR-2026-237-03".



\bibliography{references} 
\appendix
\section{ }\label{rdps-append}
\begin{figure*}[ht!]
    \centering
        \includegraphics[width=0.23\linewidth]{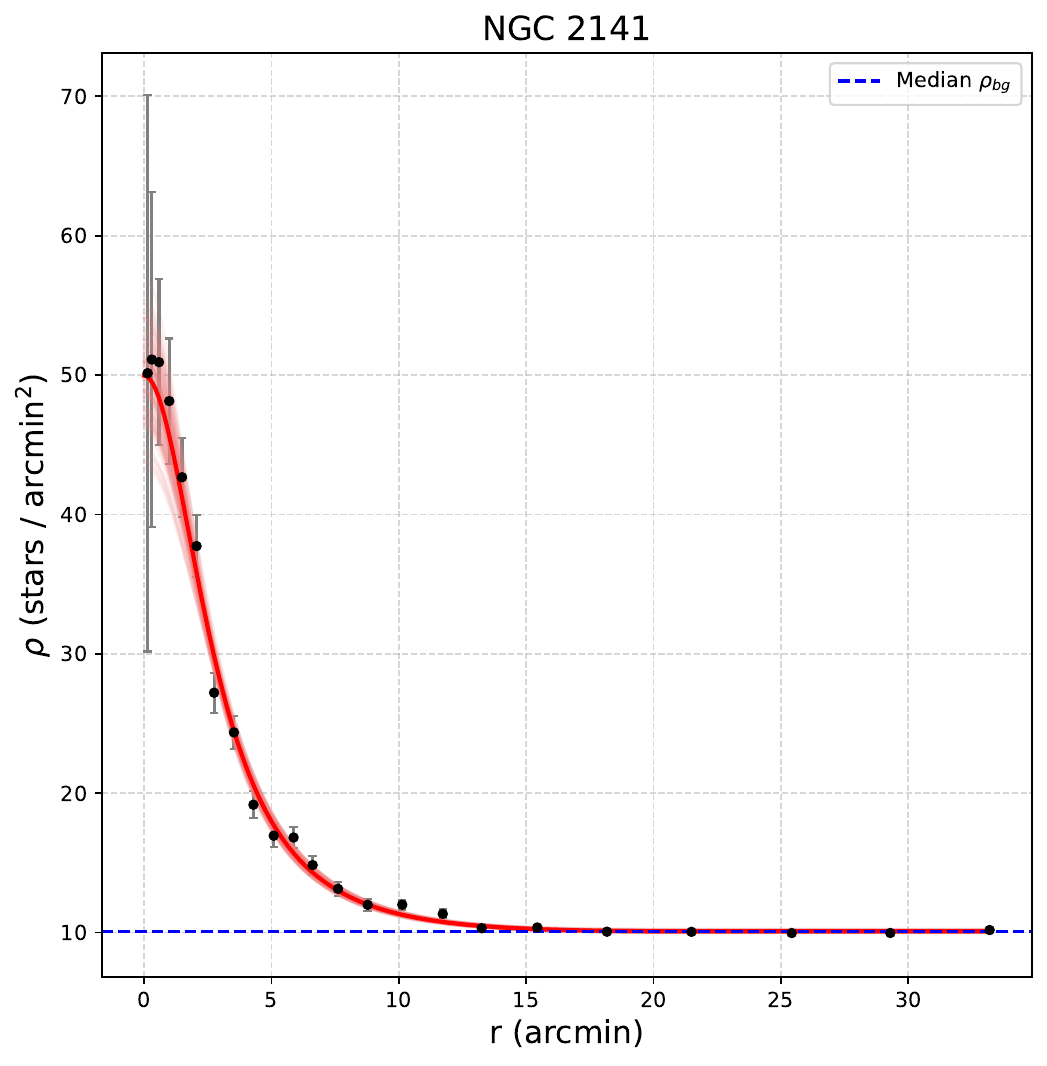}
    \includegraphics[width=0.23\linewidth]{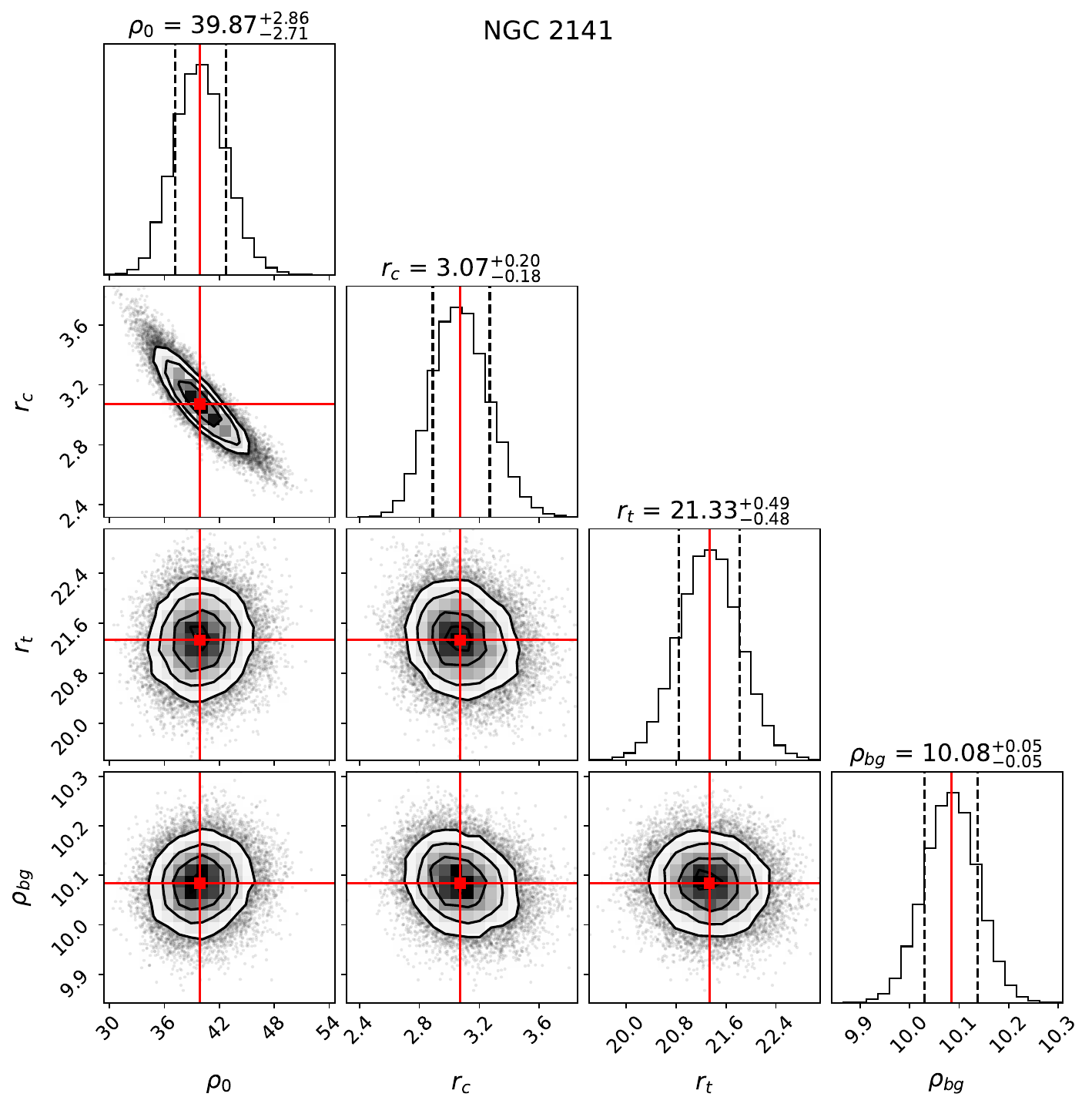}
        \includegraphics[width=0.23\linewidth]{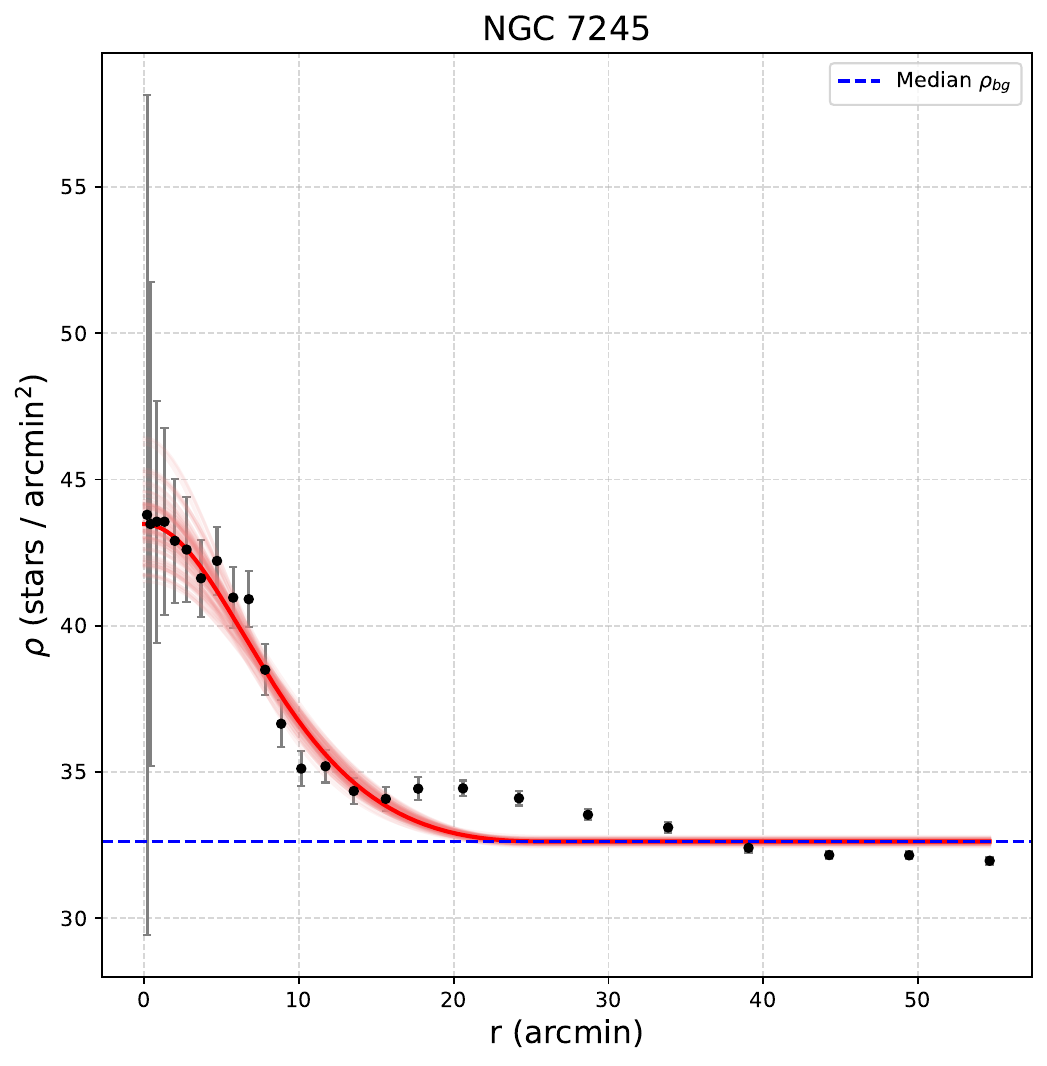}
    \includegraphics[width=0.23\linewidth]{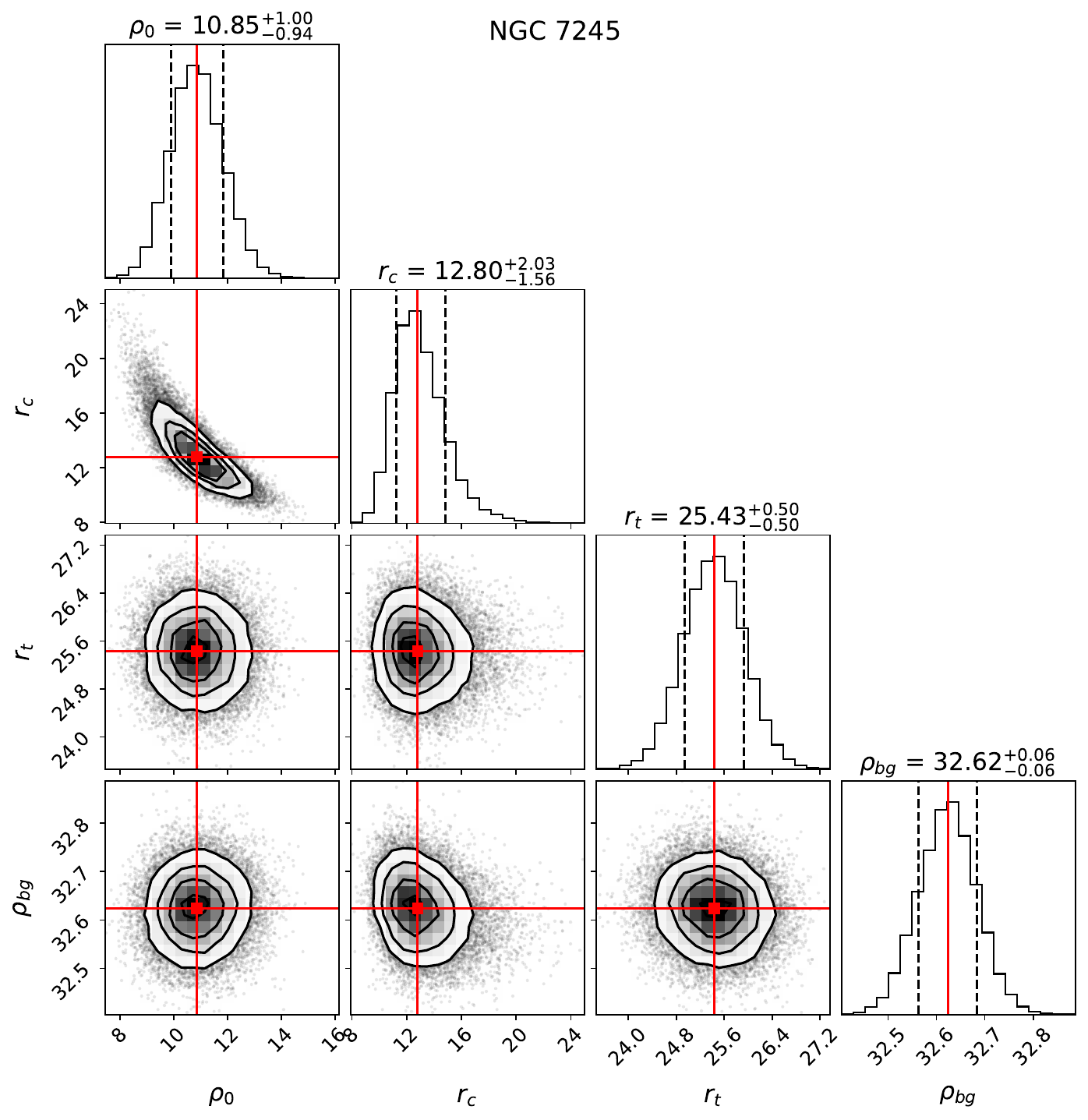}\\
        \includegraphics[width=0.23\linewidth]{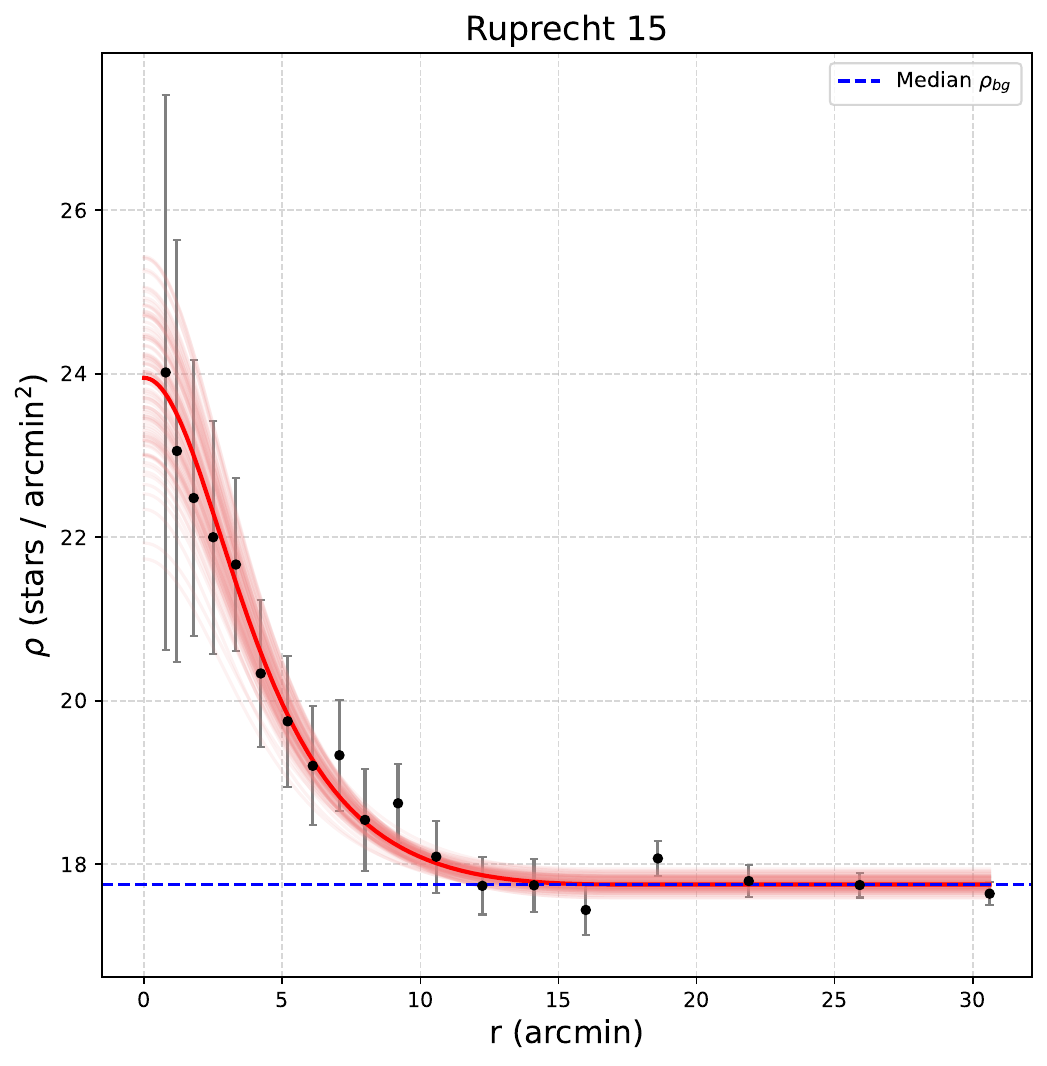}
    \includegraphics[width=0.23\linewidth]{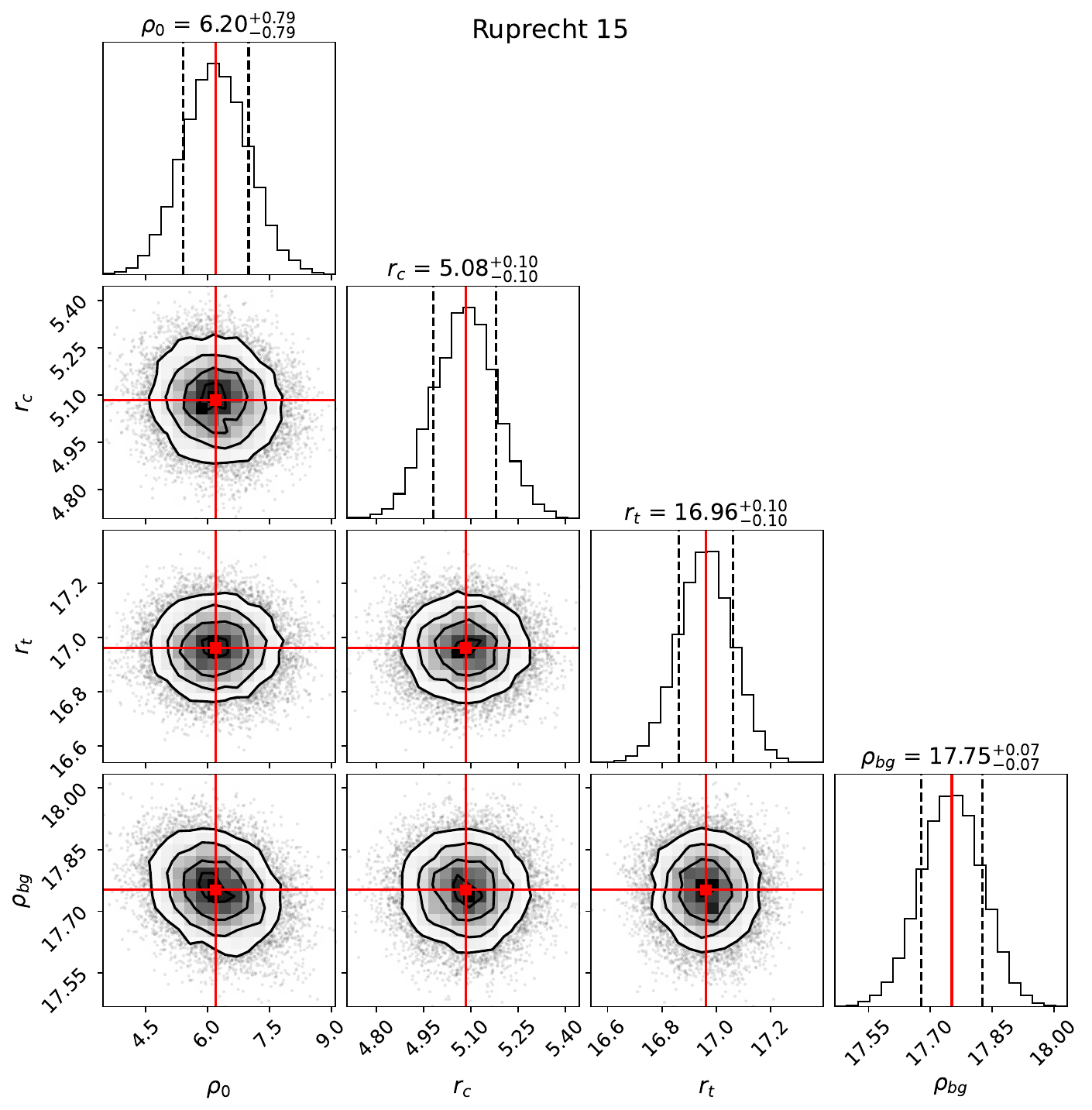}
        \includegraphics[width=0.23\linewidth]{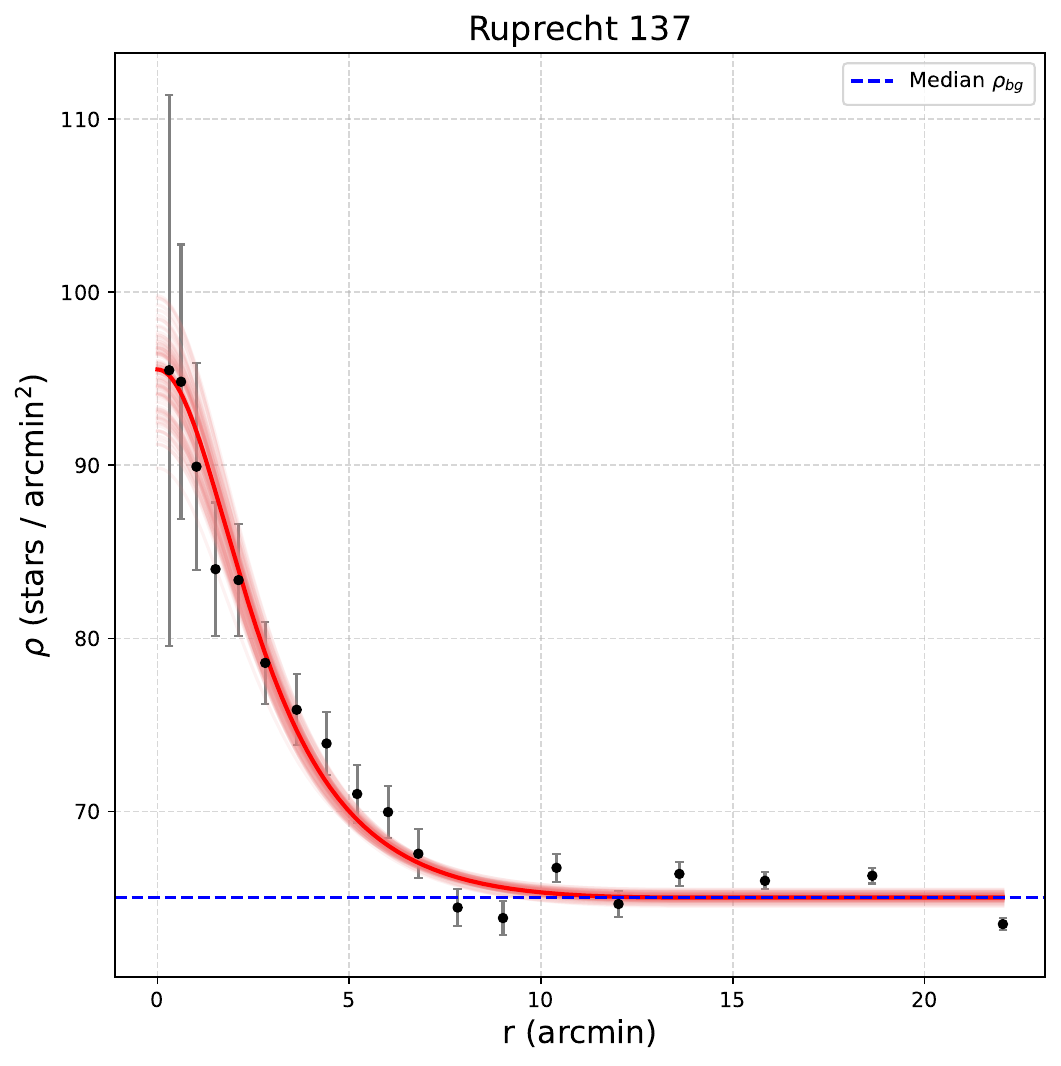}
    \includegraphics[width=0.23\linewidth]{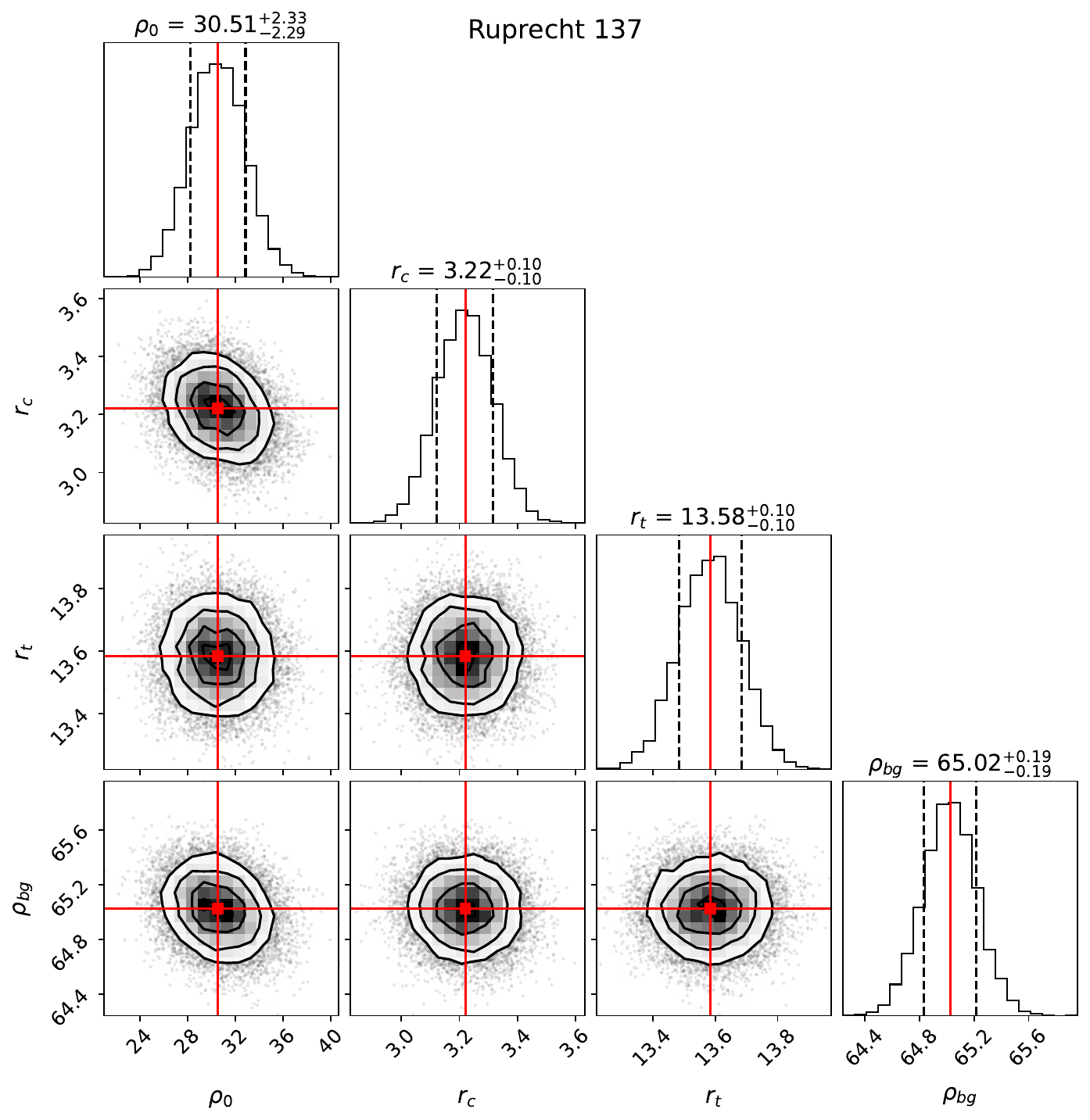}\\
        \includegraphics[width=0.23\linewidth]{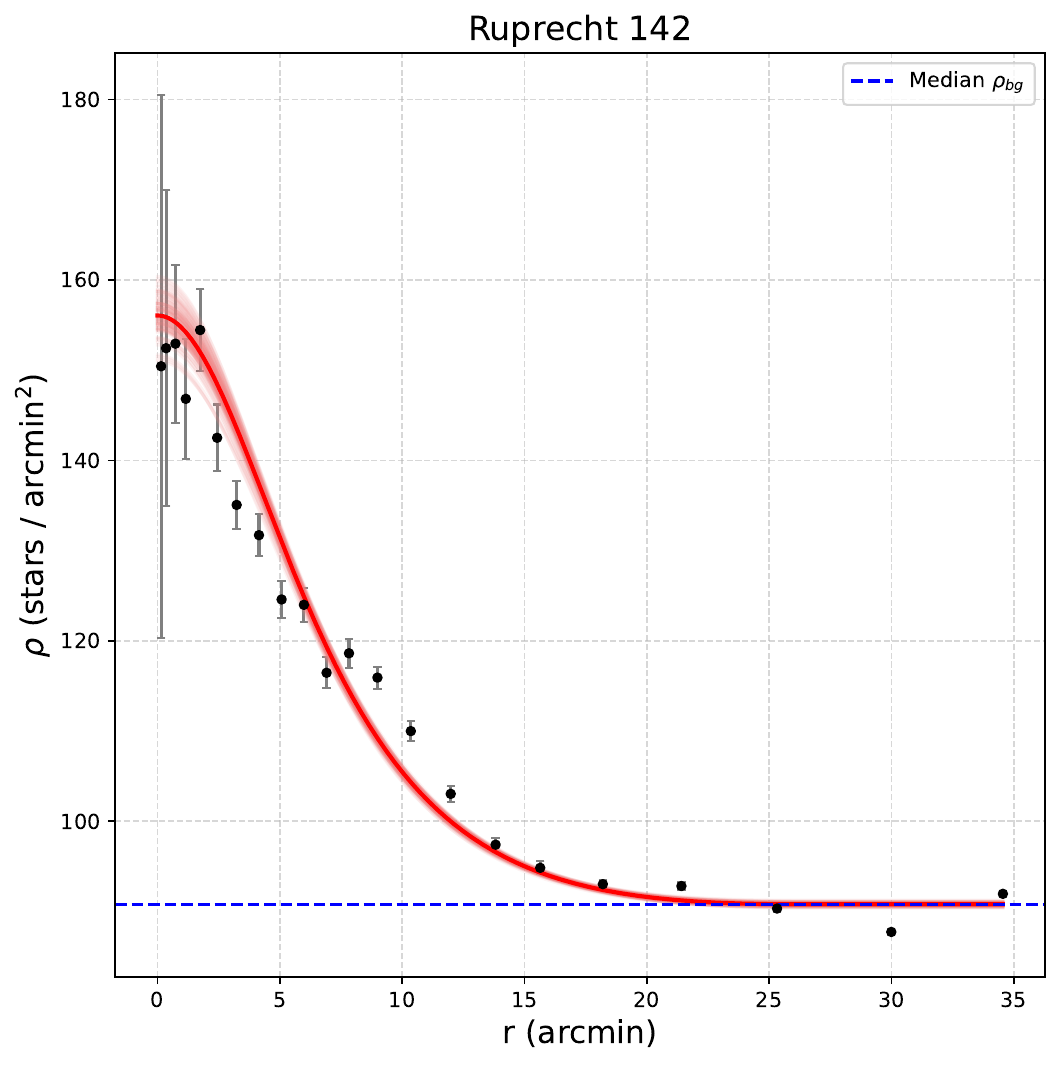}
    \includegraphics[width=0.23\linewidth]{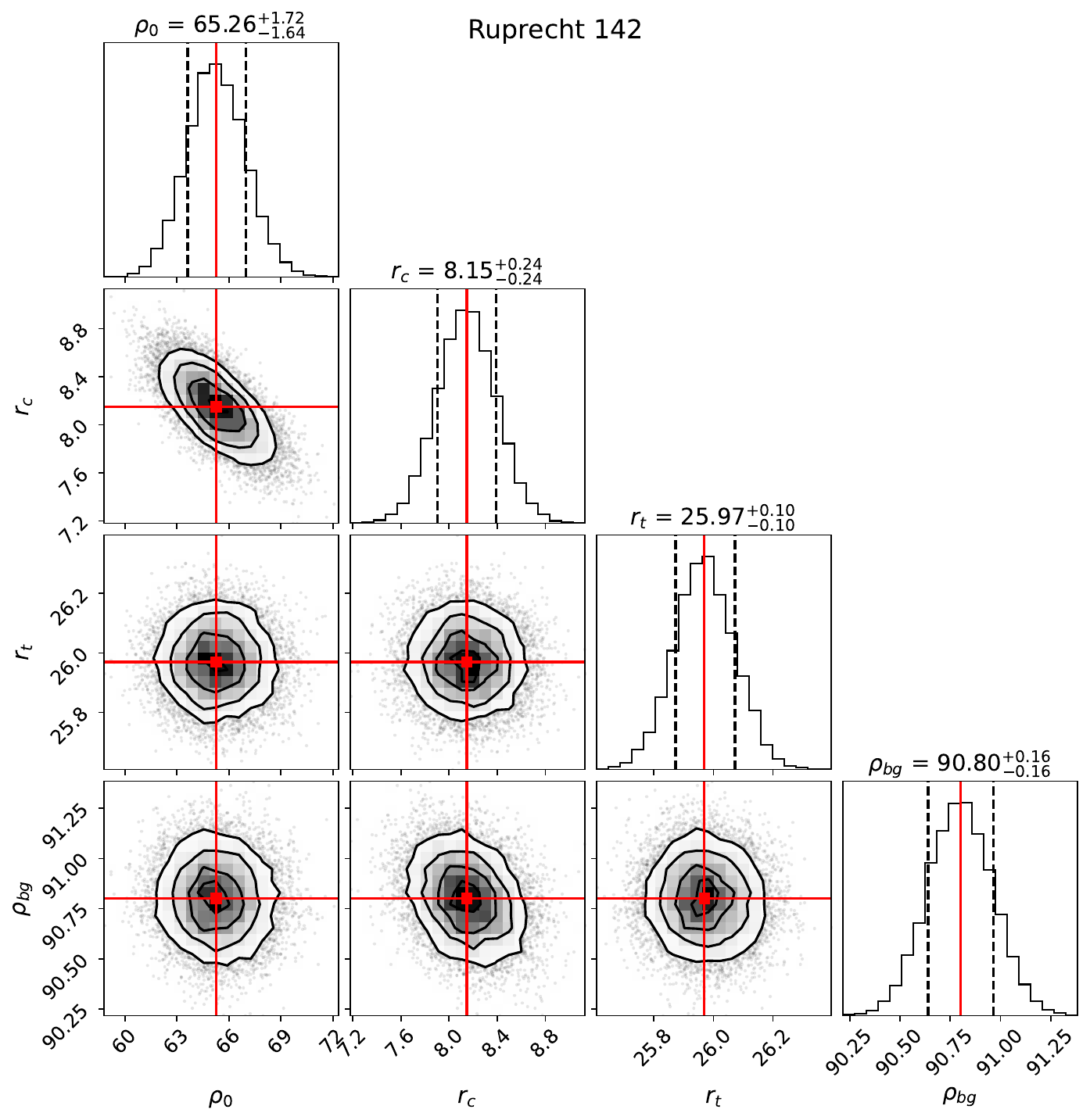}
        \includegraphics[width=0.23\linewidth]{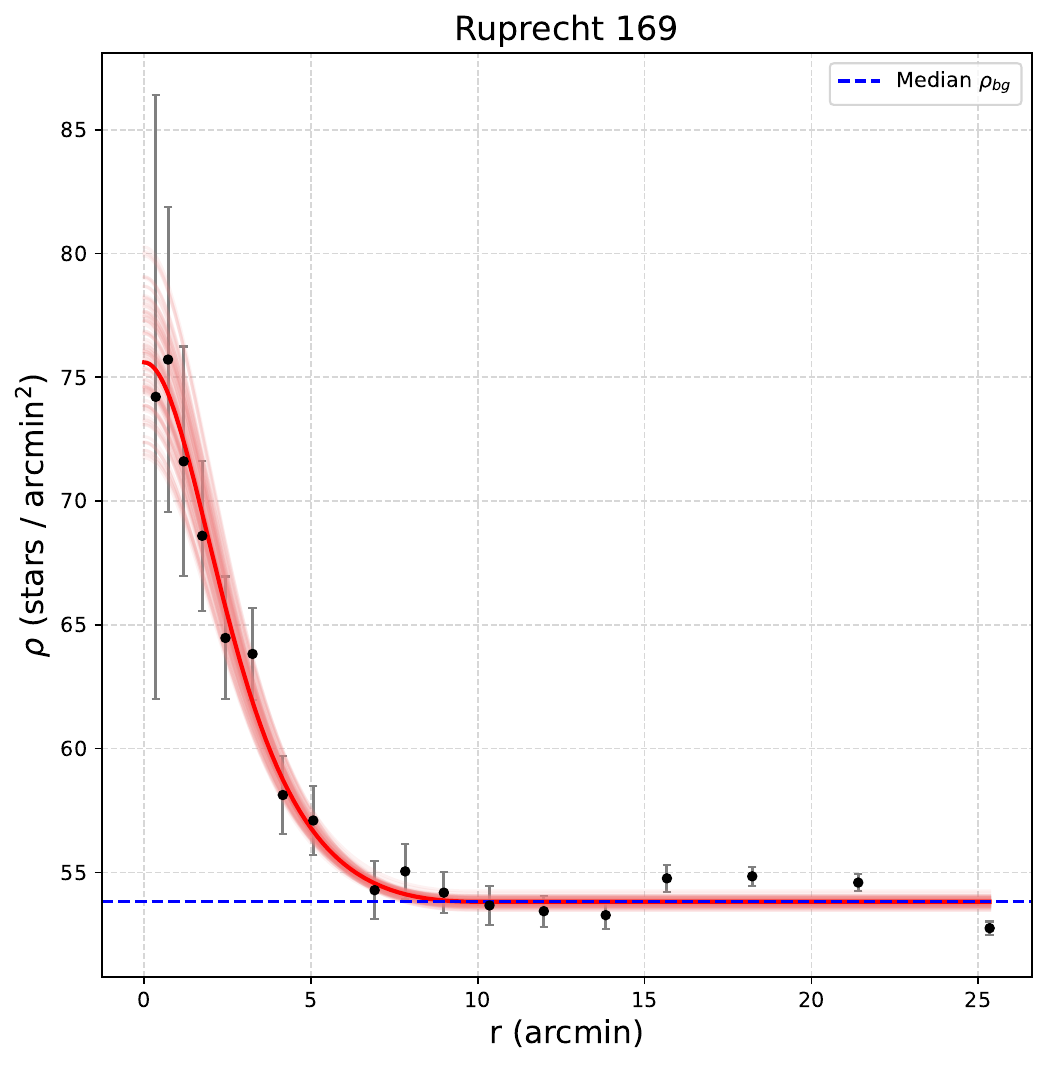}
    \includegraphics[width=0.23\linewidth]{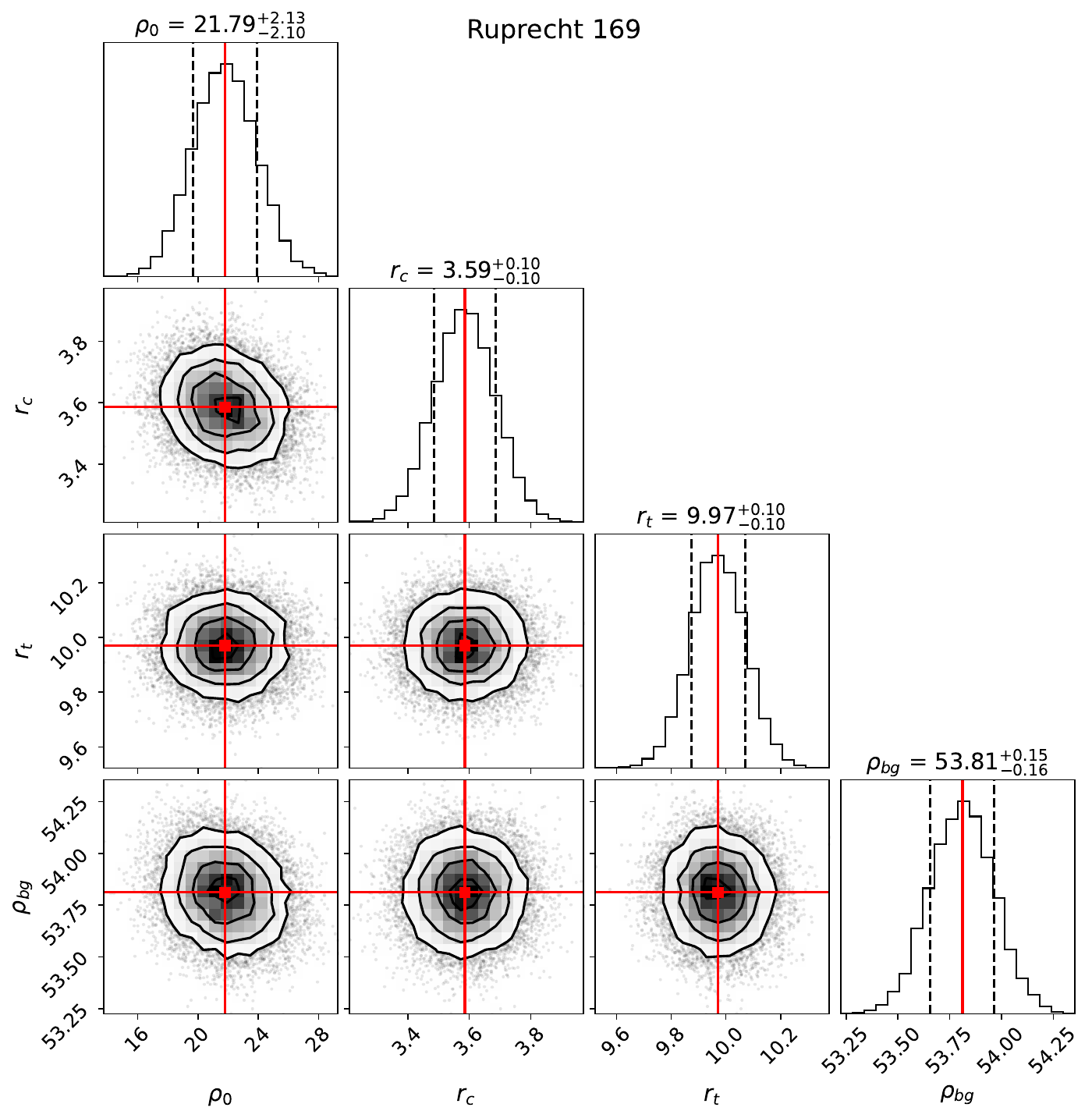}\\
    \caption{The same as Figure \ref{rdps} for  NGC 2141, NGC 7245, Ruprecht 15, Ruprecht 137, Ruprecht 142, and Ruprecht 169 OCs.}
    \label{rdps2}
\end{figure*}

\end{document}